\documentclass[12pt]{article}

\pdfoutput=1
\usepackage[utf8]{inputenc}
\usepackage{dsfont}
\usepackage{diagbox}
\usepackage{amsfonts}
\usepackage{mathtools}
\allowdisplaybreaks[4]        
\usepackage{amssymb}
\usepackage{euscript}     
\usepackage{braket}
\usepackage{starfont}
\usepackage{color,soul}         
\usepackage{tensor}        
\usepackage{amsthm}
\usepackage{graphicx}
\usepackage{slashed}
\usepackage{leftidx}
\usepackage{subfigure}
\usepackage{bbm}
\definecolor{outerspace}{rgb}{0.25, 0.29, 0.3}
\definecolor{scarlet}{rgb}{1.0, 0.13, 0.0}
\usepackage[header,title,page,titletoc]{appendix}  
\definecolor{princetonorange}{rgb}{1.0, 0.56, 0.0}
\definecolor{WildStrawberry}{rgb}{1.0, 0.26, 0.64}
\definecolor{rossocorsa}{rgb}{0.83, 0.0, 0.0}
\definecolor{navyblue}{rgb}{0.0, 0.0, 0.5}
\usepackage[numbers,sort&compress]{natbib}  
\usepackage{float}
\usepackage[paper=letterpaper,margin=1in]{geometry}
\usepackage{physics}

\newcommand{\bea}{\begin{eqnarray}}
\newcommand{\eea}{\end{eqnarray}}
\newcommand{\be}{\begin{equation}}
\newcommand{\ee}{\end{equation}}

\newtheorem{theorem}{Theorem}[section]

\usepackage{hyperref}
\hypersetup{
    colorlinks,
    citecolor=rossocorsa,
    filecolor=navyblue,
    linkcolor=navyblue,
    urlcolor=navyblue
}

\begin{document} 

\begin{titlepage}

\begin{center}

\phantom{ }
\vspace{3cm}

{\bf \Large{Optimal symmetry operators}}
\vskip 0.5cm
 Leandro Martinek${}^{\text{\Zeus}}$
\vskip 0.05in
\textit{Instituto Balseiro, Centro At\'omico Bariloche}
\vskip -.4cm
\textit{ 8400-S.C. de Bariloche, R\'io Negro, Argentina}

\begin{abstract}
    We present a constructive method to maximize the expectation value of operators that implement a symmetry on a subsystem, making use of modular tools. More generally, we study the positive cones associated with a von Neumann algebra, as defined by Araki \cite{SomePropertiesModularConjugation}. Given a reference vector, an algebra, and a state on the algebra, the purification of the state in the cone $\alpha = 0$, associated with the reference vector and the algebra, yields the unique vector whose overlap with the reference vector is maximal among all possible purifications.
    This establishes that the supremum in Uhlmann's theorem \cite{Uhlmann:1975kt} is uniquely attained by this vector, thereby providing the fidelity between the given state and the state obtained by restricting the reference vector to the algebra. Moreover, this purification can be explicitly constructed using modular tools.
    In addition, given an automorphism of the algebra, we show how to construct isometries implementing the automorphism using the positive cones. We prove that the isometry constructed from the cone $\alpha = 0$ is the one with maximal expectation value among all possible isometries implementing the automorphism.
    We illustrate these ideas with two simple examples: one involving a system of two spins, and the other in the theory of the massless scalar field in $3+1$ dimensions.
\end{abstract}
\end{center}

\small{\vspace{5cm}\noindent
${}^{\text{\Zeus}}$leandro.martinek@ib.edu.ar}

\end{titlepage}
\setcounter{tocdepth}{2}
{\parskip = .2\baselineskip \tableofcontents}

\section{Introduction}
\label{S:introduction}

Generalized symmetries prove to be useful for the identification of phases in quantum field theories (QFT). The phase is identified through the behavior of the expectation value of a certain non-local operator associated with the symmetry, e.g. a Wilson loop in the limit of large radius. However, while the symmetry operation is uniquely defined, the operator that implements the symmetry is highly ambiguous. In particular, operators that implement a symmetry with arbitrarily small expectation values are always available. To produce an anambiguous definition of order parameter one should select, given a region $R$ where the operator is localized, operators that are determined by the theory itself and the geometry, without any other external inputs. A natural choice corresponds to the non-local operator which has maximal expectation value; i.e. the \textit{optimal symmetry operator} in this sense, available in the class of operators that implement a given symmetry. 

Another approach to order parameters focuses on {\sl entropic order parameters}  associated with a region $R$ of the space \cite{casini2021entropic, Pedro}. The entropic order parameter uses the relative entropy to compare the vacuum state on the additive algebra of $R$ with the same state but in the maximal algebra of $R$, which includes the non-local operators. Upper and lower bounds of these entropic order parameters are directly related with the expectation value of the non-local operators. Hence, another application to obtaining non-local operators with maximal expectation values is to provide the tighter possible bounds.

A prominent example of the non-local operators appearing in this context are the \textit{twist} operators. They are defined to implement a global symmetry over a bounded region of the space while commuting with operators localized in the complement of a bigger region. One famous twist operator is the replica twist $\tau^{(n)}_{R}$ 
corresponding to the cyclic permutation symmmetry of a replicated QFT.  
Their expectation values compute Renyi entropies $S^{(n)}(R)$ associated with region $R$ of the space-time  through the formula 
$e^{-(n-1) S^{(n)}(R)} = \expval{\tau^{(n)}_{R}}{\Omega}$ \cite{Cardy:2007mb}. 
A rigorous definition of a twist requires two regions of the space, the region $R$ where we want to apply the twist, and the outside region $\bar{R}$ where the twist acts as the identity. Between $R$ and $\bar{R}$ there has to be some corridor in order to have a well defined operator \cite{doplicher1982local,Doplicher:1983if}. With this geometrical configuration, a standard split can be used to construct a particular twist  \cite{StandarandSplit,OnNoetherTheorem}. The entropy of the vacuum state in this standard split was called reflected entropy in \cite{dutta2021canonical}. Renyi reflected entropies can also be considered. 
Maximization of the expectation value of the Renyi replica twist will give a {\sl Renyi entropy of purification}, an analogous to the \textit{entropy of purification} (EoP) between $R$ and $\bar{R}$ \cite{Terhal:2002riz, Nguyen:2017yqw}. This quantity has been actively studied and is argue to be the holographic dual of a quantity called \textit{entanglement wedge cross section} \cite{umemoto2018entanglement}. The EoP performs a minimization over all possible purifications of a state, and it may seem difficult to arrive to a simple analytic result. This contrasts with it's holographic dual, that is a quantity with a simple geometric formula. The clarification of this issue gives another motivation to study the maximization of expectation value of twist operators.

In this work, we consider the general problem of the maximization of the expectation value of a symmetry operator in the algebraic approach to QFT. The general setup considers a Hilbert space $\mathcal{H}$, a von Neumman algebra $\mathcal{A} \subset \mathcal{B}(\mathcal{H})$ that acts on this Hilbert space and is usually associated to a region of spacetime, and some vector $\ket{\Omega}$, cyclic and separating for $\mathcal{A}$, that we will call the vacuum state. If we have some automorphism $\beta: \mathcal{A} \rightarrow \mathcal{A}$ of the algebra, it is a known result that we can always construct a \textit{standard} unitary representative $U(\beta) \in \mathcal{B}(\mathcal{H})$ that implements the automorphism over the operators of the algebra \cite{haag2012local, bratteli1979operator}. The unitary implementing the automorphism is highly non unique because we can modify this operator with a unitary   $Q^{\prime} \in \mathcal{A}^{\prime}$. The new operator $\widetilde{U}(\beta) = Q^{\prime} U(\beta)$ also implements the same automorphism over $\mathcal{A}$. More generally, we will call representatives of the automorphism to the isometries that belongs to the set
\begin{equation}
    \mathcal{I}_{\mathcal{A}}(\beta) = \{ U(\beta) \in \mathcal{B}(\mathcal{H}) \ | \ U(\beta) A = \beta(A) U(\beta) \ \   A \in \mathcal{A}, \ U(\beta)^{\dagger} U(\beta) = I \} \ .
\end{equation}
The construction of the standard representative use the notion of the \textit{natural} or \textit{standard} positive cone $\mathcal{P}_{\Omega}^{1/4}(\mathcal{A})$ associated the the vacuum vector $\ket{\Omega}$ and the algebra $\mathcal{A}$. 
In general, the positive cones $\mathcal{P}^{\alpha}_{\Omega}(\mathcal{A})$ are set of vectors \cite{SomePropertiesModularConjugation}
\begin{equation}
    \mathcal{P}^{\alpha}_{\Omega}(\mathcal{A}) = 
    \overline{\{ \Delta_{\Omega}^{\alpha} P \ket{\Omega} ; P \in \mathcal{A}^{+} \}} ,
     \quad \alpha \in \left[0 , 1/2 \right],
\end{equation}
where $\Delta_{\Omega}$ is the modular operator associated to $\ket{\Omega}$ and $\mathcal{A}$ (see Appendix \ref{A:RelativeModular}) and $\mathcal{A}^{+}$ is the set of positive elements of $\mathcal{A}$. These cones has several properties, the main one that we will used in this work is a theorem given in \cite{PositiveConesAssociated} which ensure that, given a normal state $\phi \in \mathcal{A}_{*}^{+}$ exist a unique purification $\ket{\phi_{\alpha}}$ of this state in each cone. This means that there exists a unique $\ket{\phi_{\alpha}} \in \mathcal{P}^{\alpha}_{\Omega}(\mathcal{A})$ such that $\expval{A}{\phi_{\alpha}} = \phi(A)$ for all $A \in \mathcal{A}$, for every $\alpha \in \left[0, 1/4 \right]$.

In this work we give an explicit construction of representatives of an automorphism $\beta$ using the family of cones $\mathcal{P}_{\Omega}^{\alpha}(\mathcal{A})$. We construct a unique isometry $U_{\alpha}(\beta)$ such that $U_{\alpha}(\beta) \in \mathcal{I}_{\mathcal{A}}(\beta)$ and $U_{\alpha}(\beta) \ket{\Omega} \in \mathcal{P}_{\Omega}^{\alpha}(\mathcal{A})$. We will call this operator the representative of $\beta$ over the cone $\mathcal{P}_{\Omega}^{\alpha}(\mathcal{A})$. This generalizes the construction of the standard representative given in \cite{haag2012local,bratteli1979operator}.
One of the main result of this work is that the operator with maximal expectation value is the representative of the automorphism constructed using the cone $\alpha = 0$ and we will call it \textit{optimal symmetry operator}. It achieves the supremum of the expectation values over all the possible representatives
\begin{equation}
    \text{Sup}_{U(\beta) \in \mathcal{I}_{\mathcal{A}}(\beta)} \abs{\expval{U(\beta)}{\Omega}} = \expval{U_{0}(\beta)}{\Omega} \ .
\end{equation}
Moreover, we prove a related result concerning this maximization. If we have two states $\omega, \phi \in \mathcal{A}_{*}^{+}$ \footnote{A \textit{state} is a functional $\omega: \mathcal{A} \rightarrow \mathbb{C}$ such that is positive $\omega(A^{\dagger}A) \geq 0$ and it is normalized $\omega(I) = 1$. The set of positive functionals is usually called $\mathcal{A}_{*}^{+}$.} Uhlmann's theorem \cite{Uhlmann:1975kt, alberti_note_1983} implies the fidelity can be written as
\begin{equation}
    F(\omega, \phi) = \text{Sup}_{\ket{\psi}} \abs{\braket{\Omega}{\psi}} \ ,
\label{eq:fidelityintro}
\end{equation}
where the supremum is taken over all possible purifications $\ket{\psi}$ of $\phi$, and where  $\ket{\Omega}$ is a purification of $\omega$. If we suppose that the states are normal and that we can purify $\omega$ vector $\ket{\Omega}$ which is cyclic and separating for $\mathcal{A}$, then the purification $\ket{\phi_{0}} \in \mathcal{P}_{\Omega}^{0}(\mathcal{A})$ of $\phi$ is the unique one that achieves the supremum in \eqref{eq:fidelityintro}. Also, we derive an explicit form of the Fidelity in terms of the modular tools
\begin{equation}
    F(\omega, \phi) = \braket{\Omega}{\phi_{0}} = \expval{\sqrt{\Delta_{\Omega}^{1/2} \Delta_{\phi,\Omega}\Delta_{\Omega}^{1/2}}}{\Omega}, \quad \ket{\phi_{0}} \in \mathcal{P}_{\Omega}^{0}(\mathcal{A}) ,\label{fidelidad}
\end{equation}
where $\Delta_{\phi,\Omega}$ is the relative modular operator associated to $\mathcal{A}$, the vector $\ket{\Omega}$ and the purification of $\phi$ in the standard cone $\ket{\phi} \in \mathcal{P}_{\Omega}^{1/4}(\mathcal{A})$.

One important technical tool for this work is the partial isometry that connects purification over the cones. In \cite{PositiveConesAssociated} it was proved that given a normal state $\phi \in \mathcal{A}_{*}^{+}$ we can purify it to a unique vector representative $\ket{\phi_{\alpha}}$ such that $\ket{\phi_{\alpha}} \in \mathcal{P}_{\Omega}^{\alpha}(\mathcal{A})$. In \cite{ArakiMasuda1982} it was proved that every vector $\ket{\xi} \in \mathcal{H}$ has a polar decomposition $\ket{\xi} = u_{\alpha} \ket{\xi_{\alpha}}$ where $u_{\alpha} \in \mathcal{A}^{\prime}$ is a partial isometry and $\ket{\xi_{\alpha}} \in \mathcal{P}_{\Omega}^{\alpha}(\mathcal{A})$. Following these two works, we prove that the purification of the normal state $\phi$ over the cone $\ket{\phi_{\alpha}} \in \mathcal{P}_{\Omega}^{\alpha}(\mathcal{A})$ is connected by an isometry $R_{\alpha}$ to the one in the standard cone $\ket{\phi} \in \mathcal{P}_{\Omega}^{1/4}(\mathcal{A})$ as
\begin{equation}
    R_{\alpha} \ket{\phi} = \ket{\phi_{\alpha}} \in \mathcal{P}_{\Omega}^{\alpha}(\mathcal{A}) \ .
\end{equation}
This isometry arises in the polar decomposition of the operator
\begin{equation}
    J_{\Omega} \Delta_{\phi, \Omega}^{1/2} \Delta_{\Omega}^{1/2 - 2 \alpha} J_{\Omega} = R_{\alpha}^{\dagger} P_{\alpha} \ ,
\end{equation}
where $J_{\Omega}$ is the modular conjugation associated with $\ket{\Omega}$ and $\mathcal{A}$. Thus, purifications over the cones can be constructed using known modular tools. That the polar decomposition of this operator is related to the purification over the cones has already been shown in \cite{PositiveConesLSpaces}.

The structure of this work is as follows. In section \ref{S:positive_cones}, we introduce the positive cones, and illustrate their meaning in the finite dimensional case using the language of density matrices. Then, we prove for the general case that the purifications of a state over the cones are connected by an isometry constructed using modular tools.
In section \ref{S:FidelityP0}, we show that the fidelity can be computed using formula (\ref{fidelidad}).
In particular, we show that the supremum of the Uhlmann's theorem is achieved for the purification over the cone $\mathcal{P}_{\Omega}^{0}(\mathcal{A})$.
In section \ref{S:RepresentationAutomorphism}, we show how we can construct a unique representative of an automorphism over the different cones. Some particular cases are discussed, such as the standard representative, and we explain some simplifications in the construction for the case of an inner automorphism.
In section \ref{S:Best_Operator}, we study the representative with maximal expectation value and show it is given by the representative in the cone $\mathcal{P}_{\Omega}^{0}(\mathcal{A})$.
In sections \ref{S:ExampleSpins} and \ref{S:Scalar}, we give some examples of the maximization of the expectation value of an operator implementing an automorphism on a subalgebra. One example is in the finite dimensional case of two interacting spins, and the other in the theory of the massless scalar field in $3+1$ dimensions. In both examples, the optimal operator can be computed by alternative methods, and we confirm that it agrees with the representative in the cone $\alpha = 0$.

\newpage

\section{Positive cones}
\label{S:positive_cones}

Takesaki studied particular cases of positive cones in \cite{Takesaki:1970aki}, and Araki generalized the idea in \cite{SomePropertiesModularConjugation}. One of these cones, called the \textit{standard} or \textit{natural} cone, enjoys some useful properties that have been extensively used in the literature, for example, in the representation of an automorphism by a unitary \cite{bratteli1979operator, haag2012local}, in the construction of a twist operator with the split property \cite{buchholz1986noether}, or in the study of recovery channels, relative entropies, reflected entropies, and fidelities \cite{Hollands:2020owv, Faulkner:2020iou, dutta2021canonical}. Here we review the construction of these cones and show their expression in the finite-dimensional case. For more insight into the mathematical approach to positive cones associated with von Neumann algebras see \cite{SomePropertiesModularConjugation,PositiveConesAssociated,PositiveConesLSpaces,ArakiMasuda1982,Kosaki1981}.

\subsection{Definition and properties}
\label{Ss:cone_def_prop}

Given a Hilbert space $\mathcal{H}$, a von Neumann algebra $\mathcal{A} \subset \mathcal{B}(\mathcal{H})$ \footnote{We denoted by $\mathcal{B}(\mathcal{H})$ the set of all bounded linear operators acting on a Hilbert space $\mathcal{H}$.} and a cyclic and separating vector $\ket{\Omega}$, we define the positive cone $\mathcal{P}^{\alpha}_{\Omega}(\mathcal{A})$ as the closure of the set \cite{SomePropertiesModularConjugation}
\begin{equation}
    \mathcal{P}^{\alpha}_{\Omega}(\mathcal{A}) = 
    \overline{\{ \Delta_{\Omega}^{\alpha} P \ket{\Omega} ; P \in \mathcal{A}^{+} \}} , \quad \alpha \in \left[0, 1/2 \right] \ ,
\label{eq:defpositivecone}
\end{equation}
where $\Delta_{\Omega}$ is the modular operator associated with the vector $\ket{\Omega}$ and the algebra $\mathcal{A}$ \cite{haag2012local} (see appendix \ref{A:RelativeModular}), $\mathcal{A}^{+}$ is the subset of positive operators of $\mathcal{A}$, and the overline denotes the closure of the set. The vectors belonging to this cone have the following properties:

\begin{enumerate}
    \item $\Delta_{\Omega}^{it} \mathcal{P}^{\alpha}_{\Omega}(\mathcal{A}) = \mathcal{P}^{\alpha}_{\Omega}(\mathcal{A})$ ,
    \item If $\ket{\xi} \in \mathcal{P}^{\alpha}_{\Omega}(\mathcal{A})$ then $\ket{\xi}$ is in the domain of $\Delta_{\Omega}^{1/2-2\alpha}$
    and $J_{\Omega} \ket{\xi} = \Delta_{\Omega}^{1/2-2\alpha} \ket{\xi}$ ,
    \item $J_{\Omega} \mathcal{P}^{\alpha}_{\Omega}(\mathcal{A}) = \mathcal{P}^{1/2 - \alpha}_{\Omega}(\mathcal{A})$, 
    \item $\left(\mathcal{P}^{\alpha}_{\Omega}(\mathcal{A}) \right)^{\prime} = \mathcal{P}^{1/2 -\alpha}_{\Omega}(\mathcal{A})$,
\end{enumerate}
where the dual of the cone is defined as
\begin{equation}
    \left(\mathcal{P}^{\alpha}_{\Omega}(\mathcal{A}) \right)^{\prime} =
    \{ \ket{\psi} | \braket{\xi}{\psi} \geq 0 \ \forall \ket{\xi} \in \mathcal{P}^{\alpha}_{\Omega}(\mathcal{A}) \} \ .
\end{equation}
The modular tools associated with the commutant algebra $\mathcal{A}^{\prime}$ are related with the modular tools of $\mathcal{A}$ as
\begin{equation}
    S_{\Omega}^{\prime} = J_{\Omega} S_{\Omega} J_{\Omega}, \quad
    \Delta_{\Omega}^{\prime} = J_{\Omega} \Delta_{\Omega} J_{\Omega} = \Delta_{\Omega}^{-1} \ .
\end{equation}
Then, the positive cones associated with $\mathcal{A}^{\prime}$ and $\ket{\Omega}$ are related to those associated with $\mathcal{A}$ and $\ket{\Omega}$ as
\begin{equation}
    \mathcal{P}^{\alpha}_{\Omega}(\mathcal{A}^{\prime}) = J_{\Omega} \mathcal{P}^{\alpha}_{\Omega}(\mathcal{A}) = \mathcal{P}^{1/2 - \alpha}_{\Omega}(\mathcal{A}) \ .
\end{equation}
For a more detailed exposition of the positive cones and their properties see \cite{SomePropertiesModularConjugation}.

Given a von Neumann algebra $\mathcal{A}$, a state is a linear functional $\phi: \mathcal{A} \rightarrow \mathbb{R}$
such that $\phi(A^{\dagger} A) \geq 0$ and $\phi(I) = 1$, where $I$ is the identity. The following theorem, proved in \cite{PositiveConesAssociated, ArakiMasuda1982}, states the uniqueness of the purification in the positive cones:
\begin{theorem}
        Given any normal state $\phi$ of the von Neumman algebra $\mathcal{A}$ there exists a unique purification $\ket{\phi_{\alpha}} \in \mathcal{P}^{\alpha}_{\Omega}(\mathcal{A})$ of $\phi$ with $\alpha \in [0, 1/4]$ such that $\expval{A}{\phi_{\alpha}} = \phi(A)$ for all $A \in \mathcal{A}$.
\label{T:purificationcones}
\end{theorem}
The condition that a state is normal ensures that it can be represented as a vector in the Hilbert space.
This result is the starting point of this work. We will study the question of which operator connects the purification on different cones. In the following section, we will review the expression of the positive cones in the finite-dimensional case to better understand it in the simpler language of density matrices. We will also prove the main results in this context as it is less technical and easier to understand. In section \ref{sS:connecting_cones} we will give the proof of the main results in the general case using modular tools.

\subsection{Understanding positive cones: finite dimensional case}
\label{sS:cones_finite}

For a generalization of this approach, see \cite{PositiveConesAssociated}.
Suppose that we have a finite-dimensional Hilbert space $\mathcal{H}$ of dimension $d$ and a density matrix $\rho$, with no zero eigenvalues. We can purify $\rho$ in the duplicated Hilbert space $\mathcal{H} \otimes \mathcal{H}$ by the vector
\begin{equation}
    \ket{\sqrt{\rho}} = \sum_{i = 1}^{d} \sqrt{\rho_{i}} \ket{v_{i}}\ket{v_{i}} \ ,
\label{eq:def_pur_rho}
\end{equation}
where $\rho_{i}$ and $\ket{v_{i}}$ are the eigenvalues and eigenvectors, respectively, of $\rho$. The condition that all the eigenvalues are non-zero guarantees that the vector $\ket{\sqrt{\rho}}$ is cyclic and separating for the algebra $\mathcal{B}(\mathcal{H}) \otimes I$. 

We will introduce some notation before continuing. If we have an operator $A$ that acts on $\mathcal{H}$, we can identify it with a vector in the duplicated space $\mathcal{H} \otimes \mathcal{H}$. Writing $A$ in a basis $\{ \ket{v_{i}} \}$ of $\mathcal{H}$, we define the \textit{vectorization} of $A$ as \cite{Watrous_2018}
\begin{equation}
    A = \sum_{i,j=1}^{d} A_{ij} \ketbra{v_{i}}{v_{j}}
    \rightarrow
    \ket{A} = \sum_{i,j=1}^{d} A_{ij} \ket{v_{i}} \overline{\ket{v_{j}}} \ ,
\end{equation}
where $\overline{\ket{v_{j}}}$ is the vector with its components conjugated in the canonical basis of the space $\mathcal{H}$ and $A_{ij}$ are the matrix elements of $A$ in the orthonormal basis $\{ \ket{v_{i}} \}$. Additionally, every vector in $\mathcal{H} \otimes \mathcal{H}$ can be identified with an operator acting on $\mathcal{H}$ in the converse way.
Therefore, we have a bijection between operators acting on $\mathcal{H}$ and vectors in $\mathcal{H} \otimes \mathcal{H}$.
This notation has the following properties:
\begin{equation}
    (B \otimes I) \ket{A} = \ket{BA} , \quad
    (I \otimes B) \ket{A} = \ket{AB^{T}} , \quad
    \braket{A}{B} = \Tr_{\mathcal{H}}(A^{\dagger}B) .
\end{equation}
With this, it is easy to check that $\ket{\sqrt{\rho}}$ is a purification of $\rho$ in the duplicated Hilbert space. We now take $\ket{\Omega} = \ket{\sqrt{\rho}}$ as the reference vector, the algebra $\mathcal{A} = \mathcal{B}(\mathcal{H}) \otimes I$, and construct the positive cones associated with this pair.
In this case the modular operators associated with $(\ket{\Omega}, \mathcal{A})$ are well known \cite{Witten:2018zxz,haag2012local}
\begin{equation}
    \Delta = \rho \otimes \rho^{-1}, \quad J ( Q \otimes I) J = (I \otimes Q^{*}) \ .
\label{eq:modularfinite}
\end{equation}
\indent Then, what does it mean to belong to the positive cone $\mathcal{P}^{\alpha}_{\Omega}(\mathcal{A})$?
We will give a different definition in terms of the operators representing the vectors, rather than the vectors themselves. This definition was first given in \cite{PositiveConesAssociated}. For $\alpha \in [0, 1/4]$, we have that
\begin{equation}
    \mathcal{P}^{\alpha}_{\Omega}(\mathcal{A}) =
    \{ \ket{A} \in \mathcal{H} \otimes  \mathcal{H} : \rho^{1/2 - 2 \alpha} A \geq 0, 
    \ A \in \mathcal{B}(\mathcal{H}) \} .
\label{eq:def_pos_cone_finite}
\end{equation}
We will prove that this equality holds. First, we will show that the vectors in the positive cone satisfy the property. Suppose we have $\ket{\xi} \in \mathcal{P}^{\alpha}_{\Omega}(\mathcal{A})$, then there exists a positive operator $P \geq 0$ such that $(P \otimes I) \in \mathcal{A}$ and
\begin{equation}
   \ket{\xi} = \Delta^{\alpha} (P \otimes I) \ket{\Omega}
   = \ket{\rho^{\alpha} P \rho^{1/2-\alpha}}.
\end{equation}
If we take $A = \rho^{\alpha} P \rho^{1/2-\alpha}$
\begin{equation}
    \rho^{1/2- 2\alpha} A = \rho^{1/2 - \alpha} P \rho^{1/2-\alpha} \geq 0 . 
\end{equation}
Then we conclude that $\mathcal{P}^{\alpha}_{\Omega}(\mathcal{A})$ is included in the definition \eqref{eq:def_pos_cone_finite}.
On the other hand, if we have a vector $\ket{A}$ such that the operator $A$ satisfies $\rho^{1/2 - 2 \alpha} A \geq 0$, we can take a vector in the cone $\ket{\psi} \in \mathcal{P}^{1/2 - \alpha}_{\Omega}(\mathcal{A})$ of the form $\ket{\psi} = \ket{\rho^{1/2 - \alpha} Q \rho^{\alpha}}$, with $Q \in \mathcal{B}(\mathcal{H})$ and $Q \geq 0$. Therefore we have that
\begin{equation}
    \braket{\psi}{A} 
    = \braket{\rho^{1/2 - \alpha} Q \rho^{\alpha}}{A}
    = \Tr \left( \rho^{\alpha} Q \rho^{1/2 - \alpha}  A \right)
    = \Tr \left( \rho^{\alpha} Q \rho^{\alpha} \rho^{1/2 - 2 \alpha}  A \right) \geq 0.
\end{equation}
Because this holds for every $\ket{\psi} \in \mathcal{P}^{1/2 - \alpha}_{\Omega}(\mathcal{A})$ we have that $\ket{A}$ belongs to the dual cone of $\mathcal{P}^{1/2 - \alpha}_{\Omega}(\mathcal{A})$ which is $\mathcal{P}^{\alpha}_{\Omega}(\mathcal{A})$. Thus, we have equality between the two sets. Therefore, the positive cones are sets of operators that satisfy particular positivity conditions. It is interesting to note that the \textit{standard} or \textit{natural} cone, which corresponds to $\alpha = 1/4$, is associated with the set of positive operators.

Now we study how we can purify another density matrix $\sigma$ on one of these cones. The answer is straightforward in this framework and is given by
\begin{equation}
    \ket{\sigma_{\alpha}} = \ket{\rho^{2 \alpha - 1/2} \sqrt{\rho^{1/2 - 2 \alpha} \sigma \rho^{1/2 - 2 \alpha}}} \in \mathcal{P}^{\alpha}_{\Omega}(\mathcal{A}).
\label{eq:sigmaalphafinite}
\end{equation}
To verify this, we first show that the vector belongs to the cone. Using the definition \eqref{eq:def_pos_cone_finite}, this is immediate.
\begin{equation}
\rho^{1/2 - 2 \alpha} (\rho^{2 \alpha - 1/2} \sqrt{\rho^{1/2 - 2 \alpha} \sigma \rho^{1/2 - 2 \alpha}}) = \sqrt{\rho^{1/2 - 2 \alpha} \sigma \rho^{1/2 - 2 \alpha}} \geq 0 
\Rightarrow \ket{\sigma_{\alpha}} \in \mathcal{P}^{\alpha}_{\Omega}(\mathcal{A}) .
\end{equation}
Now we need to check whether it is a purification of $\sigma$, which means that it reproduces the same expectation values in $\mathcal{A}$
\begin{equation}
\begin{split}
    \expval{(A \otimes I)}{\sigma_{\alpha}} 
    &=\Tr \left( \sqrt{\rho^{1/2 - 2 \alpha} \sigma \rho^{1/2 - 2 \alpha}} \rho^{2 \alpha - 1/2} A \rho^{2 \alpha - 1/2} \sqrt{\rho^{1/2 - 2 \alpha} \sigma \rho^{1/2 - 2 \alpha}}\right) \\
    &=\Tr \left( \rho^{1/2 - 2 \alpha} \sigma  A \rho^{2 \alpha - 1/2}\right) =\Tr \left( \sigma  A \right) .
\end{split}
\end{equation}
Thus, we have an explicit purification of $\sigma$ in the cone $\mathcal{P}^{\alpha}_{\Omega}(\mathcal{A})$.

It is a known result that different purifications of a density matrix are connected by unitary operators acting on the commutant algebra. In this case, we can show an explicit form of the unitary that connects purification in different cones. Suppose that $\sigma$ has no zero eigenvalues, and we purify it in the standard cone, giving the purification $\ket{\sqrt{\sigma}}$. To move from this purification to another in a different cone, we can use the unitary operator
\begin{equation}
    W_{\alpha} = \sqrt{\rho^{1/2 - 2 \alpha} \sigma \rho^{1/2 - 2 \alpha}} \rho^{2 \alpha - 1/2} \sigma^{-1/2}  ,
    \quad
    W_{\alpha}^{\dagger} W_{\alpha} = W_{\alpha} W_{\alpha}^{\dagger} = I \ .
\label{eq:R_def}
\end{equation}
Defining $R_{\alpha} = (I \otimes W_{\alpha}^{*}) \in \mathcal{A}^{\prime}$ we have that
\begin{equation}
    R_{\alpha} \ket{\sqrt{\sigma}} = \ket{\sqrt{\sigma}_{\alpha}}.
\end{equation}

\subsection{Connecting the cones}
\label{sS:connecting_cones}

In \cite{PositiveConesAssociated} it is shown that every normal state $\phi \in \mathcal{A}_{*}^{+}$ can be purified in a unique way to a vector in the positive cones $\mathcal{P}^{\alpha}_{\Omega}(\mathcal{A})$ for $\alpha \in \left[0, 1/4 \right]$. In \cite{ArakiMasuda1982} it is shown that every vector $\ket{\xi} \in \mathcal{H}$ always has the polar decomposition $\ket{\xi} = u_{\alpha} \ket{\xi_{\alpha}}$ where $u_{\alpha} \in \mathcal{A}^{\prime}$ is a partial isometry and $\ket{\xi_{\alpha}} \in \mathcal{P}^{\alpha}_{\Omega}(\mathcal{A})$ for $\alpha \in \left[0, 1/4 \right]$. In this section, we combine these ideas and techniques to get a constructive way to find a partial isometry $R_{\alpha}$ which connects the purification of $\phi$ over the different cones. This partial isometry can be constructed using modular tools.

In \eqref{eq:R_def} we give an explicit form for the operator $R_{\alpha}$ which connects the purification on the standard cone to a purification in another cone $\alpha$. From its explicit form it is easy to understand how it can be constructed with modular tools. Suppose that we are in the finite dimensional case again. Let
\begin{equation}
    \Delta_{\Omega} = \rho \otimes \rho^{-1}, \quad \Delta_{\phi, \Omega} = \sigma \otimes \rho^{-1} \ ,
\label{eq:relativemodularfinite}
\end{equation}
be the modular and the relative modular operator associated with $\ket{\Omega}$, $\ket{\phi}$ and the algebra $\mathcal{A}$, $\rho$ and $\sigma$ are the reduced states of $\ket{\Omega}$ and $\ket{\phi}$ over the algebra, respectively. For simplicity, we will suppose that $\rho$ and $\sigma$ are invertible. Then, if we construct the operator
\begin{equation}
    H_{\alpha} = J \Delta_{\phi, \Omega}^{1/2} \Delta_{\Omega}^{1/2-2\alpha} J = ( \rho^{-1 + 2 \alpha} \otimes \sigma^{1/2} \rho^{1/2 - 2 \alpha} )^* \ ,
\label{eq:def_Talpha1}
\end{equation}
its polar decomposition is given by
\begin{equation}
    H_{\alpha} = R_{\alpha}^{\dagger} P_{\alpha} \ .
\end{equation}
The positive part has the form
\begin{equation}
\begin{split}
    &P_{\alpha}^{2} = H_{\alpha}^{\dagger} H_{\alpha} =  J \Delta_{\Omega}^{1/2-2\alpha} \Delta_{\phi, \Omega} \Delta_{\Omega}^{1/2-2\alpha} J
    = (\rho^{-2 + 4 \alpha} \otimes \rho^{1/2 - 2 \alpha} \sigma \rho^{1/2 - 2 \alpha})^{*} \\
    & \Rightarrow P_{\alpha} = J \sqrt{\Delta_{\Omega}^{1/2-2\alpha} \Delta_{\phi, \Omega} \Delta_{\Omega}^{1/2-2\alpha}} J 
    = (\rho^{-1 + 2 \alpha} \otimes \sqrt{\rho^{1/2 - 2 \alpha} \sigma \rho^{1/2 - 2 \alpha}} )^{*} \ ,
\end{split}
\label{eq:PositivePartT}
\end{equation}
where we found $P_{\alpha}$ using the fact that $P_{\alpha}^{2}$ has a unique positive root. Now, if we want to recover the unitary \footnote{It is unitary because the density matrices are invertible.} part from the polar decomposition, we can write it as
\begin{equation}
\begin{split}
    R_{\alpha} = P_{\alpha} H_{\alpha}^{-1}
    &= J \sqrt{\Delta_{\Omega}^{1/2-2\alpha} \Delta_{\phi, \Omega} \Delta_{\Omega}^{1/2-2\alpha}}  \Delta_{\phi, \Omega}^{-1/2} \Delta_{\Omega}^{-1/2+2\alpha} J \\
    &= I \otimes (\sqrt{\rho^{1/2 - 2 \alpha} \sigma \rho^{1/2 - 2 \alpha}} \rho^{2\alpha - 1/2} \sigma^{-1/2})^{*} \\
    &= I \otimes W_{\alpha}^{*} \ .
\end{split}
\label{eq:R_def_2}
\end{equation}
Then, we recover the same form as in \eqref{eq:R_def}. Thus, a good starting point is to define the operator $H_{\alpha}$ with the modular tools in the general case. H. Kosaki has already proved the connection between the polar decomposition of the operator $J \Delta_{\phi, \Omega}^{1/2} \Delta_{\Omega}^{1/2 - 2\alpha} J$ and the purification of the state $\phi$ on the positive cones~\cite{PositiveConesAssociated, PositiveConesLSpaces}. His work used mathematical tools that are not familiar to a physics audience. In \cite{ArakiMasuda1982}, the result of H. Kosaki was generalized to the more familiar setup of a Hilbert space where the von Neumann algebra acts. We will give a proof of the relevant result for this work using results from \cite{ArakiMasuda1982}.

Suppose that we have a Hilbert space $\mathcal{H}$, a von Neumann algebra $\mathcal{A} \subset \mathcal{B}(\mathcal{H})$, a cyclic and separating vector $\ket{\Omega}$ and a normal state $\phi \in \mathcal{A}_{*}^{+}$. Because the vacuum is cyclic and separating for $\mathcal{A}$, we can define the modular operator $\Delta = \Delta_{\Omega}$ and the modular conjugation $J = J_{\Omega}$ associated with this pair.

We will denote by $\mathcal{U}(\mathcal{A})$ the set of all operators $A \in \mathcal{A}$ such that
\begin{equation}
    \tau_{z}(A) = \Delta^{z} A \Delta^{-z} \ ,
\end{equation}
is a bounded operator analytic in $z \in \mathbb{C}$ (for a definition of analytic see \cite{bratteli1979operator} definition 2.5.20). It is easy to show that the set of operators $\mathcal{U}(\mathcal{A})$ is dense in $\mathcal{A}$. To see this, suppose $A \in \mathcal{A}$, we define a sequence of operators $A_{n} \in \mathcal{A}$ such that
\begin{equation}
    A_{n} = \sqrt{\frac{n}{\pi}} \int_{- \infty}^{\infty} \Delta^{it} A \Delta^{-it} e^{-n t^{2}} dt \Rightarrow \lim_{n \rightarrow \infty} A_{n} = A \ ,
\end{equation}
where the convergence is in the weak topology. When we evolve it with the modular flow at a complex time we arrive at (see Appendix \ref{A:ModularFlowImaginary})
\begin{equation}
    \tau_{z}(A_{n}) = \int_{- \infty}^{\infty} \Delta^{it } A \Delta^{-it} e^{-n (t+ i z)^{2}} dt \ .
\end{equation}
This is a bounded operator and is analytic in $z$ (see \cite{bratteli1979operator} Proposition 2.5.22 or \cite{StratilaZsido1979} section 9.17). Therefore, $A_{n} \in \mathcal{U}(\mathcal{A})$ and $A_{n} \rightarrow A$ for a given $A \in \mathcal{A}$, so $\mathcal{U}(\mathcal{A})$ is dense in $\mathcal{A}$. Because $\mathcal{U}(\mathcal{A})$ is dense in $\mathcal{A}$ and $\mathcal{A} \ket{\Omega}$ is dense in $\mathcal{H}$, we deduce that $\mathcal{U}(\mathcal{A}) \ket{\Omega}$ is also dense in $\mathcal{H}$. We note that if $A \in \mathcal{U}(\mathcal{A})$ then $\tau_{z}(A) \in \mathcal{U}(\mathcal{A}) \subset \mathcal{A}$.

The operator $H_{\alpha}$ defined in \eqref{eq:def_Talpha1} can be defined by its action on the dense set $\mathcal{U}(\mathcal{A}) \ket{\Omega}$
\begin{equation}
    H_{\alpha} A \ket{\Omega} = \tau_{2 \alpha - 1}(A) \ket{\phi}, \quad A \in \mathcal{U}(\mathcal{A}) \ .
\label{eq:def_Halpah2}
\end{equation}
But, for the following proof, we will define a slightly different operator, in order to use some theorems proven in \cite{ArakiMasuda1982}, and thus we obtain the operator $H_{\alpha}$.

First, let $\ket{\phi_{0}} \in \mathcal{P}_{\Omega}^{0}(\mathcal{A})$ be a purification of $\phi$. By property 2 mentioned in section \ref{S:positive_cones}, we have that $\ket{\phi_{0}} \in D(\Delta^{1/2})$, where $ D(\Delta^{1/2})$ is the domain of the unbounded operator $\Delta^{1/2}$. Therefore, we also have that  $\ket{\phi_{0}} \in D(\Delta^{1/2 - 2 \alpha})$ for $\alpha \in \left[0, 1/4 \right]$ (corollary 9.15 \cite{StratilaZsido1979}). The modular conjugation has the property that $J \Delta J = \Delta^{-1}$, so that $J \ket{\phi_{0}} \in D(\Delta^{1/2 - 2 \alpha^{\prime}})$ where $\alpha^{\prime} = 1/2 - \alpha$. Let $T_{\alpha^{\prime}}$ be defined as
\begin{equation}
    T_{\alpha^{\prime}} A^{\prime} \ket{\Omega} = \tau_{2 \alpha^{\prime}}(A^{\prime}) J\ket{\phi_{0}} \ ,
\end{equation}
where $A^{\prime} \in \mathcal{U}(\mathcal{A}^{\prime})$. Lemma 3.1 in \cite{ArakiMasuda1982} states that if $\ket{\phi_{0}} \in D(\Delta^{1/2 - 2 \alpha^{\prime}})$, then $T_{\alpha^{\prime}}$ is a closable operator.
Therefore, by the cyclic property of $\ket{\Omega}$, $T_{\alpha^{\prime}}$ is a densely defined closable operator, and then it has the polar decomposition (theorem 6.1.11 \cite{kadison_ringrose_vol2})
\begin{equation}
    T_{\alpha^{\prime}} = K_{\alpha^{\prime}}^{\dagger} P_{\alpha^{\prime}} \ ,
\end{equation}
where $P_{\alpha^{\prime}}$ is a positive operator and $K_{\alpha^{\prime}}$ is a partial isometry such that
\begin{equation}
    K_{\alpha^{\prime}} K_{\alpha^{\prime}}^{\dagger} = R(P_{\alpha^{\prime}}), \quad
    K_{\alpha^{\prime}}^{\dagger} K_{\alpha^{\prime}} = R(T_{\alpha^{\prime}}) = s^{\mathcal{A}}(J\phi_{0}) \ .
\end{equation}
Here, $R(P_{\alpha^{\prime}})$ and $R(T_{\alpha^{\prime}})$ are projections onto the range of the operators $P_{\alpha^{\prime}}$ and $T_{\alpha^{\prime}}$, respectively, and $s^{\mathcal{A}}(J\phi_{0}) \in \mathcal{A}$ is the projection onto the subspace $\overline{\mathcal{A}^{\prime} J\ket{\phi_{0}}}$. Lemma 3.3 in \cite{ArakiMasuda1982} states that $K_{\alpha^{\prime}} \in \mathcal{A}$, and that
\begin{equation}
    P_{\alpha^{\prime}} A^{\prime} \ket{\Omega} = \tau_{2 \alpha^{\prime}}(A^{\prime}) P_{\alpha^{\prime}} \ket{\Omega} \ , \quad A^{\prime} \in \mathcal{U}(\mathcal{A}^{\prime}) \ .
\end{equation}
Therefore, we have $R(P_{\alpha^{\prime}}) = s^{\mathcal{A}}(P_{\alpha} \ket{\Omega})$. Lemmas 3.4 and 3.5 in \cite{ArakiMasuda1982} state that
\begin{equation}
    K_{\alpha^{\prime}} J\ket{\phi_{0}} = P_{\alpha^{\prime}} \ket{\Omega} \in \mathcal{P}^{\alpha^{\prime}}_{\Omega}(\mathcal{A}) \ .
\end{equation}
This is the content of theorem 7 (1) in \cite{ArakiMasuda1982}, which state the polar decomposition of vectors in the domain of $D(\Delta^{1/2 - 2 \alpha^{\prime}})$ as a partial isometry in $\mathcal{A}$ and a vector in the positive cone $\mathcal{P}_{\Omega}^{\alpha^{\prime}}(\mathcal{A})$.

Now, if we define $N_{\alpha} = J K_{\alpha^{\prime}} J \in \mathcal{A}^{\prime}$, we have that
\begin{equation}
    \ket{\phi_{\alpha}} = N_{\alpha} \ket{\phi_{0}} = J P_{\alpha^{\prime}} J \ket{\Omega} \in \mathcal{P}_{\Omega}^{\alpha}(\mathcal{A}) \ .
\end{equation}
This comes from the fact that $J \mathcal{P}_{\Omega}^{\frac{1}{2} - \alpha}(\mathcal{A}) = \mathcal{P}_{\Omega}^{\alpha}(\mathcal{A})$ and that $\alpha^{\prime} = \frac{1}{2} - \alpha$. Moreover, because $N_{\alpha} \in \mathcal{A}^{\prime}$ and $N_{\alpha}^{\dagger} N_{\alpha} = J K_{\alpha}^{\dagger} K_{\alpha} J = J s^{\mathcal{A}}(J \phi_{0}) J = s^{\mathcal{A}^{\prime}}(\phi_{0})$, we have 
\begin{equation}
    \expval{A}{\phi_{\alpha}} = \expval{A}{\phi_{0}} = \phi(A) \ .
\end{equation}
Therefore, $\ket{\phi_{\alpha}}$ is the unique purification of $\phi$ in the cone $\mathcal{P}_{\Omega}^{\alpha}(\mathcal{A})$.

We want to better understand how to obtain the partial isometry $N_{\alpha}$. This operator arises in the polar decomposition of $J T_{\alpha^{\prime}} J = N_{\alpha}^{\dagger} J P_{\alpha^{\prime}} J$. This operator acts as
\begin{equation}
    J T_{\alpha^{\prime}} J A \ket{\Omega} = \tau_{- 2 \alpha^{\prime}}(A) \ket{\phi_{0}} = \tau_{2 \alpha  - 1}(A) \ket{\phi_{0}} \ .
\end{equation}
By its action on the dense domain $\mathcal{U}(\mathcal{A})\ket{\Omega}$, it can be identified with
\begin{equation}
    S_{\phi_{0}, \Omega} S_{\Omega} \Delta^{2 \alpha - 1} 
    = J T_{\alpha^{\prime}} J
    = N_{\alpha}^{\dagger} J P_{\alpha^{\prime}} J \ ,
\label{eq:Talpha_ident}
\end{equation}
where $S_{\phi_{0}, \Omega}$ is the relative conjugate linear operator associated with the reference vector $\ket{\Omega}$, the purification $\ket{\phi_{0}}$, and the algebra $\mathcal{A}$, and $S_{\Omega}$ is the conjugate linear operator associated with $\ket{\Omega}$ and $\mathcal{A}$ (for a definition, see appendix \ref{A:RelativeModular}). Therefore, because the operator $T_{\alpha^{\prime}}$ has a polar decomposition, the operator \eqref{eq:Talpha_ident} also has a polar decomposition, and we obtain the partial isometry $N_{\alpha}$ that connects the purification of $\phi$ in the cone $\mathcal{P}_{\Omega}^{0}(\mathcal{A})$ with the purification in the cone $\mathcal{P}_{\Omega}^{\alpha}(\mathcal{A})$ for $\alpha \in \left[0, 1/4 \right]$.

We would like to rewrite this in terms of the purification $\ket{\phi} \in \mathcal{P}_{\Omega}^{1/4}(\mathcal{A})$, in order to recover the definition of the operator $H_{\alpha}$ given in \eqref{eq:def_Halpah2}. From theorem 7 (2) in \cite{ArakiMasuda1982}, we know that there exists a partial isometry $R_{0} \in \mathcal{A}^{\prime}$ such that $\ket{\phi_{0}} = R_{0} \ket{\phi}$, with $R_{0} R_{0}^{\dagger} = s^{\mathcal{A}^{\prime}}(\phi_{0})$ and $R_{0}^{\dagger} R_{0} = s^{\mathcal{A}^{\prime}}(\phi)$.
Then, the operator
\begin{equation}
    H_{\alpha} = R_{0}^{\dagger} J T_{\alpha^{\prime}} J \ ,
\label{eq:Halpha_def3}
\end{equation}
is precisely the one that satisfies \eqref{eq:def_Halpah2}. From \eqref{eq:Halpha_def3}, and by the uniqueness of the polar decomposition, we conclude that $H_{\alpha}$ admits the polar decomposition
\begin{equation}
    H_{\alpha} = R_{0}^{\dagger} N_{\alpha}^{\dagger} J P_{\alpha^{\prime}} J = R_{\alpha}^{\dagger} J P_{\alpha^{\prime}} J \ ,
\end{equation}
where 
\begin{equation}
    R_{\alpha} = N_{\alpha} R_{0}  \in \mathcal{A}^{\prime}, \quad
    R_{\alpha} R_{\alpha}^{\dagger} = s^{\mathcal{A}^{\prime}}(\phi_{\alpha}) \quad \text{and} \quad
    R_{\alpha}^{\dagger} R_{\alpha} = s^{\mathcal{A}^{\prime}}(\phi) \ .
\end{equation}
The partial isometry $R_{\alpha}$ is the operator that connects the purification of $\phi$ in the cone $\mathcal{P}_{\Omega}^{1/4}(\mathcal{A})$ with its purification in the cone $\mathcal{P}_{\Omega}^{\alpha}(\mathcal{A})$, for $\alpha \in [0, 1/4]$
\begin{equation}
    R_{\alpha} \ket{\phi} 
    = J P_{\alpha^{\prime}} J \ket{\Omega} 
    = \ket{\phi_{\alpha}} \in \mathcal{P}_{\Omega}^{\alpha}(\mathcal{A}) . 
\label{eq:Ralpha_cone}
\end{equation}
The relative conjugate linear operators $S_{\phi_{0}, \Omega}$ and $S_{\phi, \Omega}$ associated with $\ket{\phi_{0}}$ and $\ket{\phi}$, respectively, are related by $S_{\phi_{0}, \Omega} = R_{0} S_{\phi, \Omega}$. Hence, using \eqref{eq:Talpha_ident}, we obtain
\begin{equation}
\begin{split}
    H_{\alpha} &= R_{0}^{\dagger} S_{\phi_{0}, \Omega} S_{\Omega} \Delta^{2 \alpha - 1}
    = s^{\mathcal{A}^{\prime}}(\phi) S_{\phi, \Omega} S_{\Omega} \Delta^{2 \alpha - 1} \\
    &= S_{\phi, \Omega} S_{\Omega} \Delta^{2 \alpha - 1}
    = J \Delta^{1/2}_{\phi, \Omega} J \Delta^{1/2} \Delta^{2 \alpha - 1} \\
    &= J \Delta^{1/2}_{\phi, \Omega} \Delta^{1 / 2 - 2 \alpha} J \ .
\label{eq:Halpha_modular}
\end{split}
\end{equation}
where we have used that $s^{\mathcal{A}^{\prime}}(\phi)$ projects onto the range of $S_{\phi, \Omega}$.

In conclusion, the partial isometry $R_{\alpha}$, which connects the purifications across the cones as mentioned in \eqref{eq:Ralpha_cone}, can be obtained from the polar decomposition of
\begin{equation}
    J \Delta^{1/2}_{\phi, \Omega} \Delta^{1 / 2 - 2 \alpha} J = R_{\alpha}^{\dagger} J P_{\alpha^{\prime}} J \ .
\label{eq:modular_Ralpha_polar}
\end{equation}
Moreover, the positive part takes the form
\begin{equation}
    J P_{\alpha^{\prime}}^{2} J = J \Delta^{1 / 2 - 2 \alpha} \Delta_{\phi, \Omega} \Delta^{1 / 2 - 2 \alpha} J 
    \Rightarrow J P_{\alpha^{\prime}} J =  J \sqrt{\Delta^{1 / 2 - 2 \alpha} \Delta_{\phi, \Omega} \Delta^{1 / 2 - 2 \alpha}} J \ .
\label{eq:Halpha_P}
\end{equation}
\section{Fidelity with modular tools}
\label{S:FidelityP0}

Suppose that we have a Hilbert space $\mathcal{H}$, a von Neumann algebra $\mathcal{A} \subset \mathcal{B}(\mathcal{H})$, a state $\omega \in \mathcal{A}_{*}^{+}$ that has a purification $\ket{\Omega} \in \mathcal{H}$ which is cyclic and separating for $\mathcal{A}$, and a normal state $\phi \in \mathcal{A}_{*}^{+}$. The fidelity in this setting can be defined using Uhlmann's theorem \cite{Uhlmann:1975kt, alberti_note_1983}
\begin{equation}
    F(\omega, \phi) = \text{Sup}_{\ket{\psi}} \abs{\braket{\Omega}{\psi}} \ ,
\end{equation}
where the supremum is taken over all the possible purifications $\ket{\psi} \in \mathcal{H}$ of $\phi$.
In this section we will show that the supremum is achieved uniquely by the purification of $\phi$ in the cone $\mathcal{P}_{\Omega}^{0}(\mathcal{A})$ and that the fidelity between these two states can be written using modular tools as
\begin{equation}
    F(\omega, \phi) = \braket{\Omega}{\phi_{0}} = \expval{\sqrt{\Delta_{\Omega}^{1/2} \Delta_{\phi,\Omega}\Delta_{\Omega}^{1/2}}}{\Omega}, \quad \ket{\phi_{0}} \in \mathcal{P}_{\Omega}^{0}(\mathcal{A}) .
\label{eq:FidelityModular}
\end{equation}
Here, $\ket{\phi_{0}}$ is the purification of $\phi$ over the cone $\mathcal{P}_{\Omega}^{0}(\mathcal{A})$, $\ket{\phi}$ is the purification of $\phi$ over the cone $\mathcal{P}_{\Omega}^{1/4}(\mathcal{A})$ and $\Delta_{\Omega}$ and $\Delta_{\phi, \Omega}$ are the modular operator and the relative modular operator, respectively, associated with $\mathcal{A}$, $\ket{\Omega}$ and $\ket{\phi}$.

To understand this result, we will start with the finite-dimensional example studied in section \ref{sS:cones_finite}. The fidelity between two density matrices, $\rho$ and $\sigma$, is defined as
\begin{equation}
    F(\rho, \sigma) = \Tr \left(\sqrt{\sqrt{\rho} \sigma \sqrt{\rho}} \right).
\end{equation}
But the following overlap has the same value
\begin{equation}
    \braket{\sqrt{\rho}}{\sqrt{\sigma}_{0}} = \Tr \left(\sqrt{\sqrt{\rho} \sigma \sqrt{\rho}} \right).
\end{equation}
where $\ket{\sqrt{\sigma}_{0}}$ is the purification of the density matrix $\sigma$ over the cone $\alpha = 0$ associated with $\rho$, and is given by \eqref{eq:sigmaalphafinite}.
Therefore, we find a purification of $\sigma$ such that its overlap with a purification of $\rho$ gives the fidelity. This is directly related to Uhlmann's theorem \cite{Uhlmann:1975kt}.
Also, we know that in the finite-dimensional case the modular operator and relative modular operator are given by \eqref{eq:modularfinite} and \eqref{eq:relativemodularfinite}. Then, it is easy to show that
\begin{equation}
    \expval{\sqrt{\Delta_{\rho}^{1/2} \Delta_{\sigma,\rho}\Delta_{\rho}^{1/2}}}{\sqrt{\rho}}
    = \expval{\sqrt{\sqrt{\rho} \sigma \sqrt{\rho}} \otimes \rho^{-1}}{\sqrt{\rho}}
    = \Tr \left( \sqrt{\sqrt{\rho} \sigma \sqrt{\rho}} \right) \ .
\end{equation}
Now, we proceed with the general case. We have a normal state $\phi$ over the algebra $\mathcal{A}$, which we can purify to a vector $\ket{\phi} \in \mathcal{P}_{\Omega}^{1/4}(\mathcal{A})$. Let the functional $f_{\Omega, \phi}$ over the algebra $\mathcal{A}^{\prime}$ be
\begin{equation}
    f_{\Omega, \phi}(A^{\prime}) = \bra{\Omega} A^{\prime} \ket{\phi} \ .
\end{equation}
It was proved in the lemma 5.3 of \cite{ArakiMasuda1982} that there exist a vector $\ket{\xi} \in \mathcal{P}_{\Omega}^{1/4}(\mathcal{A^{\prime}}) = \mathcal{P}_{\Omega}^{1/4}(\mathcal{A})$ and an isometry $V \in \mathcal{A}^{\prime}$ such that
\begin{equation}
    f_{\Omega, \phi}(A^{\prime}) = \expval{A^{\prime}V}{\xi}, \quad
    \norm{f_{\Omega, \phi}} = \braket{\xi} = \norm{\xi}^{2} \quad \text{and} \quad
    V J \Delta_{\xi, \Omega} J \ket{\Omega} = \ket{\phi} \ .
\label{eq:theorem53}
\end{equation}
Here, $\Delta_{\xi, \Omega}$ is the relative modular operator between $\ket{\Omega}$ and $\ket{\xi}$ associated with the algebra $\mathcal{A}$, and it is related to the relative modular operator $\Delta_{\xi, \Omega}^{\prime}$ associated with the algebra $\mathcal{A}^{\prime}$ by $\Delta_{\xi, \Omega}^{\prime} = J \Delta_{\xi, \Omega} J$, because $\ket{\xi} \in \mathcal{P}_{\Omega}^{1/4}(\mathcal{A})$. The operator $K = V J \Delta_{\xi, \Omega} J$ acts on $\mathcal{U}(\mathcal{A}) \ket{\Omega}$ as follows
\begin{equation}
    K A \ket{\Omega} = \tau_{-1}(A) \ket{\phi} \ .
\end{equation}
This definition is the same as the one used for the operator $H_{\alpha}$ in \eqref{eq:def_Halpah2} for the case $\alpha = 0$. Therefore $K = H_{0}$. Using \eqref{eq:Halpha_modular}, \eqref{eq:Halpha_P} and the uniqueness of the polar decomposition, we have that
\begin{equation}
    J \Delta_{\xi, \Omega} J = \abs{K} = \abs{H_{0}} = J P_{1/2} J = J \sqrt{\Delta_{\Omega}^{1/2} \Delta_{\phi, \Omega} \Delta_{\Omega}^{1/2}} J \ .
\label{eq:PKModular}
\end{equation}
Using the property of the conjugate linear operator $S_{\xi, \Omega} \ket{\Omega} = \ket{\xi}$ (see Appendix \ref{A:RelativeModular}) and \eqref{eq:theorem53} we have that $\norm{f_{\Omega,\phi}}$ is equal to
\begin{equation}
    \norm{f_{\Omega,\phi}} = \braket{\xi} = \expval{\Delta_{\xi, \Omega}}{\Omega} = \expval{\sqrt{\Delta_{\Omega}^{1/2} \Delta_{\phi, \Omega} \Delta_{\Omega}^{1/2}}}{\Omega} \ .
\end{equation}
Moreover, by a well known expression for the fidelity \cite{alberti_note_1983} and the definition of the norm of a functional, we have
\begin{equation}
    \norm{f_{\Omega,\phi}} = F(\omega, \phi) = \text{Sup} \{ \bra{\Omega} V \ket{\phi} \ | \ V \in \mathcal{A}^{\prime}, \norm{V} = 1 \}  \ .
\end{equation}
Therefore
\begin{equation}
    F(\omega, \phi) = \expval{\sqrt{\Delta_{\Omega}^{1/2} \Delta_{\phi, \Omega} \Delta_{\Omega}^{1/2}}}{\Omega} \ .
\end{equation}
This is the same value that as the overlap $\braket{\Omega}{\phi_{0}}$, where $\ket{\phi_{0}}$ is the 
purification of $\phi$ in the positive cone $\alpha = 0$. From \eqref{eq:Ralpha_cone}, we know that $\ket{\phi_{0}} = J P_{1/2} J \ket{\Omega}$, and by \eqref{eq:Halpha_P} we get
\begin{equation}
    \braket{\Omega}{\phi_{0}} = \expval{\sqrt{\Delta_{\Omega}^{1/2} \Delta_{\phi, \Omega} \Delta_{\Omega}^{1/2}}}{\Omega} = F(\omega, \phi) \ .
\end{equation}
Now we will show that this is the unique purification that achieves the supremum. Suppose that we achieve the supremum with another purification $\ket{\Phi}$ of $\phi$ and that $\braket{\Omega}{\Phi} > 0$. Then we can define the function
\begin{equation}
    f(s) = \abs{\bra{\Omega} e^{i H s} \ket{\Phi}}^{2} \ , \ H \in \mathrm{H}(\mathcal{A}^{\prime}) \ .
\label{eq:fsupfidelity}
\end{equation}
Here, $\mathrm{H}(\mathcal{A}^{\prime})$ represents the set of all Hermitian operators that live in $\mathcal{A}^{\prime}$. This function must have a maximum at $s=0$ for all $H \in \mathrm{H}(\mathcal{A}^{\prime})$ because, with the unitary operator $e^{i H s} \in \mathcal{A}^{\prime}$, we are changing the purification of $\phi$. But, $\ket{\Phi}$ is a purification that gives the maximal overlap. Therefore, at $s=0$, we have a maximum.
Now, if we take the first and second derivatives and impose on the function the conditions for a maximum, $f^{\prime}(0) = 0$ and $f^{\prime\prime}(0) \leq 0$, we obtain the conditions that, for all $H \in \mathrm{H}(\mathcal{A}^{\prime})$, we have
\begin{equation}
    \bra{\Omega} H \ket{\Phi} = \bra{\Omega} H \ket{\Phi}^{*} \ ,
\label{eq:cond1fidelity1}
\end{equation}
\begin{equation}
    \bra{\Omega} H^{2} \ket{\Phi} + \bra{\Omega} H^{2} \ket{\Phi}^{*} - 2 \frac{\abs{\bra{\Omega} H \ket{\Phi}}^{2}}{\braket{\Omega}{\Phi}} \geq 0 \ .
\label{eq:cond1fidelity2}
\end{equation}
The condition \eqref{eq:cond1fidelity1} implies that the number $\bra{\Omega} H \ket{\Phi}$ has to be real for all Hermitian operators. Using this, we can rewrite \eqref{eq:cond1fidelity2} as
\begin{equation}
    \bra{\Omega} H^{2} \ket{\Phi} \geq \frac{\abs{\bra{\Omega} H \ket{\Phi}}^{2}}{\braket{\Omega}{\Phi}} \geq 0 \ \forall \ H \in \mathrm{H}(\mathcal{A}^{\prime}) \ .
\label{eq:cond1fidelity3}
\end{equation}
This last condition is very strong. Whenever we have a Hermitian operator $H$, we have that $H^{2}$ is a positive operator, and the converse is also true: if we have a positive operator $P$, we can always write it as the square of a Hermitian matrix, in particular, as the square of $\sqrt{P}$. The positive cone $\mathcal{P}_{\Omega}^{0}(\mathcal{A}^{\prime})$ is generated by the closure of the set $\{ P \ket{\Omega} \mid P \in (\mathcal{A}^{\prime})^{+} \}$. Therefore, the condition \eqref{eq:cond1fidelity3} tells us that the purification $\ket{\Phi}$ must belong to the dual cone $\ket{\Phi} \in (\mathcal{P}_{\Omega}^{0}(\mathcal{A}^{\prime}))^{\prime} = \mathcal{P}_{\Omega}^{1/2}(\mathcal{A}^{\prime}) = \mathcal{P}_{\Omega}^{0}(\mathcal{A})$. But, the purification of $\phi$ over the cones is unique by \eqref{T:purificationcones}, therefore, $\ket{\Phi} = \ket{\phi_{0}}$.

In conclusion, given two normal states $\omega$ and $\phi$ over the algebra $\mathcal{A}$, if there is some cyclic and separating purification $\ket{\Omega}$ of $\omega$, the supremum in Uhlmann's theorem is achieved by the purification $\ket{\phi_{0}} \in \mathcal{P}_{\Omega}^{0}(\mathcal{A})$ of $\phi$, and this purification is unique up to a phase.

That the supremum in Uhlmann's theorem is achieved is easy to show in the finite-dimensional case (see the proof of Theorem 9.4 in \cite{Nielsen:2012yss}). A proof for the general case is given in \cite{alberti2000bures, araki1972bures} in a more abstract form. A connection between the fidelity and the work \cite{ArakiMasuda1982} has already been pointed out in \cite{Faulkner:2020iou, Hollands:2020owv}.

The use of positive cones to obtain the fidelity invites us to define a generalized fidelity as
\begin{equation}
    F_{\alpha}(\omega,\phi) = \braket{\Omega}{\phi_{\alpha}} = \expval{\sqrt{\Delta_{\Omega}^{1/2 - 2 \alpha} \Delta_{\phi, \Omega} \Delta_{\Omega}^{1/2 - 2 \alpha}}}{\Omega}.
\label{eq:gen_fidelity}
\end{equation}
In the finite-dimensional case is written as
\begin{equation}
    F_{\alpha}(\rho,\sigma) = \braket{\sqrt{\rho}}{\sqrt{\sigma}_{\alpha}}
    = \Tr \left( \rho^{2 \alpha} \sqrt{\rho^{1/2 - 2 \alpha} \sigma \rho^{1/2 - 2 \alpha}} \right) .
\end{equation}
In \cite{Hollands:2020owv}, another type of generalized fidelity was defined, and it was proven that this generalization satisfies the desired properties, such as the data processing inequality (DPI). It would be interesting to check whether our generalized fidelity also satisfies these desired properties. We leave this for future investigation.

\newpage

\section{Representation of automorphisms in positive cones}
\label{S:RepresentationAutomorphism}
Let $\mathcal{H}$ be a Hilbert space, $\mathcal{A} \subset \mathcal{B}(\mathcal{H})$ an von Neumann algebra, and $\beta: \mathcal{A} \to \mathcal{A}$ an automorphism of $\mathcal{A}$. We call an isometry $U(\beta) \in \mathcal{B}(\mathcal{H})$ a representative of $\beta$ if for a $A \in \mathcal{A}$ we have
\begin{equation}
    U(\beta) A = \beta(A) U(\beta) \ \text{and} \ U(\beta)^{\dagger} U(\beta) = I \ .
\end{equation}
It is well known that, given an automorphism $\beta$, one can construct a global unitary operator $U(\beta) \in \mathcal{B}(\mathcal{H})$ that implements the automorphism on the algebra $\mathcal{A}$ \cite{haag2012local, bratteli1979operator}. Such representatives are far from unique, since $U(\beta)$ may be multiplied by any unitary $R^{\prime} \in \mathcal{A}^{\prime}$, yielding to the isometry $\overline{U}(\beta) = R^{\prime} U(\beta)$, which still implements the automorphism on $\mathcal{A}$. In this section, we show how to construct representatives of the automorphism using the cones $\mathcal{P}_{\Omega}^{\alpha}(\mathcal{A})$ and analyze their properties.

\subsection{General definition}
\label{S:purificationautomorphism_def}
In theorem 2.5.32 of \cite{bratteli1979operator} (see also theorem 2.2.4 of \cite{haag2012local}) it is shown how to construct a representative of the automorphism $\beta$ using the standard cone $\alpha = 1/4$. We will call this representative the standard representative. Here, we generalize this construction to the cones with $\alpha \in \left[0, 1/4 \right]$.

Given a Hilbert space $\mathcal{H}$, a von Neumann algebra $\mathcal{A} \subset \mathcal{B}(\mathcal{H})$, a cyclic and separating vector $\ket{\Omega}$ and an automorphism $\beta: \mathcal{A} \rightarrow \mathcal{A}$, we can construct, using the different cones $\mathcal{P}_{\Omega}^{\alpha}(\mathcal{A})$, a unique isometry $U_{\alpha}(\beta)$ such that it implements the automorphism on the operators of $\mathcal{A}$ and satisfies $U_{\alpha}(\beta) \ket{\Omega} \in \mathcal{P}_{\Omega}^{\alpha}(\mathcal{A})$. We will call this the representative of $\beta$ in the cone $\mathcal{P}_{\Omega}^{\alpha}(\mathcal{A})$.
Let $\omega(A) = \expval{A}{\Omega}$ with $A \in \mathcal{A}$ be the reduced state over $\mathcal{A}$ of $\ket{\Omega}$. We define the normal state $\phi \in \mathcal{A}_{*}^{+}$ as
\begin{equation}
    \phi(A) = \omega(\beta^{-1}(A)) \ .
\label{eq:state_phi}
\end{equation}
This state is faithful, if there exists some $A \in \mathcal{A}$ such that $\phi(A^{\dagger} A) = 0$, this implies that
\begin{equation}
    0 = \phi(A^{\dagger} A) = \expval{A^{\dagger} A}{\Omega} = \norm{A \ket{\Omega}}^{2} \ .
\end{equation}
Then, by the separability of $\ket{\Omega}$ with respect to $\mathcal{A}$, we have that $A = 0$. We can purify $\phi$ over a cone $\mathcal{P}_{\Omega}^{\alpha}(\mathcal{A})$ with $\alpha \in [0, 1/4]$ to obtain the unique representative vector $\ket{\phi_{\alpha}}$ by theorem \ref{T:purificationcones}.
Using the cyclic property of $\ket{\Omega}$, we can define the operator $U_{\alpha}(\beta)$ as
\begin{equation}
    U_{\alpha}(\beta) A \ket{\Omega} = \beta(A) \ket{\phi_{\alpha}} \ , \ \forall \ A \in \mathcal{A} \ .
\label{eq:def_U_alpha}
\end{equation}
This operator must be an isometry
\begin{equation}
\begin{split}
    \norm{U_{\alpha}(\beta) A \ket{\Omega}}^{2} &= \expval{\beta(A^{\dagger} A)}{\phi_{\alpha}} = \expval{A^{\dagger} A}{\Omega} = \norm{A \Omega}^{2} \\
    &\Rightarrow U_{\alpha}(\beta)^{\dagger} U_{\alpha}(\beta) = I \ .
\end{split}
\label{eq:uisometry}
\end{equation}
Additionally, we can show that
\begin{equation}
    U_{\alpha}(\beta) A B \ket{\Omega} 
    = \beta(AB) \ket{\phi_{\alpha}}
    = \beta(A) U_{\alpha}(\beta) B \ket{\Omega} \ \forall \ A, B \in \mathcal{A} \ .
\end{equation}
Then, using the cyclic property of $\ket{\Omega}$, we arrive at the identity
\begin{equation}
     U_{\alpha}(\beta) A = \beta(A) U_{\alpha}(\beta)\ .
\label{eq:U_auto_cond}
\end{equation}
We can also show that the identity \eqref{eq:U_auto_cond} and the condition $U_{\alpha}(\beta)\ket{\Omega} \in \mathcal{P}_{\Omega}^{\alpha}(\mathcal{A})$ determine this operator uniquely. Suppose that there exists another operator $\overline{U}_{\alpha}(\beta)$ such that it implements the automorphism as in \eqref{eq:U_auto_cond}, is an isometry $\overline{U}_{\alpha}(\beta)^{\dagger} \overline{U}_{\alpha}(\beta) = I$, and satisfies $\overline{U}_{\alpha}(\beta)\ket{\Omega} = \ket{\psi} \in \mathcal{P}_{\Omega}^{\alpha}(\mathcal{A})$.
Then the vector $\ket{\psi}$, when reduced to the algebra $\mathcal{A}$, gives the state
\begin{equation}
\begin{split}
    \expval{A}{\psi} 
    &= \bra{\overline{U}_{\alpha}(\beta) \Omega} A \overline{U}_{\alpha}(\beta) \ket{\Omega} \\
    &= \bra{\overline{U}_{\alpha}(\beta) \Omega} \overline{U}_{\alpha}(\beta) \beta^{-1}(A)  \ket{\Omega} \\
    &= \bra{\Omega}\beta^{-1}(A)  \ket{\Omega} = \phi(A) \ .
\end{split}
\end{equation}
Therefore, $\ket{\psi}$ is a purification of $\phi$ in the cone $\mathcal{P}_{\Omega}^{\alpha}(\mathcal{A})$. Since this purification is unique, it follows that $\ket{\psi} = \ket{\phi_{\alpha}}$ \cite{PositiveConesAssociated}. Then
\begin{equation}
    (U_{\alpha}(\beta) - \overline{U}_{\alpha}(\beta)) A \ket{\Omega} = (\beta(A) - \beta(A)) \ket{\phi_{\alpha}} = 0 \ \forall A \in \mathcal{A} \ .
\end{equation}
By the cyclic property of the vacuum, we arrive at
\begin{equation}
    \overline{U}_{\alpha}(\beta) = U_{\alpha}(\beta) \ .
\end{equation}

\subsection{Standard representative}
\label{sS:std_rep}

In the case where we represent the automorphism $\beta$ in the standard cone ($\alpha = 1/4$), we get a unitary operator $U_{1/4}(\beta)$ that commutes with the modular conjugation $J$ associated with $\ket{\Omega}$ and $\mathcal{A}$ (theorem 2.5.32 \cite{bratteli1979operator})
\begin{equation}
    J U_{1/4}(\beta) J = U_{1/4}(\beta) \ .
\label{eq:UcommuteJ}
\end{equation}
That it is unitary comes from the fact that the state $\phi$, defined in \eqref{eq:state_phi}, is faithful, as we proved in \ref{S:purificationautomorphism_def}. Because $\phi$ is faithful, any purification $\ket{\Phi}$ is separable for $\mathcal{A}$. If there exists some $A \in \mathcal{A}$ such that $A \ket{\Phi} = 0$ then
\begin{equation}
    0 = \norm{A \ket{\Phi}}^{2} = \expval{A^{\dagger} A}{\Phi} = \phi(A^{\dagger} A) \ .
\end{equation}
But because $\phi$ is faithful, we have that $A = 0$. Therefore
$\ket{\Phi}$ is separable for $\mathcal{A}$. Now, if we purify $\phi$
in the standard cone $\ket{\phi} \in \mathcal{P}_{\Omega}^{1/4}(\mathcal{A})$, by property 2 mentioned in section \ref{S:positive_cones} this vector satisfies $J\ket{\phi} = \ket{\phi}$.
With this property, we can deduce that it is also separable for $\mathcal{A}^{\prime}$, therefore, it is cyclic and separating for both $\mathcal{A}$ and $\mathcal{A}^{\prime}$. Because $\ket{\phi}$ is cyclic for $\mathcal{A}$, the range of $U_{1/4}(\beta)$ is dense in $\mathcal{H}$. We can then define its inverse in the obvious way, obtaining that it is unitary $U_{1/4}(\beta)^{\dagger} = U_{1/4}(\beta)^{-1}$.

The property that $U_{1/4}(\beta)$ commutes with the modular conjugation $J$ can be deduced from the fact that 
\begin{equation}
\begin{split}
    S_{\Omega} U_{1/4}(\beta)^{\dagger} A \ket{\phi}
    &= S_{\Omega} \beta^{-1}(A) \ket{\Omega}
    = \beta^{-1}(A^{\dagger}) \ket{\Omega} \\
    &= U_{1/4}(\beta)^{\dagger} A^{\dagger} \ket{\phi}
    = U_{1/4}(\beta)^{\dagger} S_{\phi} A \ket{\phi} \ ,
\end{split}
\end{equation}
where $S_{\Omega}$ and $S_{\phi}$ are the conjugate linear operators associated to the algebra $\mathcal{A}$ and the cyclic vectors $\ket{\Omega}$ and $\ket{\phi}$, respectively. By the cyclic property of $\ket{\phi}$, we conclude that 
\begin{equation}
    S_{\Omega} U_{1/4}(\beta)^{\dagger} = U_{1/4}(\beta)^{\dagger} S_{\phi} \Rightarrow S_{\phi} = U_{1/4}(\beta) S_{\Omega} U_{1/4}(\beta)^{\dagger} \ .
\end{equation}
Because $\ket{\phi} \in \mathcal{P}_{\Omega}^{1/4}(\mathcal{A})$, we have that $J_{\phi} = J_{\Omega}$ \cite{SomePropertiesModularConjugation} and, by the uniqueness of the polar decomposition, we arrive at
\begin{equation}
\begin{split}
    &J \Delta_{\phi}^{1/2} = U_{1/4}(\beta) J U_{1/4}(\beta)^{\dagger} \ U_{1/4}(\beta) \Delta_{\Omega}^{1/2} U_{1/4}(\beta)^{\dagger} \ , \\
    \Rightarrow 
    J &= U_{1/4}(\beta) J U_{1/4}(\beta)^{\dagger} , \quad
    \Delta_{\phi}^{1/2} = U_{1/4}(\beta) \Delta_{\Omega}^{1/2} U_{1/4}(\beta)^{\dagger} \ .
\end{split}
\end{equation}
In particular, we arrive at the identity \eqref{eq:UcommuteJ}.

The properties that $U_{1/4}(\beta)$ commutes with $J$ and that it implements the automorphism on the algebra $\mathcal{A}$, as mentioned in \eqref{eq:U_auto_cond}, uniquely determine it if the algebra is a factor (which is usually the case of interest in QFT). To show this, suppose that there exists another isometry $\widetilde{U}(\beta)$ such that it implements the automorphism on $\mathcal{A}$, $\widetilde{U}(\beta)^{\dagger} \widetilde{U}(\beta) = I$, and commutes with $J$. Then
\begin{equation}
    \widetilde{U}(\beta) U_{1/4}(\beta)^{\dagger} A = A \widetilde{U}(\beta) U_{1/4}(\beta)^{\dagger} \ \forall A \in \mathcal{A} 
    \Rightarrow \widetilde{U}(\beta) U_{1/4}(\beta)^{\dagger} \in \mathcal{A}^{\prime} \ .
\end{equation}
But we have that $J \widetilde{U}(\beta) U_{1/4}(\beta)^{\dagger} J = \widetilde{U}(\beta) U_{1/4}(\beta)^{\dagger}$, therefore, $U_{1/4}(\beta) \widetilde{U}(\beta)^{\dagger} \in \mathcal{A} \cap \mathcal{A}^{\prime}$. If the algebra $\mathcal{A}$ is a factor, then $\mathcal{A} \cap \mathcal{A}^{\prime} = \{ \lambda I \mid \lambda \in \mathbb{C} \}$, where $I$ is the identity operator. Using the fact that the standard representative $U_{1/4}(\beta)$ is unitary, we arrive at
\begin{equation}
     \widetilde{U}(\beta)U_{1/4}(\beta)^{\dagger} = e^{i \theta} I \Rightarrow  \widetilde{U}(\beta) = e^{i \theta} U_{1/4}(\beta) \ .
\end{equation}
\indent From the fact that the standard representative commutes with $J$, we can also understand its action on the commutant algebra
\begin{equation}
\begin{split}
    U_{1/4}(\beta) \ A \ U_{1/4}(\beta)^{\dagger} = \beta(A)  \quad  &\text{if} \ A \in \mathcal{A} \ , \\
    U_{1/4}(\beta) \ A^{\prime} \ U_{1/4}(\beta)^{\dagger} = J \beta(J A^{\prime} J) J \quad &\text{if} \ A^{\prime} \in \mathcal{A}^{\prime} \ .
\end{split}
\end{equation}
\indent If the automorphism is inner, the standard representative can be constructed explicitly. An inner automorphism is defined by a unitary $U_{\mathcal{A}} \in \mathcal{A}$, such that $\beta(A) = U_{\mathcal{A}} A U_{\mathcal{A}}^{\dagger}$. Therefore, the standard representative is
\begin{equation}
    U_{1/4}(\beta) = U_{\mathcal{A}} J U_{\mathcal{A}} J \ .
\label{eq:repinnerstandard}
\end{equation}
It follows that this is the standard representative because it implements the automorphism over $\mathcal{A}$ and commutes with $J$.

\subsection{Representative in the cone $\mathcal{P}_{\Omega}^{\alpha}(\mathcal{A})$}
\label{S:Representativeinthecone}

In this section, we will show that all the representatives of an automorphism over the cones are connected by the partial isometry $R_{\alpha} \in \mathcal{A}^{\prime}$ studied in section \ref{sS:connecting_cones}. Taking again the state \eqref{eq:state_phi}, which is faithful, we have the purification $\ket{\phi} \in \mathcal{P}_{\Omega}^{1/4}(\mathcal{A})$ and $\ket{\phi_{\alpha}} \in \mathcal{P}_{\Omega}^{\alpha}(\mathcal{A})$. These purifications are connected by the partial isometry $R_{\alpha}$
\begin{equation}
    R_{\alpha} \ket{\phi} = \ket{\phi_{\alpha}} \ .
\end{equation}
Because $\ket{\phi}$ is cyclic for $\mathcal{A}$, as we proved in section \ref{sS:std_rep}, we have that $R_{\alpha}$ is an isometry
\begin{equation}
    R_{\alpha}^{\dagger} R_{\alpha} = s^{\mathcal{A}^{\prime}}(\phi) = I \ .
\end{equation}
We recall that $s^{\mathcal{A}^{\prime}}(\phi)$ is the projection onto the subspace $\overline{\mathcal{A} \ket{\phi}} = \mathcal{H}$.
Now, if we use the definition of the standard representative of the automorphism, we can multiply it by $R_{\alpha}$ and obtain the definition of the representative over the cone $\alpha$
\begin{equation}
    R_{\alpha} U_{1/4}(\beta) A \ket{\Omega} =  \beta(A) R_{\alpha} \ket{\phi} = \beta(A) \ket{\phi_{\alpha}} \ ,
\end{equation}
\begin{equation}
    \Rightarrow  U_{\alpha}(\beta) = R_{\alpha} U_{1/4}(\beta) \ .
\label{eq:rep_auto_conealpha_R}
\end{equation}
Then, we have that the representative of the automorphism $\beta$ over the cone $\mathcal{P}_{\Omega}^{\alpha}(\mathcal{A})$ can be written using the standard representative and the isometry $R_{\alpha}$.

\subsection{Trivial cases}

It is easy to find trivial examples of this construction. Suppose that we have a QFT with some internal symmetry that is implemented by the global unitary operator $U$, and a vacuum vector $\ket{\Omega}$ which is cyclic and separating for every algebra associated with bounded regions of the space-time. Given an algebra $\mathcal{A}$ associated with a region, $U$ applies the automorphism $\beta(A) = U A U^{\dagger}$ over the operators $A \in \mathcal{A}$. If the symmetry is unbroken, we have that
\begin{equation}
    U \ket{\Omega} = \ket{\Omega} \in \mathcal{P}_{\Omega}^{\alpha}(\mathcal{A}) \quad \forall \ \alpha \in \left[0, 1/4 \right]\ .
\end{equation}
Therefore, the global unitary is the representative of $\beta$ in every cone, and in particular, $\expval{U}{\Omega} = 1$. Therefore, in looking for representatives of a global symmetry acting on a ball region in QFT, the non-trivial examples arise when the symmetry is broken. More generally, we have a similar situation if the automorphism $\beta$ leaves the state $\ket{\Omega}$ on $\mathcal{A}$ invariant. In this case, $U_{\alpha}(\beta) = U_{1/4}(\beta)$.

\subsection{Representative of an inner automorphism}
\label{sS:RepInner}
We will consider the case where we have a unitary $U_{\mathcal{A}} \in \mathcal{A}$ and we are considering the automorphism of $\mathcal{A}$ given by $\beta(A) = U_{\mathcal{A}} A U_{\mathcal{A}}^{\dagger}$.
The standard representative that implements the automorphism is $U_{1/4}(\beta) = U_{\mathcal{A}} J U_{\mathcal{A}} J$.
We want to find the form of $R_{\alpha}$ in this case to get a representative of the automorphism in another cone \eqref{eq:def_U_alpha}.
In this case, the relative modular operator between $\ket{\phi} = U_{1/4}(\beta)  \ket{\Omega}$ and $\ket{\Omega}$ is
\begin{equation}
    S_{\phi, \Omega} A \ket{\Omega} = A^{\dagger} \ket{\phi}
    = A^{\dagger} U_{\mathcal{A}} J U_{\mathcal{A}} J \ket{\Omega}
    = J U_{\mathcal{A}} J A^{\dagger} U_{\mathcal{A}}  \ket{\Omega}
    = J U_{\mathcal{A}} J S_{\Omega} U_{\mathcal{A}}^{\dagger} A  \ket{\Omega} \ .
\end{equation}
Then
\begin{equation}
    S_{\phi, \Omega} = J U_{\mathcal{A}} J S U_{\mathcal{A}}^{\dagger} = J U_{\mathcal{A}} \Delta^{\frac{1}{2}}_{\Omega} U_{\mathcal{A}}^{\dagger} \ .
\end{equation}
Therefore, by the uniqueness of the polar decomposition,
\begin{equation}
    \Delta_{\phi, \Omega} = U_{\mathcal{A}} \Delta_{\Omega} U_{\mathcal{A}}^{\dagger} \ .
\label{eq:relativemodlaroperatorinner}
\end{equation}
Replacing this expression in \eqref{eq:modular_Ralpha_polar} we get
\begin{equation}
    J \Delta_{\phi, \Omega}^{\frac{1}{2}} \Delta_{\Omega}^{\frac{1}{2} - 2 \alpha} J
    = J U_{\mathcal{A}} \Delta^{\frac{1}{2}}_{\Omega} U_{\mathcal{A}}^{\dagger} \Delta_{\Omega}^{\frac{1}{2} - 2 \alpha} J 
    = R_{\alpha}^{\dagger} P_{\alpha} \ .
\label{eq:Tinner}
\end{equation}
It is useful to write the polar decomposition of
\begin{equation}
    J \Delta^{\frac{1}{2} - 2 \alpha}_{\Omega} U_{\mathcal{A}} \Delta_{\Omega}^{\frac{1}{2}} J =
    \widetilde{P}_{\alpha} W_{\alpha} \ .
\label{eq:polardescompositioninner}
\end{equation}
Replacing this polar decomposition in \eqref{eq:Tinner} we obtain
\begin{equation}
    R^{\dagger}_{\alpha} P_{\alpha} = J U_{\mathcal{A}} J W_{\alpha}^{\dagger} \widetilde{P}_{\alpha} \ .
\end{equation}
Because the polar decomposition is unique, we have that
\begin{equation}
    R_{\alpha} = W_{\alpha} J U_{\mathcal{A}}^{\dagger} J \ , \quad \widetilde{P}_{\alpha} = P_{\alpha} \ .
\end{equation}
From the fact that the isometry $R_{\alpha}$ belongs to $\mathcal{A}^{\prime}$, we deduce that $W_{\alpha} \in \mathcal{A}^{\prime}$ and is also an isometry. Then, we have that the representative of the automorphism $\beta$ over the cone $\alpha$ is given by \eqref{eq:rep_auto_conealpha_R}
\begin{equation}
    U_{\alpha}(\beta) = R_{\alpha} U_{1/4}(\beta) = W_{\alpha} U_{\mathcal{A}} \ .
\end{equation}
It follows that $W_{\alpha}$ complete the inner automorphism with a isometry in $\mathcal{A}^{\prime}$ to get the representative in the cone $\mathcal{P}_{\Omega}^{\alpha}(\mathcal{A})$. It is useful that $W_{\alpha}$ can be calculated directly from the polar decomposition \eqref{eq:polardescompositioninner}.

\subsection{Representative of an inner one-parameter group}
\label{sS:InnerOneparametergroup}
We will study the representatives of an inner automorphism in the case where the inner automorphism is of the form
\begin{equation}
    U_{\mathcal{A}}(\epsilon) = e^{i \epsilon H} \ ,
\end{equation}
where $H \in \mathcal{A}$ is a Hermitian operator. We will seek to obtain the representative of the automorphism to the lowest order in $\epsilon$. This case will be useful for the discussion of the scalar field in $3+1$ dimensions, as it provides a straightforward way to find the representative of a coherent operator in different cones.

In the following, we assume that all operators are bounded so that we do not need to worry about domains. From \eqref{eq:polardescompositioninner}, we see that we only need to perform the polar decomposition
\begin{equation}
    J \Delta^{\frac{1}{2} - 2 \alpha} U_{\mathcal{A}}(\epsilon) \Delta^{\frac{1}{2}} J = P_{\alpha}(\epsilon) W_{\alpha}(\epsilon) \ .
\label{eq:inneroneparameterpolardescopmosition}
\end{equation}
When $\epsilon = 0$ we have that
\begin{equation}
    J \Delta^{\frac{1}{2} - 2 \alpha} U_{\mathcal{A}}(0) \Delta^{\frac{1}{2}} J 
    = \Delta^{2 \alpha - 1} \geq 0
    \Rightarrow P_{\alpha}(0) = \Delta^{2 \alpha - 1}, \quad W_{\alpha}(0) = I \ .
\end{equation}
Taking derivatives with respect to $\epsilon$ on both sides of \eqref{eq:inneroneparameterpolardescopmosition} and
evaluating at $\epsilon = 0$, we get
\begin{equation}
    - i J \Delta^{\frac{1}{2}- 2 \alpha} H \Delta^{\frac{1}{2}} J = P_{\alpha}^{\prime} - i \Delta^{2 \alpha - 1} Q_{\alpha} \ ,
\label{eq:dersus}
\end{equation}
where we define $P_{\alpha}^{\prime} = \left. \frac{d P_{\alpha}(\epsilon)}{d \epsilon} \right|_{\epsilon = 0}$, and
\begin{equation}
    \left. \frac{d W_{\alpha} (\epsilon)}{d \epsilon} \right|_{\epsilon = 0} = - i Q_{\alpha} \ .
\label{eq:derivadaW}
\end{equation}
We also have another identity for $P_{\alpha}^{\prime}$. We know that 
\begin{equation}
    P_{\alpha}(\epsilon)^{2} = J \Delta^{\frac{1}{2} - 2 \alpha} U(\epsilon) \Delta U^{\dagger}(\epsilon) \Delta^{\frac{1}{2} - 2 \alpha} J  \ .
\end{equation}
Differentiating this expression with respect to $\epsilon$ on both sides and evaluating at $\epsilon = 0$ gives
\begin{equation}
    P_{\alpha}^{\prime} \Delta^{2 \alpha - 1} + \Delta^{2 \alpha - 1} P_{\alpha}^{\prime}
    = -i J (\Delta^{\frac{1}{2} - 2 \alpha} H \Delta^{\frac{3}{2} - 2 \alpha} -  \Delta^{\frac{3}{2} - 2 \alpha} H \Delta^{\frac{1}{2} - 2 \alpha})  J \ .
\label{eq:derp2}
\end{equation}
From \eqref{eq:dersus}, we have $P_{\alpha}^{\prime} = - i J \Delta^{\frac{1}{2} - 2 \alpha} H \Delta^{\frac{1}{2}} J + i \Delta^{2 \alpha -1} Q_{\alpha}$, replacing this in \eqref{eq:derp2}, we arrive at an expression that relates $H$ with $Q_{\alpha}$. After some algebraic work over that expression (see Appendix \ref{ap:AlgebraicWork}), we arrive at
\begin{equation}
    \Delta^{1/2 - \alpha} Q_{\alpha} \Delta^{- (1/2 - \alpha)} + \Delta^{-(1/2 - \alpha)} Q_{\alpha} \Delta^{1/2 - \alpha} 
    = J ( \Delta^{\alpha} H \Delta^{-\alpha} + \Delta^{-\alpha} H \Delta^{\alpha} )J \ .
\label{eq:relHQ}
\end{equation}
When we replace $\alpha = 0$, we get the same formula as the one used in \cite{bratteli1979operator} to prove the Tomita-Takesaki theorem. In that case, they developed an inversion formula which gives $Q_{0}$ in terms of $H$. Inspired by that inversion formula, we generalize it for arbitrary $\alpha$.
If we define
\begin{equation}
    D_{\alpha}(A) = \Delta^{\alpha} A \Delta^{-\alpha} + \Delta^{-\alpha} A \Delta^{\alpha} \ ,
\end{equation}
and we want an operator $A_{\alpha} \in \mathcal{A}$ such that $D_{\alpha}(A_{\alpha}) = B$ for some given operator $B \in \mathcal{A}$, then $A_{\alpha}$ is given by the following inversion formula (see Appendix \ref{A:InversionFormula})
\begin{equation}
    A_{\alpha} = I_{\alpha}(B) = \frac{1}{4 \alpha} \int_{-\infty}^{\infty} \ dt \ \frac{\Delta^{it} B \Delta^{-it}}{\cosh(\frac{\pi t}{2 \alpha})}
    \Rightarrow D_{\alpha}(I_{\alpha}(B)) = B \ .
\end{equation}
In our case, we want an operator $Q_{\alpha}$ such that $D_{\frac{1}{2} - \alpha}(Q_{\alpha}) = J ( \Delta^{\alpha} H \Delta^{-\alpha} + \Delta^{-\alpha} H \Delta^{\alpha} ) J$. Then $Q_{\alpha}$ is given by
\begin{equation}
    J Q_{\alpha} J = \frac{1}{4 \alpha^{\prime}} \int_{-\infty}^{\infty} \ dt \ \frac{\Delta^{it} \left(  \Delta^{\alpha} H \Delta^{-\alpha} +  \Delta^{-\alpha} H \Delta^{\alpha} \right) \Delta^{-it}}{\cosh(\frac{\pi t}{2 \alpha^{\prime}})} \ .
\end{equation}
where we define $\alpha^{\prime} = \frac{1}{2} - \alpha$ to make the notation more compact. If we have a function $f(t)$ such that its Fourier transform $\hat{f}(\omega)$ satisfies that $e^{\beta \omega} \hat{f}(\omega)$ is in $L^{1}$ for some $\beta$, then we can define
\begin{equation}
    H(f) = \int_{-\infty}^{\infty} dt \ f(t) \Delta^{it} H \Delta^{-it}  \ .
\end{equation}
This operator can be evolved with modular flow at imaginary time $\beta$ and gives another bounded operator, as explained in section \ref{sS:connecting_cones}
\begin{equation}
    \Delta^{\beta} H(f) \Delta^{-\beta} = H(f_{\beta}), \quad 
    f_{\beta}(t) = \frac{1}{\sqrt{2 \pi}} \int_{-\infty}^{\infty} d \omega \ e^{i \omega (t + i \beta)} \hat{f}(\omega) = f(t + i \beta) \ .
\end{equation}
In our case, we have that
\begin{equation}
f(t) = \frac{1}{4 \alpha^{\prime}} \frac{1}{\cosh(\frac{\pi t}{2 \alpha^{\prime}})} \Rightarrow 
\hat{f}(\omega) = \frac{1}{2 \sqrt{2 \pi}} \frac{1}{\cosh(\alpha^{\prime} \omega )} \ .
\end{equation}
We have to evolve $H$ at imaginary time $\pm \alpha$, therefore, the function has to satisfy that
\begin{equation}
    \lim_{\omega \rightarrow \pm \infty} \frac{1}{2 \sqrt{2 \pi}} \frac{e^{\pm \alpha \omega}}{\cosh(\alpha^{\prime} \omega )} = 0 \quad
    \Rightarrow \alpha \leq \alpha^{\prime} = \frac{1}{2} - \alpha 
    \Rightarrow \alpha \leq \frac{1}{4} \ ,
\end{equation}
which is the range of values of $\alpha$ within which we are working.
In conclusion, we can give a more compact and simpler form of $Q_{\alpha}$
\begin{equation}
    Q_{\alpha} = \int_{- \infty}^{\infty} dt \ L_{\alpha}(t) \Delta^{it} J H J \Delta^{-it},
\label{eq:Qalpha}
\end{equation}
where
\begin{equation}
\begin{split}
    L_{\alpha}(t) 
    &= \frac{1}{2 \pi} \int_{-\infty}^{\infty} d\omega \ \frac{\cosh(\alpha \omega)}{\cosh \left( \left(\frac{1}{2} - \alpha \right) \omega \right)} e^{i \omega t}  \\
    &= \frac{1}{2 (1 - 2 \alpha)} \left[\frac{1}{\cosh(\frac{\pi}{1-2\alpha}(t + i \alpha))} + \frac{1}{\cosh(\frac{\pi}{1-2\alpha} (t - i \alpha))} \right] \ .
\label{eq:L_fun}
\end{split}
\end{equation}
Finally, we have that the representative in the cone $\mathcal{P}_{\Omega}^{\alpha}(\mathcal{A})$ of the automorphism given by the inner $U_{\mathcal{A}}(\epsilon) = e^{i\epsilon H}$ on the algebra $\mathcal{A}$ is, to first order in $\epsilon$,
\begin{equation}
    U_{\alpha}(\beta_{\epsilon}) = U_{\mathcal{A}}(\epsilon) W_{\alpha}(\epsilon)
    = I + i (H - Q_{\alpha}) \epsilon + O(\epsilon^{2})\ .
\label{eq:firstordere}
\end{equation}
\indent It is interesting to note that when $\alpha = 1/4$, we have that $L_{1/4}(t) = \delta(t)$ and therefore $Q_{1/4} = J H J$, which recovers the general result mentioned in \eqref{eq:repinnerstandard}.

\newpage

\section{Optimal symmetry operators}
\label{S:Best_Operator}

Suppose that we have a Hilbert space $\mathcal{H}$, a von Neumann algebra $\mathcal{A} \subset \mathcal{B}(\mathcal{H})$, an automorphism $\beta: \mathcal{A} \rightarrow \mathcal{A}$, and a vector $\ket{\Omega}$ which is cyclic and separating for $\mathcal{A}$. 
We discuss in section \ref{S:RepresentationAutomorphism} that there are many isometries $U(\beta)$ that implement the automorphism on $\mathcal{A}$. Moreover, we show how to construct a family of them. 
In this section, we explore which of all the possible representatives of $\beta$ has the maximal expectation value, and we find that the representative in the cone $\mathcal{P}_{\Omega}^{0}(\mathcal{A})$ is the one that achieves this.

Let $\mathcal{I}_{\mathcal{A}}(\beta)$ be the set of all representatives of $\beta$
\begin{equation}
    \mathcal{I}_{\mathcal{A}}(\beta) = \{ U(\beta) \in \mathcal{B}(\mathcal{H}) \ | \ U(\beta) A = \beta(A) U(\beta) \ \   A \in \mathcal{A}, \ U(\beta)^{\dagger} U(\beta) = I \} \ .
\end{equation}
The expectation values of the representatives are bounded
\begin{equation}
    0 \leq \abs{\expval{U(\beta)}{\Omega}} \leq 1 \quad \forall \ U(\beta) \in \mathcal{I}_{\mathcal{A}}(\beta) \ .
\end{equation}
Then, the absolute value of the expectation values of the representatives have a supremum. If this supremum is attained by an operator $U_{0}(\beta) \in \mathcal{I}_{\mathcal{A}}(\beta)$, we can, without loss of generality, assume that $\expval{U_{0}(\beta)}{\Omega} > 0$. When such an operator $U_{0}(\beta)$ exists, we call it the \textit{optimal symmetry operator} associated with the automorphism $\beta$, the algebra $\mathcal{A}$ and the vector $\ket{\Omega}$.  Using the results developed in section \ref{S:FidelityP0}, it is straightforward to show that this supremum is achieved by the representative of $\beta$ in the cone $\alpha = 0$.

Let $U(\beta) \in \mathcal{I}_{\mathcal{A}}(\beta)$ be an arbitrary representative, and let $\ket{\Phi} = U(\beta) \ket{\Omega}$. For every representative, the reduction of $\ket{\Phi}$ over $\mathcal{A}$ gives the state
\begin{equation}
\begin{split}
  \phi(A) 
  &= \expval{A}{\Phi} \\
  &= \expval{U(\beta)^{\dagger} A U(\beta)}{\Omega} 
  = \expval{\beta^{-1}(A)}{\Omega} 
  = \omega(\beta^{-1}(A)) \quad \text{for} \ A \in \mathcal{A} \ ,
\end{split}
\end{equation}
where $\omega(A) = \expval{A}{\Omega}$ is the reduction of $\ket{\Omega}$ on $\mathcal{A}$.
Therefore, because the reduced state does not depend on the chosen representative, we can consider every $\ket{\Phi}$ as a purification of $\phi$.
Then, by Uhlmann's theorem \cite{Uhlmann:1975kt}, the fidelity between $\omega$ and $\phi$ provides an upper bound for the expectation value of the representatives
\begin{equation}
    \expval{U(\beta)}{\Omega} = \braket{\Omega}{\Phi} \leq F(\Omega, \phi) \ .
\end{equation}
In section \ref{S:FidelityP0}, we show that the fidelity between $\omega$ and $\phi$ is given by the purification $\ket{\phi_{0}} \in \mathcal{P}_{\Omega}^{0}(\mathcal{A})$ of $\phi$. In section \ref{S:RepresentationAutomorphism}, we show that we can always construct a representative $U_{0}(\beta) \in \mathcal{I}_{\mathcal{A}}(\beta)$ such that $U_{0}(\beta) \ket{\Omega} \in \mathcal{P}_{\Omega}^{0}(\mathcal{A})$. Therefore, we have that $U_{0}(\beta) \ket{\Omega} = \ket{\phi_{0}}$ and
\begin{equation}
    \expval{U_{0}(\beta)}{\Omega} = \braket{\Omega}{\phi_{0}} = F(\omega, \phi) \ .
\end{equation}
In conclusion, the supremum of the expectation values of the operators in $\mathcal{I}_{\mathcal{A}}(\beta)$ is achieved by the representative $U_{0}(\beta)$ constructed using the cone $\mathcal{P}_{\Omega}^{0}(\mathcal{A})$ and is unique, as we showed in section \ref{S:RepresentationAutomorphism}. In other words, we have that the \textit{optimal symmetry operator} is the representative $U_{0}(\beta)$ of $\beta$ in the cone $\mathcal{P}_{\Omega}^{0}(\mathcal{A})$.

In section \ref{S:Representativeinthecone} we showed that, using the standard purification and the isometry connecting the cones, we can construct the \textit{optimal symmetry operator} using modular tools. In sections \ref{S:ExampleSpins} and \ref{S:Scalar}, we give some examples of these constructions.

\section{Example: Two interacting spins}
\label{S:ExampleSpins}

Our Hilbert space will be $\mathcal{H} = \mathbb{C}^{2} \otimes \mathbb{C}^{2}$, and our algebra of interest will be $\mathcal{A} = \mathcal{M}_{2}(\mathbb{C}) \otimes I$. The canonical basis of the Hilbert space is taken to be $\{ \ket{00},\ket{01},\ket{10},\ket{11}\}$. Let the Pauli matrices be
\begin{equation}
    \sigma_x = 
    \begin{pmatrix}
    0 & 1 \\
    1 & 0
    \end{pmatrix}, \quad
    \sigma_y = 
    \begin{pmatrix}
    0 & i \\
    -i & 0
    \end{pmatrix}, \quad
    \sigma_z = 
    \begin{pmatrix}
    1 & 0 \\
    0 & -1
    \end{pmatrix} \ ,
\end{equation}
and the spin operators 
\begin{equation}
    \vec{S}_{1} = (\sigma_{x} \otimes I, \sigma_{y} \otimes I, \sigma_{z} \otimes I) \ , \quad
    \vec{S}_{2} = (I \otimes \sigma_{x}, I \otimes \sigma_{y}, I \otimes \sigma_{z}), \quad
    \vec{S}_{T} = \vec{S}_{1} + \vec{S}_{2} \ .
\end{equation}
The problem to analyze consists of two spins interacting via a Hamiltonian
\begin{equation}
    H = - k  \vec{S}_{1} \cdot \vec{S}_{2} 
    = - \frac{k}{2} (S_{T}^{2} - S_{1}^{2} - S_{2}^{2}) 
    = \frac{k}{2} (6 - S_{T}^{2}) \ ,
\end{equation}
where $k>0$. This Hamiltonian can be diagonalized using the eigenstates
\begin{equation}
    \ket{E} = \frac{\ket{01} - \ket{10}}{\sqrt{2}}, \
    \ket{\Omega_{1}} = \ket{00}, \
    \ket{\Omega_{2}} = \ket{11} \ ,
    \ket{\Omega_{3}} = \frac{\ket{01} + \ket{10}}{\sqrt{2}}, \
\end{equation}
where $\ket{E}$ is a vector with eigenvalue $3k$, and  $\{ \ket{\Omega_{i}} \}$ span the eigenspace of the eigenvalue $-k$, which is the lower one. Then we have that the fundamental state is degenerated. We set our reference vector as
\begin{equation}
    \ket{\Omega} = \sqrt{a} \ket{00} + \sqrt{1-a} \ket{11} \ .
\label{eq:def_vacumm_qubits}
\end{equation}
This theory is rotationally invariant. Therefore, a symmetry of the system
is a rotation about the x-axis of the spins. A rotation by an angle of $\pi / 2$ on one spin is given by
\begin{equation}
    R_{x} = e^{i \frac{\pi}{4} \sigma_{x}} = \frac{I + i \sigma_{x}}{\sqrt{2}} \ .
\end{equation}
The global symmetry operator can be written as
\begin{equation}
    U_{x} = R_{x} \otimes R_{x}, \quad \left[H, U_{x} \right] = 0 \ .
\end{equation}
The reference vector is not invariant under the action of the symmetry
\begin{equation}
    \ket{\xi} = U_{x} \ket{\Omega} = \sqrt{a} \ket{1_{y} 1_{y}} + \sqrt{1-a} \ket{0_{y} 0_{y}} \ ,
\end{equation}
where $\ket{0_{y}} = (i \ket{0} + \ket{1})/\sqrt{2}$ and $\ket{1_{y}} =  (\ket{0} + i \ket{1})/\sqrt{2}$.

Now we want to use this example to construct different representatives of an automorphism over the cones $\mathcal{P}_{\Omega}^{\alpha}(\mathcal{A})$. The automorphism $\beta$ will be the action of the symmetry $U_{x}$ on the algebra $\mathcal{A} = \mathcal{M}_{2}(\mathbb{C}) \otimes I$. Because we are in the finite-dimensional case, all automorphisms are inner, and in this case, the inner unitary that implements it is $R_{x} \otimes I \in \mathcal{A}$.

To get a non-trivial example, the reference vector cannot be invariant under the global symmetry. If this is not the case and the symmetry keeps the reference vector unchanged, we have that $U_{x} \ket{\Omega} = \ket{\Omega} \in \mathcal{P}_{\Omega}^{\alpha}(\mathcal{A})$ for 
any $\alpha \in \left[0, 1/4 \right]$. Therefore, $U_{x}$ is the representative in all the cones.

To continue with the example, we obtain the reduced density matrix $\rho$ over $\mathcal{A}$ as
\begin{equation}
    \rho = \Tr_{2} (\ketbra{\Omega}{\Omega}) = \begin{pmatrix} a & 0 \\ 0 & 1-a \\ \end{pmatrix} \Rightarrow \ket{\Omega} = \ket{\sqrt{\rho}} \ ,
\label{eq:density_matrix}
\end{equation}
where we used the notation explained in section \ref{sS:cones_finite}.
The modular tools associated with the reference vector $\ket{\Omega}$ are \cite{Witten:2018zxz}
\begin{equation}
    \Delta = \rho \otimes \rho^{-1}, \quad J = \text{SWAP} *, \quad J(Q \otimes I)J = I \otimes Q^{*} \ ,
\end{equation}
where the SWAP operator is the operator defined by $\text{SWAP} \ket{ij} = \ket{ji}$, $*$ denotes the complex conjugation in the canonical basis, and $Q \in \mathcal{M}_{2}(\mathbb{C})$ is any 2x2 matrix. To construct the representative of the automorphism over the different cones, we have to identify the state defined in \eqref{eq:state_phi}. We have that
\begin{equation}
\begin{split}
    \phi(A) &= \expval{\beta^{-1}(A \otimes I)}{\Omega}
    = \expval{R_{x}^{\dagger} A R_{x} \otimes I}{\Omega} \\
    &= \Tr_{1,2} ( \ketbra{\Omega}{\Omega} (R_{x}^{\dagger} A R_{x} \otimes I))
    = \Tr( \rho \ R_{x}^{\dagger} A R_{x}) \\
    &= \Tr( R_{x} \rho R_{x}^{\dagger} \ A )
    = \Tr( \sigma A ) \ .
\end{split}
\end{equation}
Therefore, the state $\phi$ is associated with the density matrix $\sigma = R_{x} \rho R_{x}^{\dagger}$. This density matrix has eigenvectors $\ket{0_{y}}$ with eigenvalue $1-a$ and $\ket{1_{y}}$ with eigenvalue $a$. Following section \ref{sS:cones_finite}, the purification of this state on the standard cone $\mathcal{P}_{\Omega}^{1/4}(\mathcal{A})$ is given by
\begin{equation}
    \ket{\phi} = \ket{\sqrt{\sigma}} = - i (\sqrt{a} \ket{1_{y} 0_{y}} + \sqrt{1-a} \ket{0_{y} 1_{y}}) \ ,
\end{equation}
Using the fact that the automorphism is inner, we construct the standard representative as we show in \eqref{eq:repinnerstandard}
\begin{equation}
    U_{x, 1/4} = (R_{x} \otimes I) J (R_{x} \otimes I) J = R_{x} \otimes R_{x}^{*} \ .
\end{equation}
It verifies that
\begin{equation}
    U_{x, 1/4} \ket{\Omega} = \ket{\phi} \in \mathcal{P}_{\Omega}^{1/4}(\mathcal{A}) \ .
\end{equation}
\indent To construct the representatives in the cones $\alpha$, we can use \eqref{eq:def_Talpha1} or the explicit form given in \eqref{eq:R_def_2} to get the operator $R_{\alpha}$ that connects the standard representative with the other representatives $U_{x, \alpha} = R_{\alpha} U_{x, 1/4}$. It is interesting to note that these representatives do not commute with the Hamiltonian, meaning that they are not global symmetries. We will not show the explicit form of the matrices $R_{\alpha}$ or $U_{x,\alpha}$ here because it has a complicated, uninformative expression.
Using this explicit form, we calculate the vector $\ket{\phi_{\alpha}} = U_{x, \alpha} \ket{\Omega}$, and verify that $\ket{\phi_{\alpha}} \in \mathcal{P}_{\Omega}^{\alpha}(\mathcal{A})$ using the definition \eqref{eq:def_pos_cone_finite} for the cones.

Following the results from section \ref{S:Best_Operator}, we can explicitly verify that the representative $U_{x, 0}$ is the one with the maximal expectation value among all the representatives. To numerically check this, we use the parametrization of the unitary matrices of size $2 \times 2$
\begin{equation}
    U_{p}(a,b,c,d) = e^{i a}
    \begin{pmatrix}
    e^{i b} \cos c & e^{i d} \sin c \\
    - e^{-i d} \sin c & e^{-i b} \cos c
    \end{pmatrix} \ a,b,c,d \in \mathbb{R} \ .
\end{equation}
We construct the general representative of the automorphism as
\begin{equation}
    U_{x,G}(a, b, c,d) = R_{x} \otimes U_{p}(a,b,c,d) \ .
\end{equation}
With this expression, we can cover all the possibilities, and we check that the representative $U_{x, 0}$ in the cone $\alpha = 0$ gives the representative with the maximal expectation value.

\begin{figure}[H]
    \centering
    \includegraphics[width=0.7\linewidth]{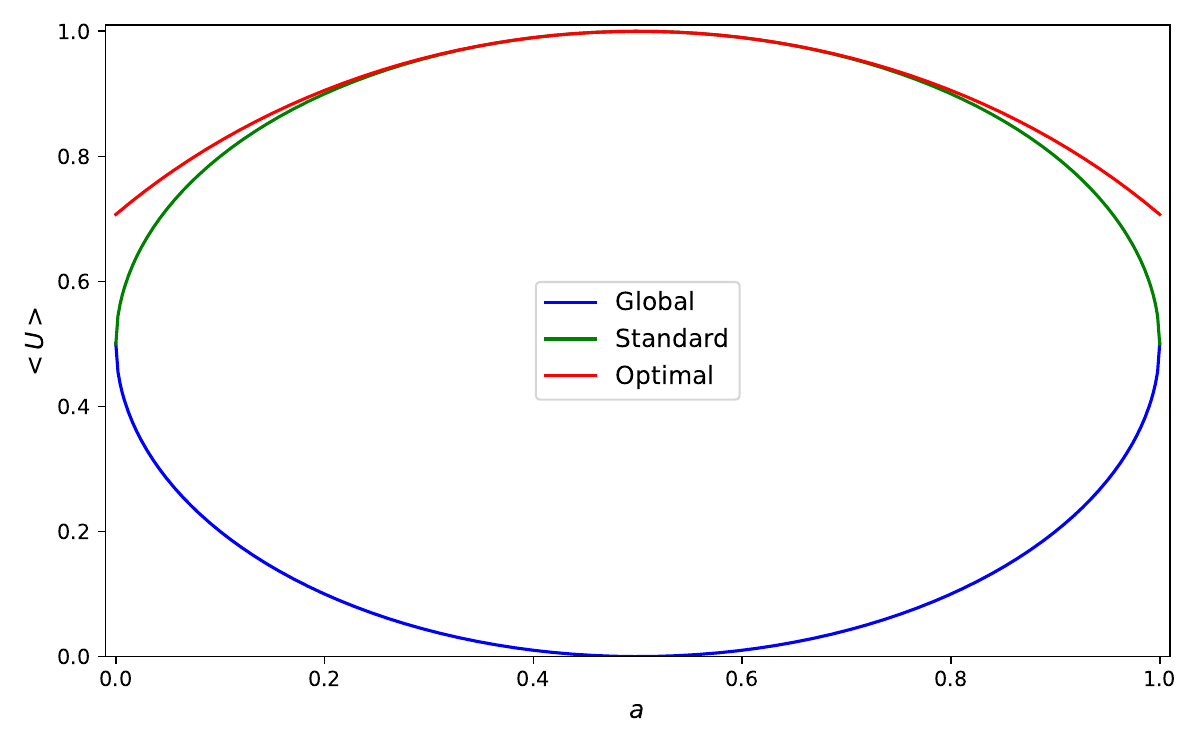}
    \caption{Expectation value of the global $U_{x}$, standard $U_{x, 1/4}$ and the optimal $U_{x, 0}$ in function of the parameter $a$.}
    \label{fig:Plot_U_vacumm}
\end{figure}

For completeness, the expectation value of the global, standard, and the optimal unitary are
\begin{equation}
\begin{split}
    &\expval{U_{x}}{\Omega} = \frac{1}{2} - \sqrt{a (1-a)} \ , \\
    &\expval{U_{x, 1/4}}{\Omega} = \frac{1}{2} + \sqrt{a (1-a)} \ , \\
    &\expval{U_{x, 0}}{\Omega} = \frac{1}{2} \left( \sqrt{1 + (1 - 2a) \sqrt{1 + 4a - 4a^2}} + \sqrt{1 - (1 - 2a) \sqrt{1 + 4a - 4a^2}} \right) \ .
\end{split}
\end{equation}
In Figure \ref{fig:Plot_U_vacumm}, we can see a plot of these functions. For $a = 1/2$, the standard and the optimal achieve the value $\expval{U_{0,x}}{\Omega} = \expval{U_{1/4,x}}{\Omega} = 1$. In this case, the standard representative leaves the reference vector invariant, $U_{1/4,x} \ket{\Omega} = \ket{\Omega}$, and thus it coincides with all the representatives over the cones.

\begin{figure}[H]
    \centering
    \includegraphics[width=0.7\linewidth]{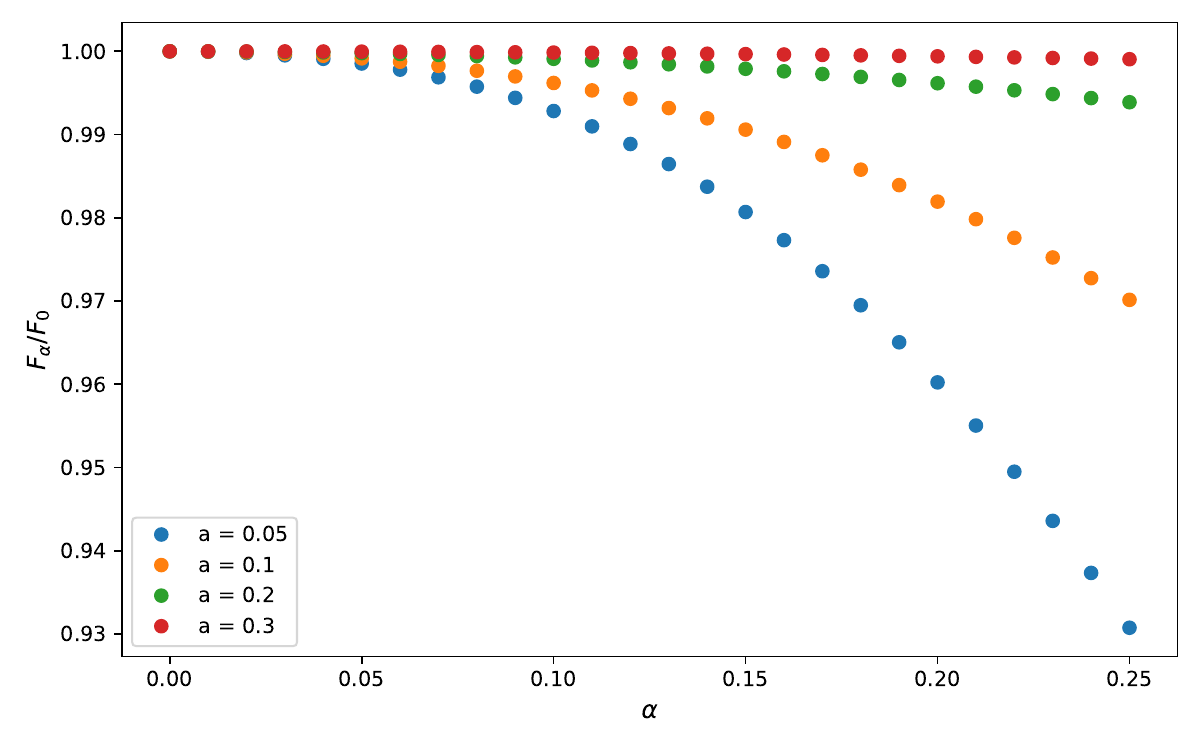}
    \caption{
    Generalize fidelity normalized $F_{\alpha}(\rho,\sigma) / F(\rho, \sigma)$ as a function of the parameter $\alpha$ for different values of the parameter $a$.
    }
    \label{fig:plot_G_Fidelity}
\end{figure}

Furthermore, in this example, we can study the expectation value of the different representatives over the cones $\alpha \in \left[0, 1/4 \right]$. This is equal to the generalize fidelity between $\rho$ and $\sigma$ proposed in \eqref{eq:gen_fidelity}
\begin{equation}
    \expval{U_{x, \alpha}}{\Omega} = F_{\alpha}(\rho, \sigma) = \Tr \left( \rho^{2 \alpha} \sqrt{\rho^{1/2 - 2 \alpha} \sigma \rho^{1/2 - 2 \alpha}} \right) \ .
\end{equation}
In figure \ref{fig:plot_G_Fidelity}, we can see the form of this generalized fidelity as a function of $\alpha$ for different values of $a$. We observe that, as the parameter $a$ approaches $1/2$, the generalized fidelity approaches the same value for all $\alpha$, as we discussed before. After $a = 1/2$, all the values are repeated.

\newpage

\section{Example: Massless scalar field in 3+1}
\label{S:Scalar}
We will work with the theory of a massless scalar field $\phi(x)$ in $3+1$ dimensions. This is a conformal field theory (CFT). The field has dimension $\Delta_{\phi} = 1$ and has a two-point function given by
\begin{equation}
    \expval{\phi(x) \phi(y)} = \frac{1}{8 \pi \abs{x-y}^{2}} \ ,
\end{equation}
where $x$ and $y$ are spacetime coordinates.
At the slice $t=0$, we have that the theory can be written with the operators $\phi(\vec{x}) = \phi(0, \vec{x})$ and $\pi(\vec{y}) = \partial_{t} \phi(t, \vec{y}) |_{t=0}$ \footnote{For details of the mathematical formulation of the free field algebra on the $t = 0$ surface, see \cite{araki1964neumann, Araki:1963klf}. For a compact review, see \cite{Casini:2019qst}.}, which satisfy the following commutation relations
\begin{equation}
    \left[\phi(\vec{x}), \pi(\vec{y}) \right] = i \delta(\vec{x} - \vec{y}) \ , \quad
    \left[\phi(\vec{x}), \phi(\vec{y}) \right] = \left[\pi(\vec{x}), \pi(\vec{y}) \right] = 0 \ .
\end{equation}
We define the coherent operator
\begin{equation}
    U(v) = e^{i v \phi(F)}, \quad 
    \phi(F) = \int_{\mathbb{R}^{3}} dx^{3} \ F(\vec{x}) \ \phi(\vec{x}) \ ,
\end{equation}
where $F(\vec{x})$ is a function with compact support inside the sphere of radius $R$, which we will call $\mathbb{B}$, and $v \in \mathbb{R}$. The operator $U(v)$ is a unitary operator that belongs to the algebra $\mathcal{A}$ of the sphere $\mathbb{B}$ and, over $\mathcal{A}$, it applies the automorphism $\beta: \mathcal{A} \rightarrow \mathcal{A}$ 
\begin{equation}
\begin{split}
    \beta(\phi(\vec{x})) &= U(v) \phi(\vec{x}) U(v)^{\dagger} = \phi(\vec{x}) \ , \\
    \beta(\pi(\vec{x})) &= U(v) \pi(\vec{x}) U(v)^{\dagger} = \pi(\vec{x})  - F(\vec{x}) v \ ,
\end{split}
\end{equation}
for $\phi(\vec{x}), \pi(\vec{x}) \in \mathcal{A}$.
The goal of this example is to calculate, for this automorphism, the representatives over the cones $\mathcal{P}_{\Omega}^{\alpha}(\mathcal{A})$. Another goal is to check that the representative in the cone $\mathcal{P}_{\Omega}^{0}(\mathcal{A})$ is the one that gives the maximal expectation value. It is shown in Appendix \ref{A:CoherentRepresentative} that, because the modular operator $\Delta$ is quadratic in the fields, the representative operators over the different cones are also coherent operators and linear in the parameter $v$ of the transformation. 

In this example, we will take $F(\vec{x})$ as follows
\begin{equation}
    F(\vec{x}) =  \left\{ \begin{matrix} 1 & \abs{\vec{x}} \leq R \\ 0 & \abs{\vec{x}} > R  \\ \end{matrix}\right. \ .
\label{eq:smearingfunction}
\end{equation}

\newpage

\subsection{Dimensional reduction}
\label{sS:dimensionalReduction}

The scalar field in 3+1 dimensions can be decomposed using spherical harmonics
\begin{equation}
    \phi(t, \vec{x}) = \sum_{l,m} \phi_{l,m}(t, r) Y_{l}^{m}(\varphi, \theta) \ ,
\end{equation}
Integrating out the angular part, the action decomposes in an action for every mode $S = \sum_{l,m} S_{l,m}$ where each mode $\phi_{l,m}(t,r)$ lives on the half-line in $1+1$ dimensions \cite{Huerta:2022tpq, Srednicki:1993im}. Therefore, every field $\phi_{l,m}(t,r)$ is independent of the others.

Because our function is spherically symmetric $F(\vec{x}) = F(r)$, where $r = \abs{\vec{x}}$, when we integrate against the field we will obtain
\begin{equation}
    \phi(F) = \int_{0}^{\infty} dr \int d\Omega  \ r^{2} F(r) \phi(0, r, \Omega)
    = \int_{0}^{\infty} dr \ \sqrt{4\pi}  r F(r) \phi_{0}(0,r) = \phi_{0}(f)
\end{equation}
where we define the zero-mode field ($l=0, m=0$) as
\begin{equation}
    \phi_{0}(t,r) = \frac{r}{\sqrt{4 \pi}} \int d\Omega \ \phi(t, \vec{x}) \ .
\label{eq:def_zeromode}
\end{equation}
The smearing function in the $l=0$ theory is obtained from the one in the 3+1 theory as
\begin{equation}
    f(r) = \sqrt{4 \pi} r F(r) \ .
\label{eq:zeroto4functions}
\end{equation}
The new field $\phi_{0}(t, r)$ lives in a theory on the half-line and satisfies Dirichlet boundary condition $\phi_{0}(t,0) = 0$.
Because the fields for different modes are independent, $U(v) = e^{i v \phi(F)} = e^{i v \phi_{0}(f)}$ is a unitary operator in the $l=0$ theory. In \cite{Huerta:2022tpq}, it was shown that the modular tools also decompose into every mode, like the action. Therefore, the representatives in different cones will belong to the $l = 0$ mode theory as well, because they are constructed with the modular tools. Hence, we will restrict our attention to the $l=0$ theory.

We can expand the field $\phi(t, \vec{x})$ in modes as
\begin{equation}
    \phi(t, \vec{x}) = \int \frac{d^{3}\vec{p}}{\sqrt{2 p (2 \pi )^{3}}} \left( a_{\vec{p}} \ e^{-i(p t - \vec{p} \cdot \vec{x})} + a_{\vec{p}}^{\dagger} \ e^{i (p t - \vec{p} \cdot \vec{x})} \right) \ ,
\end{equation}
where $p = \abs{\vec{p}}$, and $a_{\vec{p}}$ and $a_{\vec{p}}^{\dagger}$
are the annihilation and creation operators, respectively.
Explicitly integrating out the angular part, the $l=0$ mode field takes the form
\begin{equation}
    \phi_{0}(t, r) = \frac{r}{\sqrt{4 \pi}}\int_{0}^{2 \pi} \int_{0}^{\pi} d\varphi d\theta \sin(\theta) \ \phi(t, \vec{x}) 
    = \frac{1}{\sqrt{\pi}} \int_{0}^{\infty} dp \ \frac{1}{\sqrt{p}} \sin(p r) (a_{p} e^{-ipt} + a_{p}^{\dagger} e^{ipt}) \ ,
\label{eq:phitimefourier}
\end{equation}
where we define the creation and annihilation operators of the $l=0$ mode theory as
\begin{equation}
    a_{p} = \frac{p}{\sqrt{4 \pi}} \int_{0}^{2 \pi} \int_{0}^{\pi} d\varphi d\theta \sin(\theta) \ a_{\vec{p}}, \quad
    \left[ a_{p}, a_{p^{\prime}}^{\dagger} \right] = \delta(p - p^{\prime}) \ .
\end{equation}
Another useful relation is the identity
\begin{equation}
    \int_{0}^{\infty} dx \sin(u x) \sin(v x) = \frac{\pi}{2} \delta(u - v) \ .
\end{equation}
The action of this theory is \cite{Huerta:2022tpq}
\begin{equation}
    S_{0} = \int dt dr \frac{1}{2} \left[  - (\partial_{t} \phi_{0}(t,r))^{2} + (\partial_{r} \phi_{0}(t,r))^{2} \right] \ .
\label{eq:actionhalfline}
\end{equation}
On the slice $t=0$, we have the operators $\phi_{0}(r)$ and $\pi_{0}(r) = \partial_{t} \phi_{0}(t,r)|_{t=0}$ given by
\begin{equation}
    \phi_{0}(r) = \frac{1}{\sqrt{\pi}} \int_{0}^{\infty} dp \ \frac{1}{\sqrt{p}} \sin(p r) (a_{p} + a_{p}^{\dagger}) \ , \quad
    \pi_{0}(r) = - \frac{i}{\sqrt{\pi}} \int_{0}^{\infty} dp \ \sqrt{p} \sin(p r) (a_{p} - a_{p}^{\dagger}) \ ,
\label{eq:fieldsfourier}
\end{equation}
that have commutation relations and two-point functions
\begin{equation}
\begin{split}
    & \left[ \phi_{0}(r), \pi_{0}(r^{\prime}) \right] = i \delta(r - r^{\prime}), \quad 
    \left[ \phi_{0}(r), \phi_{0}(r^{\prime}) \right] 
    = \left[ \pi_{0}(r), \pi_{0}(r^{\prime}) \right] = 0 \ ,
    \\
    &\expval{\phi_{0}(r) \phi_{0}(r^{\prime})}{\Omega} = \frac{1}{2\pi} \ln\left( \abs{ \frac{r + r^{\prime}}{r - r^{\prime}} } \right), \\
    &\expval{\phi_{0}(r) \pi_{0}(r^{\prime})}{\Omega} = \frac{i}{2} \delta(r - r^{\prime}), \\
    &\expval{\pi_{0}(r) \pi_{0}(r^{\prime})}{\Omega} = - \frac{2 r r^{\prime}}{\pi (r^{2} - \left. r^{\prime}\right.^{2})^{2}} \ .
\end{split}
\end{equation}
\indent This angular dimensional reduction was used to calculate the entropy of the sphere in \cite{Srednicki:1993im} and also the modular Hamiltonian was obtained for all the modes in \cite{Huerta:2022tpq}.

\subsection{Standard operator}
\label{sS:scalarstandard}
The modular tools for a sphere in a CFT are well known due to the Hislop-Longo theorem \cite{Hislop:1981uh} (a brief derivation is given in \cite{Casini:2011kv}). Let $J_{\mathbb{B}}$ be the modular conjugation associated with the algebra of the sphere $\mathbb{B}$ and the vacuum vector $\ket{\Omega}$. Its action on a scalar field with $r < R$ at $t=0$ is given by
\begin{equation}
    J_{\mathbb{B}} \phi(r, \theta, \varphi) J_{\mathbb{B}} = \left(\frac{R}{r}\right)^{2} \phi(R^{2}/r, \theta, \varphi) \ .  
\end{equation}
Its action on the $l=0$ mode field is derived from \eqref{eq:def_zeromode}
\begin{equation}
    J_{\mathbb{B}} \phi_{0}(r) J_{\mathbb{B}} = \phi_{0}(R^{2}/r) \ .
\end{equation}
The representative of the automorphism in the standard cone $\alpha = 1/4$ can be directly calculated using \eqref{eq:repinnerstandard}
\begin{equation}
     U_{\text{std}}(v) = U(v) J_{\mathbb{B}} U(v) J_{\mathbb{B}} =  e^{i v (\phi_{0}(f) - \phi_{0}(g_{\text{std}}))} \ .
\end{equation}
where
\begin{equation}
    \phi_{0}(g_{\text{std}}) = J_{\mathbb{B}} \phi_{0}(f) J_{\mathbb{B}} = \int_{0}^{R} \ dr \ \sqrt{4 \pi} r J_{\mathbb{B}} \phi_{0}(r) J_{\mathbb{B}} = \int_{R}^{\infty} dr \sqrt{4 \pi} r \left( \frac{R}{r} \right)^{4} \phi_{0}(r) \ \\
\end{equation}
\begin{equation}
    \Rightarrow g_{\text{std}}(r) = \sqrt{4 \pi} r \left( \frac{R}{r} \right)^{4} \ .
\label{eq:beta_standard}
\end{equation}
It is the standard one because $U_{\text{std}}(v)$ applies the automorphism on $\mathcal{A}$ and commutes with the modular conjugation $J_{\mathbb{B}}$, as discussed in section \ref{sS:std_rep}.

\subsection{Representative in cone $\mathcal{P}_{\Omega}^{\alpha}(\mathcal{A})$}
\label{sS:RepresentativeConeScalar}

The automorphism is an inner one-parameter group $U(v) = e^{i v \phi_{0}(v)} \in \mathcal{A}$. In section \ref{sS:RepInner}, we show that an inner is always completed with a unitary $W_{\alpha}(v) \in \mathcal{A}^{\prime}$ to give the representative in the cone $\alpha$. In Appendix \ref{A:CoherentRepresentative}, it is proved that this unitary is also a coherent operator $W_{\alpha}(v) = e^{-i v (\phi_{0}(g_{\alpha}) - \pi_{0}(h_{\alpha}))}$, where $g_{\alpha}(r)$ and $h_{\alpha}(r)$ are smearing functions with support outside the ball $\mathbb{B}$. Therefore, the representative of the automorphism $\beta$ in the cone $\alpha$ will be a coherent operator with parameter $v$
\begin{equation}
    U_{\alpha}(v) 
    = e^{iv (\phi_{0}(f) - \phi_{0}(g_{\alpha}) - \pi_{0}(h_{\alpha}))}
    = I + i \left[ 
        \phi_{0}(f) 
        - \phi_{0}(g_{\alpha}) 
        - \pi_{0}(h_{\alpha})
    \right] v + O(v^{2})\ ,
\label{eq:repcoherentanzat}
\end{equation}
In section \ref{sS:InnerOneparametergroup}, we show that if the automorphism is an inner one-parameter group, the coefficient of the linear term in $v$ in the Taylor expansion of $U_{\alpha}(v)$ is given by \eqref{eq:firstordere}.
However, because the representative has the form given in \eqref{eq:repcoherentanzat}, we can obtain a direct expression for the smearing functions using the formula \eqref{eq:Qalpha}
\begin{equation}
    \phi_{0}(g_{\alpha}) + \pi_{0}(h_{\alpha}) = \int_{-\infty}^{\infty} \ ds \ L_{\alpha}(s) \Delta_{\mathbb{B}}^{is} J_{\mathbb{B}} \phi_{0}(f) J_{\mathbb{B}} \Delta_{\mathbb{B}}^{-is} \ .
\label{eq:representativescalar}
\end{equation}
The function $L_{\alpha}(s)$ is given in \eqref{eq:L_fun}.
From this expression, we can obtain a computable form of $g_{\alpha}(r)$ and show that $h_{\alpha}(r) = 0$.

First, we will work out the expression
\begin{equation}
    \int_{-\infty}^{\infty} \ ds \ L_{\alpha}(s) \Delta_{\mathbb{B}}^{is} \phi_{0}(f) \Delta_{\mathbb{B}}^{-is}
    = \int_{-\infty}^{\infty} \int_{0}^{R} \ ds dr \ L_{\alpha}(s) f(r) \Delta_{\mathbb{B}}^{is} \phi_{0}(r) \Delta_{\mathbb{B}}^{-is} \ .
\label{eq:intLphimodular}
\end{equation}
For simplicity, we will rename $\Delta_{\mathbb{B}} = \Delta$. The field $\phi_{0}(r)$ evolved with the modular flow of the algebra of the sphere is \cite{Hislop:1981uh, Casini:2011kv}
\begin{equation}
    \Delta^{i s} \phi_{0}(r) \Delta^{-i s} = \phi_{0}(\widetilde{t},\widetilde{r}) \ ,
\end{equation}
where we use the shorthand $\widetilde{t} = \widetilde{t}(r,s)$ and $\widetilde{r} = \widetilde{r}(r,s)$ for 
\begin{equation}
\begin{split}
    \widetilde{t}(r,s) &= - \frac{R (R^{2} - r^{2}) \sinh(2 \pi s)}{R^2 + r^2 + (R^2-r^2) \cosh(2 \pi s)}, \\
    \widetilde{r}(r,s) &= \frac{2rR^{2}}{R^2 + r^2 + (R^2-r^2) \cosh(2 \pi s)} \ .
\end{split}
\end{equation}
Note that there is no conformal factor because the field has zero dimension. We can write the field at time $t$ in terms of operators on the slice $t=0$ as
\begin{equation}
    \phi_{0}(t,r) = \int_{0}^{\infty} dr^{\prime} \left( K(t, r, r^{\prime}) \phi_{0}(r^{\prime}) + T(t, r, r^{\prime}) \pi_{0}(r^{\prime}) \right) \ ,
\label{eq:fieldexpt0}
\end{equation}
\begin{equation}
    K(t, r, r^{\prime}) = -i \left[\phi_{0}(t,r), \pi_{0}(r^{\prime}) \right], \quad T(t, r, r^{\prime}) = i \left[\phi_{0}(t,r), \phi_{0}(r^{\prime}) \right] \ .
\end{equation}
Using the expressions \eqref{eq:phitimefourier}, \eqref{eq:fieldsfourier} and the commutation
relation of the creation and annihilation operators, we arrive at 
\begin{equation}
\begin{split}
    &K(t, r, r^{\prime}) = \frac{2}{\pi} \int_{0}^{\infty} dp \ \sin(p r) \sin(p r^{\prime}) \cos(p t) \ , \\
    &T(t, r, r^{\prime}) = \frac{2}{\pi} \int_{0}^{\infty} dp 
    \frac{1}{p} \sin(p r) \sin(p r^{\prime}) \sin(p t) \ .
\end{split}
\end{equation}
Then the integral \eqref{eq:intLphimodular} can be rewritten with fields at $t=0$ as
\begin{equation}
    \int_{-\infty}^{\infty} \int_{0}^{R} \int_{0}^{\infty} \ ds dr dr^{\prime} \ L_{\alpha}(s) f(r) \left( K(\widetilde{t}, \widetilde{r}, r^{\prime}) \phi_{0}(r^{\prime}) + T(\widetilde{t}, \widetilde{r}, r^{\prime}) \pi_{0}(r^{\prime}) \right) \ .
\end{equation}
This expression can be simplified using the following property 
\begin{equation}
    \widetilde{r}(r, -s) = \widetilde{r}(r, s) \ , \quad
    \widetilde{t}(r, -s) = -\widetilde{t}(r, s) \  \ .
\end{equation}
Then, we note that $T(\widetilde{t}, \widetilde{r}, r^{\prime})$ is odd in $s$, but $L_{\alpha}(s)$ is even in $s$, and we have to perform the integral
\begin{equation}
    \int_{-\infty}^{\infty} ds \ L_{\alpha}(s) T(\widetilde{t}(r,s), \widetilde{r}(r,s), r^{\prime}) \ .
\end{equation}
Therefore, the contribution corresponding to the field $\pi(r)$ at $t=0$ is zero, which implies that $h_{\alpha}=0$. Thus, the integral \eqref{eq:intLphimodular} reduces to
\begin{equation}
\begin{split}
    \int_{-\infty}^{\infty} ds \ L_{\alpha}(s) \Delta^{is} \phi_{0}(f) \Delta^{-is}
    &= \int_{0}^{R} dr^{\prime} \ 
        \left(\int_{-\infty}^{\infty} \int_{0}^{R} ds dr \ f(r) L_{\alpha}(s)
            K(\widetilde{t}, \widetilde{r}, r^{\prime}) 
        \right) \phi_{0}(r^{\prime}) \\
    &= \int_{0}^{R} dr^{\prime} \xi_{\alpha}(r^{\prime}) \phi_{0}(r^{\prime}) \ ,
\end{split}
\end{equation}
where we call
\begin{equation}
    \xi_{\alpha}(r^{\prime}) = 
        \int_{-\infty}^{\infty} \int_{0}^{R} ds dr \ f(r) L_{\alpha}(s)  
            K(\widetilde{t}, \widetilde{r}, r^{\prime}) \ .
\end{equation}
It is not difficult to calculate the function $\xi_{\alpha}(r)$.  We can express $K(t,r,r^{\prime})$ as 
\begin{equation}
\begin{split}
    K(t,r,r^{\prime}) 
    &= \frac{2}{\pi} \int_{0}^{\infty} dp \ \sin(p r) \sin(p r^{\prime}) \cos(p t) \\ 
    &= \frac{1}{2} \left[ \delta(r^{\prime} - (r - t)) + \delta(r^{\prime} - (r + t)) - \delta(r^{\prime} + (r - t)) - \delta(r^{\prime} + (r + t)) \right] \ .
\end{split}
\end{equation}
If we use the property $\widetilde{r}(-r,s) = - \widetilde{r}(r,s)$ and define $f_{-}(r)$ as the odd extension of $f(r)$ to the full line, we can rewrite $\xi_{\alpha}(r^{\prime})$ as
\begin{equation}
    \xi_{\alpha}(r^{\prime}) 
    = 
        \int_{-\infty}^{\infty} \int_{0}^{R} ds dr \ f(r) L_{\alpha}(s)   
            K(\widetilde{t}, \widetilde{r}, r^{\prime})
    = 
        \int_{-\infty}^{\infty} \int_{-R}^{R} ds dr \ f_{-}(r) L_{\alpha}(s)   
            \widetilde{K}(\widetilde{t}, \widetilde{r}, r^{\prime}) \ ,
\label{eq:xiequation}
\end{equation}
where
\begin{equation}
    \widetilde{K}(t, r, r^{\prime}) = \frac{1}{2} \left[ \delta(r^{\prime} - (r - t)) + \delta(r^{\prime} - (r + t)) \right] \ .
\end{equation}
To reduce the notation, we define the null coordinates
\begin{equation}
    \widetilde{r}^{\pm}(r,s) = \widetilde{r}(r,s) \pm \widetilde{t}(r,s) = R \frac{ \left( R + r \right) - e^{\mp 2 \pi s} \left( R - r \right) }{ \left( R + r \right) + e^{\mp 2 \pi s} \left( R - r \right) } \ .
\end{equation}
This is useful to express $\widetilde{K}(\widetilde{t}, \widetilde{r}, r^{\prime})$ as
\begin{equation}
    \widetilde{K}(\widetilde{t}, \widetilde{r}, r^{\prime}) = \frac{1}{2} \left[\delta(r^{\prime} - \widetilde{r}^{-}(r,s)) + \delta(r^{\prime} - \widetilde{r}^{+}(r,s)) \right] \ .
\end{equation}
Because we are integrating with respect to $r$ in \eqref{eq:xiequation}, we will have that 
\begin{equation}
\begin{split}
    \xi_{\alpha}(r^{\prime}) 
    &= 
        \int_{-\infty}^{\infty} \int_{-R}^{R} ds dr \ f_{-}(r) L_{\alpha}(s)   
            \frac{1}{2} \left[\delta(r^{\prime} - \widetilde{r}^{-}(r,s)) + \delta(r^{\prime} - \widetilde{r}^{+}(r,s)) \right]) \\ 
    &= 
        \frac{1}{2} \int_{-\infty}^{\infty} ds \ L_{\alpha}(s)  
        \left[ \frac{f_{-}(\widetilde{r}_{\text{inv}}^{-}(r^{\prime},s))}{\abs{\partial_{r} \widetilde{r}^{-}(\widetilde{r}_{\text{inv}}^{-}(r^{\prime},s), s)}} + \frac{f_{-}(\widetilde{r}_{\text{inv}}^{+}(r^{\prime},s))}{\abs{\partial_{r} \widetilde{r}^{+}(\widetilde{r}_{\text{inv}}^{+}(r^{\prime},s), s)}} \right] \ ,
\end{split}
\label{eq:xiequation2}
\end{equation}
where $\widetilde{r}_{\text{inv}}^{\pm}(r^{\prime},s)$ is the inverse of the function $\widetilde{r}^{\pm}(r,s)$ with respect to the variable $r$
\begin{equation}
    \widetilde{r}^{\pm}(\widetilde{r}_{\text{inv}}^{\pm}(r^{\prime},s),s) = r^{\prime} \ .
\end{equation}
From \eqref{eq:xiequation2}, we can calculate explicitly $\xi_{\alpha}(r)$. Thus, to summarize
\begin{equation}
    \int_{-\infty}^{\infty} ds \ L_{\alpha}(s) \Delta^{is} \phi_{0}(f) \Delta^{-is} 
    = \int_{0}^{R} dr \ \xi_{\alpha}(r) \phi_{0}(r) \ .
\end{equation}
To get the function $g_{\alpha}(r)$ in \eqref{eq:representativescalar}, we still have to apply the modular conjugation $J_{\mathbb{B}}$, which acts on the field as we explain in section \ref{sS:scalarstandard}
\begin{equation}
\begin{split}
    \phi_{0}(g_{\alpha}) 
    &= \int_{-\infty}^{\infty} ds \ L_{\alpha}(s)  \Delta^{is} J_{\mathbb{B}} \phi_{0}(f) J_{\mathbb{B}} \Delta^{-is} 
    = \int_{0}^{R} dr \ \xi_{\alpha}(r) J_{\mathbb{B}} \phi_{0}(r) J_{\mathbb{B}} \\
    &= \int_{0}^{R} dr \ \xi_{\alpha}(r) \phi_{0}(R^{2}/r)
    = \int_{R}^{\infty} dr \ \left( \frac{R^{2}}{r^{2}} \right)\xi_{\alpha}\left(R^{2}/r\right) \phi_{0}(r) \ .
\end{split}
\end{equation}
Hence, the function $g_{\alpha}(r)$ can be written in terms of the function $\xi_{\alpha}(r)$ obtained in \eqref{eq:xiequation2} as
\begin{equation}
    g_{\alpha}(r) = \left( \frac{R}{r}\right)^{2} \xi_{\alpha} \left( \frac{R^{2}}{r} \right) \ .
\label{eq:beta_modular}
\end{equation}

\subsection{Optimal symmetry operator: Variational approach}
\label{sS:bestvar}
In section \ref{S:Best_Operator}, we show that the optimal symmetry operator associated with an automorphism $\beta$ is given by its representative in the cone $\alpha=0$. In section \ref{sS:RepresentativeConeScalar}, we find an explicit form
to calculate the representatives on the different cones for the automorphism. Now, we want to check that the representative in the cone $\alpha = 0$ is the one that gives the maximal expectation value. For this, we construct the optimal symmetry operator with another approach, and we will show in section \ref{sS:resultexamplescalar} that it coincides with the representative obtained from the cone $\alpha = 0$.

We will assume that the optimal symmetry operator is also coherent. Therefore, we can write it as 
\begin{equation}
    U_{\text{opt}}(v) = e^{i v (\phi_{0}(f) - \phi_{0}(g_{\text{opt}}) - \pi_{0}(h_{\text{opt}}))} \ .
\end{equation}
The expectation value of the optimal operator is
\begin{equation}
    \expval{U_{\text{opt}}(v)}{\Omega} = e^{- \frac{1}{2} v^{2} c_{\text{opt}}} \ ,
\end{equation}
where
\begin{equation}
\begin{split}
    c_{\text{opt}} &= \expval{(\phi_{0}(f) - \phi_{0}(g_{\text{opt}}) - \pi_{0}(h_{\text{opt}}))^{2}}{\Omega} \\
    &= \expval{\phi_{0}(f) \phi_{0}(f)}{\Omega} + \expval{\phi_{0}(g_{\text{opt}}) \phi_{0}(g_{\text{opt}})}{\Omega} \\
    &+ \expval{\pi_{0}(h_{\text{opt}}) \pi_{0}(h_{\text{opt}})}{\Omega} - 2 \expval{\phi_{0}(f) \phi_{0}(g_{\text{opt}})}{\Omega} \ .
\end{split}
\end{equation}
Several terms disappear because of the commutation relations and the structure of the two-point functions. The condition to maximize the expectation value of $U_{\text{opt}}(v)$  translates to the condition that $c_{\text{opt}}$ is minimized with respect to variations of the functions $g_{\text{opt}}$ and $h_{\text{opt}}$. It is clear that $h_{\text{opt}} = 0$ because $\expval{\pi_{0}(h_{\text{opt}}) \pi_{0}(h_{\text{opt}})}{\Omega} > 0$ for every nonzero $h_{\text{opt}}$. Therefore, we only consider variations over $g_{\text{opt}}$. If we define 
\begin{equation}
\begin{split}
    L(r, r^{\prime}) &= \expval{\phi_{0}(r) \phi_{0}(r^{\prime})}{\Omega} \quad \text{for} \quad r, r^{\prime} \in \left[0, R \right] \ , \\
    M(r, r^{\prime}) &= \expval{\phi_{0}(r) \phi_{0}(r^{\prime})}{\Omega} \quad \text{for} \quad r \in \left[R, \infty \right], r^{\prime} \in \left[0, R \right] \ , \\
    N(r, r^{\prime}) &= \expval{\phi_{0}(r) \phi_{0}(r^{\prime})}{\Omega} \quad \text{for} \quad r, r^{\prime} \in \left[R, \infty \right] \ ,
\end{split}
\end{equation}
and using the notation $(L \cdot f)(r) = \int_{0}^{\infty} dr^{\prime} L(r, r^{\prime}) f(r^{\prime})$
and $g \cdot f = \int_{0}^{\infty} dr g(r)f(r)$,
we can write each term of $c_{\text{opt}}$ as
\begin{equation}
    c_{\text{opt}} = f \cdot L \cdot f + g_{\text{opt}} \cdot N \cdot g_{\text{opt}} - 2 g_{\text{opt}} \cdot M \cdot f \ .
\end{equation}
Taking the variation with respect to $g_{\text{opt}}$ gives
\begin{equation}
    \frac{\delta c_{\text{opt}}}{\delta g_{\text{opt}}(r)} = 
    2 (N \cdot g_{\text{opt}})(r) - 2 (M \cdot f)(r) = 0 \quad \forall \ r \in \left[R, \infty \right] \ .
\end{equation}
This implies that $g_{\text{opt}}(r)$ has to satisfy
\begin{equation}
    (N \cdot g_{\text{opt}})(r) = (M \cdot f)(r) \ .
\label{eq:MaximizeBeta}
\end{equation}
In the $l=0$ mode theory, we have
\begin{equation}
    \expval{\phi_{0}(r) \phi_{0}(r^{\prime})}{\Omega} = \frac{1}{2 \pi} \ln\left( \abs{ \frac{r + r^{\prime}}{r - r^{\prime}} } \right) \ ,
\end{equation}
then $N(r,r^{\prime})$, $M(r,r^{\prime})$ and $f(r)$ are known in \eqref{eq:MaximizeBeta}. Therefore, we need to find a function $g_{\text{opt}}(r)$ that satisfies the following equation
\begin{equation}
    \int_{R}^{\infty} dr^{\prime} \ N(r, r^{\prime}) g_{\text{opt}}(r^{\prime}) = (M \cdot f)(r) \ .
\end{equation}
We introduce the change of variable $r^{\prime} = R^{2}/u$ on the left side. After manipulating the expression, we arrive at
\begin{equation}
\begin{split}
    &\int_{0}^{R} du \ N(r, R^2/u) \left( \frac{R}{u} \right)^{2} g_{\text{opt}}(R^2/u) = (M \cdot f)(r) \ , \\
    &\int_{0}^{R} du \ \expval{\phi(r) \phi(R^{2}/u)}{\Omega} \xi_{\text{opt}}(u) = (M \cdot f)(r) \ , \\
    &\int_{0}^{R} du \ \expval{\phi(R^{2}/r) \phi(u)}{\Omega} \xi_{\text{opt}}(u) = (M \cdot f)(r) \ , \\
    &\int_{0}^{R} du \ \expval{\phi(v) \phi(u)}{\Omega} \xi_{\text{opt}}(u) = (M \cdot f)(R^{2}/v) \ , \\
    & \int_{0}^{R} du \ \ln(\abs{\frac{v - u}{v + u}}) \xi_{\text{opt}}(u) = G(v)  \ . \\
\end{split}
\end{equation}
Where 
\begin{equation}
    \xi_{\text{opt}}(u) = \left( \frac{R}{u} \right)^{2} g_{\text{opt}}(R^2/u), \quad
    v = R^{2}/r, \quad 
    G(v) = - 2 \pi (M \cdot f)(R^{2}/v) \ ,
\end{equation}
and we use that $\expval{\phi(r) \phi(R^{2}/r^{\prime})}{\Omega} = \expval{\phi(R^{2}/r) \phi(r^{\prime})}{\Omega}$.
Extending the function $\xi_{\text{opt}}(u)$ from $\left[0, R\right]$ to $\left[-R, R\right]$ as an odd function helps to write the integral as 
\begin{equation}
\begin{split}
    \int_{0}^{R} du \ln(\abs{\frac{v-u}{v+u}}) \xi_{\text{opt}}(u) 
    = \int_{-R}^{R} du \ \ln(\abs{v-u}) \xi_{\text{opt}}(u)
    = G(v) \ .
\end{split}
\end{equation}
This last equality is the Carleman's equation. The solution to this problem is given by \cite{HandbookIntegralEquations}
\begin{equation}
    \xi_{\text{opt}}(u) = \frac{1}{\pi^{2} \sqrt{R^2-u^2}} \int_{-R}^{R} dv \ \frac{\sqrt{R^2-v^2}}{v-u} \frac{dG(v)}{dv} \ ,
\end{equation}
which verifies that $\xi_{\text{opt}}(u)$ is an odd function. 
Once we have the function $\xi_{\text{opt}}(u)$, we can recover the function $g_{\text{opt}}(r)$ as
\begin{equation}
    g_{\text{opt}}(r) = \left( \frac{R}{r} \right)^{2} \xi_{\text{opt}} \left( \frac{R^{2}}{r} \right) \ .
\label{eq:beta_var}
\end{equation}

\newpage

\subsection{Explicit results}
\label{sS:resultexamplescalar}

To perform explicit calculations in this example, we take $R=1$ as the radius of the sphere $\mathbb{B}$. Due to scale invariance, the expectation values of coherent operators inside the sphere of radius $R$ are related to those for radius $1$ by
\begin{equation}
    \expval{e^{i v \phi_{0}(f)}}{\Omega} = e^{- \frac{1}{2} v^{2} R^{4} c(\widetilde{f})} \ .
\end{equation}
where $c(\widetilde{f}) = \expval{\phi_{0}(\widetilde{f}) \phi_{0}(\widetilde{f})}{\Omega}$ and $\widetilde{f}(r) = f(r R)$ is the function $f(r)$ rescaled to a function on the sphere of radius $1$.

\begin{figure}[H]
    \centering
    \includegraphics[width=0.85\linewidth]{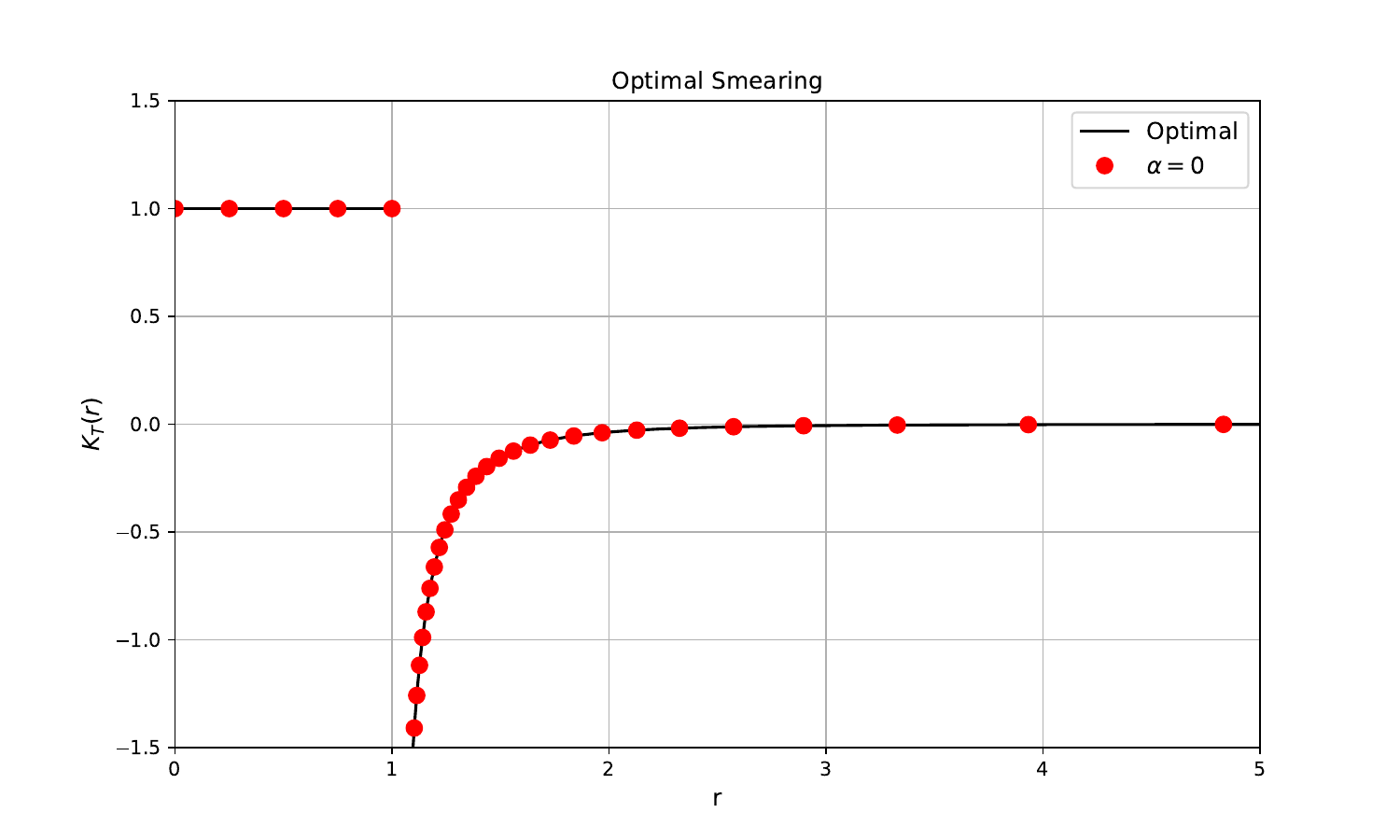}
    \caption{
        Plot, as a function of $r = \abs{\vec{x}}$, of the smearing function $K_{0}(\vec{x})$, obtained from the representative in the cone $\alpha = 0$ (see section \ref{sS:RepresentativeConeScalar}), shown as red dots, and $K_{\text{opt}}(\vec{x})$, obtained using the variational approach (see section \ref{sS:bestvar}), shown as a black solid line. The sphere radius was set to $R=1$.
    }
    \label{fig:alphabest}
\end{figure}

\begin{figure}[H]
    \centering
    \includegraphics[width=0.85\linewidth]{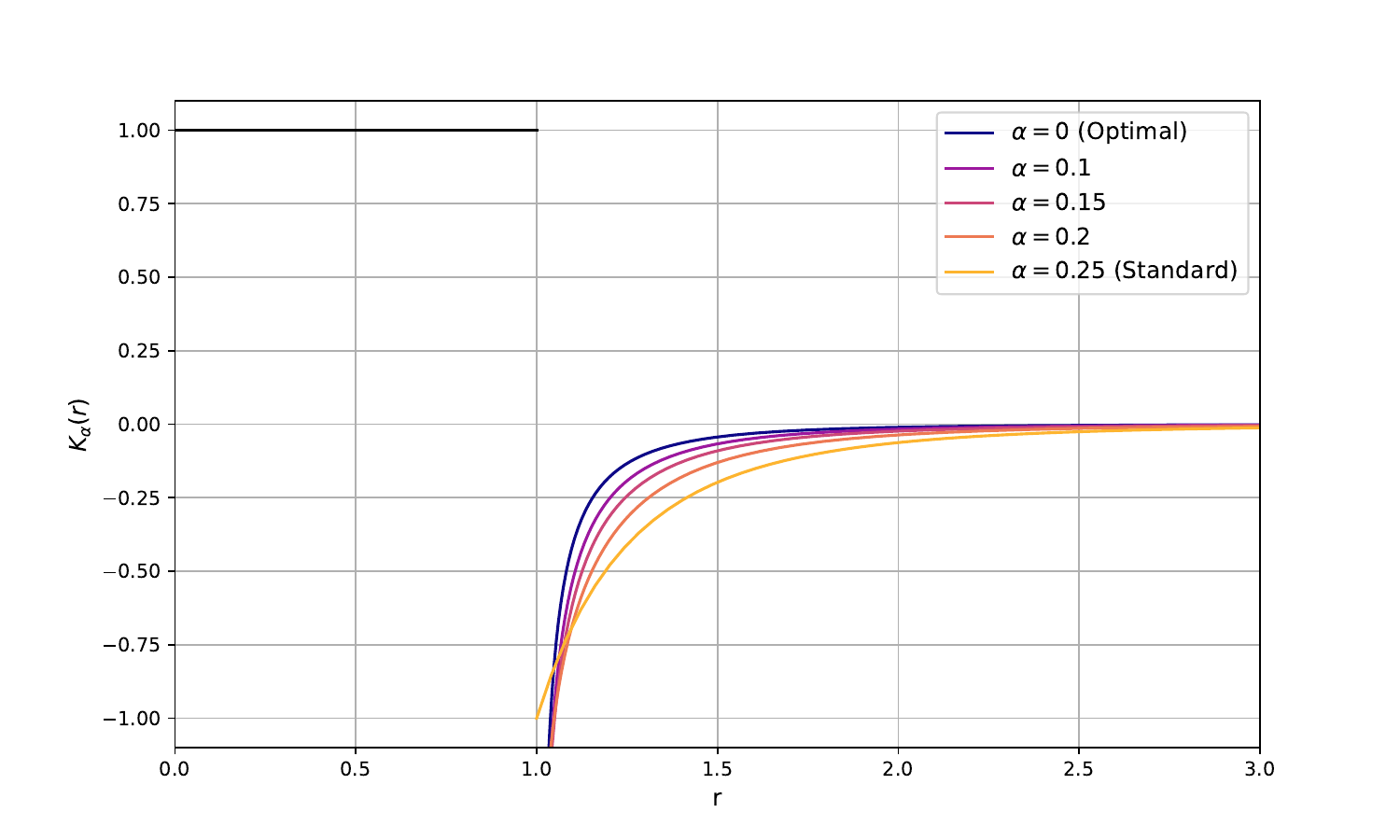}
    \caption{
        Plot, as function of $r = \abs{\vec{x}}$, of the smearing function $K_{\alpha}(r)$ obtained from the representatives in the cones (see section \ref{sS:RepresentativeConeScalar}) for several values of $\alpha$. Inside the sphere $r \in \left[0, 1 \right]$ all the functions coincide and are represented by a black line.
        The sphere radius was set to $R=1$.
    }
    \label{fig:alpharepresentatives}
\end{figure}

In previous sections, we found that the representatives of interest can always be written as
\begin{equation}
    U_{T}(v) = e^{i v (\phi_{0}(f) - \phi_{0}(g_{T}))} = e^{i v \phi_{0}(k_{T})} \ .
\end{equation}
Here, $g_{T}(r)$ denote the functions introduced in sections \ref{sS:scalarstandard}, \ref{sS:RepresentativeConeScalar}, and \ref{sS:bestvar}, with $g_{\text{std}}(r)$ corresponding to the standard representative, $g_{\alpha}(r)$ to the representative in the cone $\alpha$, and $g_{\text{opt}}(r)$ to the optimal one. Recall that $f(r) = \sqrt{4\pi} r F(r)$, with $F(r)$ defined in \eqref{eq:smearingfunction}, and that $k_{T}(r) = f(r) - g_{T}(r)$ gives the full smearing function. From the $l=0$ theory, the total smearing function in the $3+1$ theory can then be recovered using \eqref{eq:zeroto4functions}
\begin{equation}
    K_{T}(\vec{x}) = \frac{1}{\sqrt{4 \pi} \abs{\vec{x}}} k_{T}(\abs{\vec{x}}) =  \left\{ \begin{matrix} 1 & \abs{\vec{x}} \leq 1 \\ 
    -  \frac{1}{\sqrt{4 \pi} \abs{\vec{x}}} g_{T}(\abs{\vec{x}}) & \abs{\vec{x}} > 1  \\ \end{matrix}\right. \ .
\end{equation}
First, we want to verify whether the representative in the cone $\alpha = 0$ is the optimal symmetry operator. To do this, we compare $K_{0}(\vec{x})$, obtained from the modular tools in section \ref{sS:RepresentativeConeScalar}, with the smearing function $K_{\text{opt}}(\vec{x})$, obtained from the variational approach in section \ref{sS:bestvar}. Figure \ref{fig:alphabest} shows a plot of both smearing functions. We confirm that the representative of the automorphism in the cone $\alpha = 0$ yields the optimal symmetry operator.
\begin{equation}
    K_{0}(\vec{x}) = K_{\text{opt}}(\vec{x}) \Rightarrow U_{\text{opt}}(v) = U_{0}(v) \ .
\end{equation}
\indent In figure \ref{fig:alpharepresentatives}, we show the smearing functions $K_{\alpha}(\vec{x})$ corresponding to the different representatives in the cones $\mathcal{P}_{\Omega}^{\alpha}(\Omega)$ for various values of $\alpha \in \left[0, 1/4 \right]$. These smearing functions are constructed using the functions $g_{\alpha}(r)$ from section \ref{sS:RepresentativeConeScalar}.

\begin{figure}[H]
    \centering
    \includegraphics[width=1\linewidth]{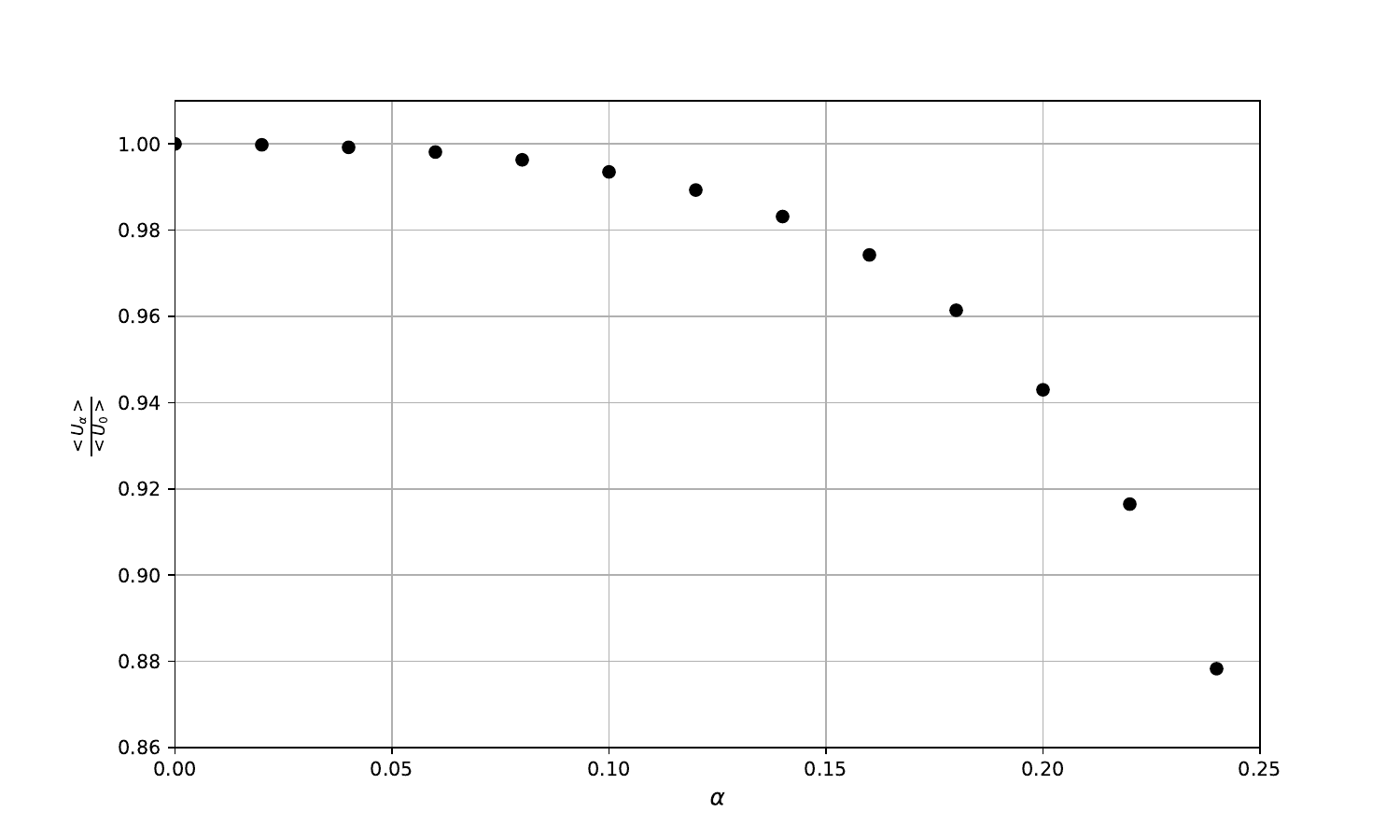}
    \caption{
        Plot of $\expval{U_{\alpha}(v)}{\Omega} / \expval{U_{0}(v)}{\Omega}$ for several values of $\alpha$. The sphere radius was set to $R=1$, and the transformation parameter was set to $v=1$.
    }
    \label{fig:alphaexpval}
\end{figure}

Now, we want to analyze the expectation values of the different representatives in the cones
\begin{equation}
    U_{\alpha}(v) = e^{i v \phi_{0}(k_{\alpha})} \Rightarrow \expval{U_{\alpha}(v)}{\Omega} = e^{-\frac{1}{2} v^{2} c(k_{\alpha})} \ ,
\end{equation}
where
\begin{equation}
    c(k_{\alpha}) = \expval{\phi_{0}(k_{\alpha})\phi_{0}(k_{\alpha})}{\Omega} \ .
\end{equation}
In figure \ref{fig:alphaexpval}, we plot how $\expval{U_{\alpha}(v)}{\Omega}$ varies with $\alpha \in \left[0 , 1/4 \right]$, taking $v=1$. Its behavior is similar to that observed in section \ref{S:ExampleSpins}.
To conclude, we give the values of the coefficients that determine the expectation values of the inner operator $U(v) = e^{i v \phi(F)}$, the standard representative $U_{\text{std}}(v) = e^{i v \phi_{0}(k_{std})}$ and the optimal representative $U_{\text{opt}}(v) = e^{i v \phi_{0}(k_{opt})}$
\begin{equation}
    c(F) = 1, \quad c(k_{\text{std}}) \approx 0.935, \quad c(k_{\text{opt}}) \approx 0.616 \ .
\end{equation}

\newpage

\section{Conclusions}
The positive cones $\mathcal{P}_{\Omega}^{\alpha}(\mathcal{A}) \subset \mathcal{H}$, with $\alpha \in \left[0, 1/4 \right]$, associated with a von Neumann algebra $\mathcal{A}$ and a reference vector $\ket{\Omega}$ have useful properties \cite{SomePropertiesModularConjugation}. For example, every cone contains a unique vector representative, or purification, of a given normal state $\phi: \mathcal{A} \rightarrow \mathbb{C}$ \cite{ArakiMasuda1982,PositiveConesAssociated,PositiveConesLSpaces}.

In this work, we show that the positive cone associated with $\alpha = 0$ plays a key role in certain maximization problems. Uhlmann's theorem \cite{Uhlmann:1975kt, alberti_note_1983} states that the fidelity between two states $\omega, \phi \in \mathcal{A}_{*}^{+}$ can be calculated as
\begin{equation}
    F(\omega, \phi) = \text{Sup}_{\ket{\psi}} \abs{\braket{\Omega}{\psi}} \ ,
\end{equation}
where the supremum is taken over all possible purifications of $\phi$, and $\ket{\Omega}$ is a purification of $\omega$. These purifications are taken over some Hilbert space $\mathcal{H}$ where $\mathcal{A} \subset \mathcal{B}(\mathcal{H})$.
In section \ref{S:FidelityP0}, we show that if we can purify $\omega$ to a cyclic and separating vector $\ket{\Omega}$, then the supremum is achieved for the purification $\ket{\phi_{0}} \in \mathcal{P}_{\Omega}^{0}(\mathcal{A})$ of $\phi$, which is unique. Also, we give an explicit form of the fidelity in terms of the modular tools.
\begin{equation}
    F(\omega, \phi) = \braket{\Omega}{\phi_{0}} = \expval{\sqrt{\Delta_{\Omega}^{1/2} \Delta_{\phi,\Omega}\Delta_{\Omega}^{1/2}}}{\Omega} \ , \quad \ket{\phi_{0}} \in \mathcal{P}_{\Omega}^{0}(\mathcal{A}) ,
\end{equation}
where $\Delta_{\Omega}$ and $\Delta_{\phi, \Omega}$ are the modular and relative modular operators associated with the algebra $\mathcal{A}$ and the vectors $\ket{\Omega}$, $\ket{\phi} \in \mathcal{P}_{\Omega}^{1/4}(\mathcal{A})$. We propose a new \textit{generalized fidelity}, which is given by
\begin{equation}
    F_{\alpha}(\omega,\phi) = \braket{\Omega}{\phi_{\alpha}}, \quad \ket{\phi_{\alpha}} \in \mathcal{P}_{\Omega}^{\alpha}(\mathcal{A}) \ ,
\end{equation}
where $\ket{\phi_{\alpha}}$ is the purification of $\phi$ in the cone $\mathcal{P}_{\Omega}^{\alpha}(\mathcal{A})$.
It has to be studied if this quantity has nice properties, as the data-processing inequality. Another similar generalize fidelity was defined in \cite{Hollands:2020owv}.

On the other hand, we start with the question of which is the non-local operator associated with a region $A$ that has maximal expectation value among a class. We rephrase this in clearer language as the operator $U(\beta)$ such that it implements a given automorphism $\beta$ on the algebra $\mathcal{A}$, associated with the region $A$, and has maximal expectation value. To tackle this question, we study the construction of such operators using the positive cones. In \cite{haag2012local,bratteli1979operator}, it is shown that using the cone $\mathcal{P}_{\Omega}^{1/4}(\mathcal{A})$ one can construct a unitary representative $U(\beta)$ of the automorphism $\beta$ such that it implements the automorphism on the algebra and satisfies $U(\beta) \ket{\Omega} \in \mathcal{P}_{\Omega}^{1/4}(\mathcal{A})$. These conditions determine this unitary operator uniquely. We generalize this construction to every cone $\mathcal{P}_{\Omega}^{\alpha}(\mathcal{A})$, with $\alpha \in \left[0 , 1/4\right]$, to obtain a isometry $U_{\alpha}(\beta)$ such that it implements the automorphism and $U_{\alpha}(\beta) \ket{\Omega} \in \mathcal{P}_{\Omega}^{\alpha}(\mathcal{A})$. Again, these two properties determine this operator uniquely. In Section \ref{S:Best_Operator}, we define the set of all the representatives of the automorphism $\beta$ and show that the supremum of the expectation value of these operators is achieved by the representative $U_{0}(\beta)$ constructed with the cone $\mathcal{P}_{\Omega}^{0}(\mathcal{A})$.

In section \ref{S:positive_cones}, we review results from \cite{PositiveConesAssociated, PositiveConesLSpaces, ArakiMasuda1982}  that are fundamental to this work. We prove that, given a state $\phi$, every purification of the state over the positive cones $\ket{\phi_{\alpha}} \in \mathcal{P}_{\Omega}^{\alpha}(\mathcal{A})$ is connected to the standard one $\ket{\phi} \in \mathcal{P}_{\Omega}^{1/4}(\mathcal{A})$ by an isometry $R_{\alpha}$ for $\alpha \in [0, 1/4]$
\begin{equation}
    R_{\alpha} \ket{\phi} = \ket{\phi_{\alpha}} \in \mathcal{P}_{\Omega}^{\alpha}(\mathcal{A}) \ .
\end{equation}
Remarkably, this operator can be constructed using modular tools. It arises in the polar decomposition of the unbounded operator
\begin{equation}
    J \Delta_{\phi, \Omega}^{1/2} \Delta_{\Omega}^{1/2 - 2 \alpha} J = R_{\alpha}^{\dagger} P_{\alpha} \ ,
\end{equation}
where $ \Delta_{\Omega}$ and $J$ are the modular operator associated with the cyclic and separating vector $\ket{\Omega}$ and the algebra $\mathcal{A}$, and $\Delta_{\phi, \Omega}$ is the relative modular operator associated with $\ket{\Omega}, \ket{\phi}$ and $\mathcal{A}$.

In sections \ref{S:ExampleSpins} and \ref{S:Scalar} we show a finite-dimensional example and a quantum field theory example, respectively, of the constructions of the different representatives of an automorphism in the cones. In both examples, we show that the optimal symmetry operator is achieved with the representative in the cone $\mathcal{P}_{\Omega}^{0}(\mathcal{A})$. We also analyze the behavior of the generalized fidelity.

A future direction of this work is to apply these tools to known maximization problems. One example is the entanglement of purification (EoP) \cite{Terhal:2002riz}. Given a system $AB$, the EoP is defined as the minimum entropy $S(A \bar{A})$ achieved by varying over all possible purifications of $AB$ in $AB\bar{A}\bar{B}$.
A more general notion is the Renyi entanglement of purification (REoP) \cite{Akers:2023obn, Chen:2025cga}, where the $n$th Renyi entropy is minimized over the possible purifications. The Renyi entropies can be calculated using the replica twist operator \cite{Cardy:2007mb}. By understanding this operator as the one that implements the permutation automorphism between replicas in a subalgebra $\mathcal{A}$ and acting as the identity on a subalgebra $\mathcal{B} \subset \mathcal{A}^{\prime}$, one can provide a lower bound for the REoP of $\mathcal{A} \vee \mathcal{B}$. Moreover, from the expectation value of the standard representative of the permutation automorphism, one can extract the $n$th Renyi reflected entropy defined in \cite{Dutta:2019gen}. For more information about this quantity and its connections with holography, see \cite{umemoto2018entanglement, Nguyen:2017yqw, Caputa:2018xuf, Hirai:2018jwy, PhysRevD.111.L021902, Jiang:2024xqz}

\section*{Acknowledgments}

We are grateful for insightful discussions with Horacio Casini. This work was partially supported by CONICET, CNEA and Universidad Nacional de Cuyo, Argentina.

\appendix

\newpage

\section{Modular tools}
\label{app:modular_generator}

\subsection{Modular operator and relative modular operator}
\label{A:RelativeModular}

In this appendix, we will review the definition of the modular operators.

Suppose that we have a Hilbert space $\mathcal{H}$, a von Neumann algebra $\mathcal{A} \subset \mathcal{B}(\mathcal{H})$, and a vector $\ket{\Omega} \in \mathcal{H}$ which is cyclic and separating for $\mathcal{A}$. Using the cyclic property, the conjugate linear operator $S_{\Omega}$ is defined as \cite{haag2012local}
\begin{equation}
    S_{\Omega} A \ket{\Omega} = A^{\dagger} \ket{\Omega} \ .
\end{equation}
It is a closable operator and, therefore, has a polar decomposition
\begin{equation}
    S_{\Omega} = J_{\Omega} \Delta_{\Omega}^{1/2} \ .
\end{equation}
The positive part $S_{\Omega}^{\dagger} S_{\Omega} = \Delta_{\Omega}$ is called the modular operator, and the unitary part $J_{\Omega}$ is the modular conjugation. The modular operator implements an automorphism on the algebra known as modular evolution
\begin{equation}
    \Delta_{\Omega}^{it} \mathcal{A} \Delta_{\Omega}^{-it} = \mathcal{A}, \quad
    \Delta_{\Omega}^{it} \mathcal{A}^{\prime} \Delta_{\Omega}^{-it} = \mathcal{A}^{\prime} .
\end{equation}
The modular conjugation sends an operator to the commutant algebra
\begin{equation}
    J_{\Omega} \mathcal{A} J_{\Omega} = \mathcal{A}^{\prime} \ .
\end{equation}
Now, if we have another cyclic and separating vector $\ket{\phi}$, we can define the conjugate linear operator $S_{\phi, \Omega}$
\begin{equation}
    S_{\phi, \Omega} A \ket{\Omega} = A^{\dagger} \ket{\phi} \ .
\end{equation}
It is a closable operator and, therefore, has a polar decomposition
\begin{equation}
    S_{\phi, \Omega} = J_{\phi, \Omega} \Delta_{\phi, \Omega}^{1/2} \ .
\end{equation}
The positive part $S_{\phi, \Omega}^{\dagger} S_{\phi, \Omega} = \Delta_{\phi, \Omega}$ is called the relative modular operator. Let $A \in \mathcal{A}$ and $A^{\prime} \in \mathcal{A}^{\prime}$. This operator acts on the algebras via a modular evolution given by
\begin{equation}
    \Delta_{\phi, \Omega}^{it} A \Delta_{\phi, \Omega}^{-it} 
    = \Delta_{\phi}^{it} A \Delta_{\phi}^{-it}, \quad
    \Delta_{\phi, \Omega}^{it} A^{\prime} \Delta_{\phi, \Omega}^{-it} 
    = \Delta_{\Omega}^{it} A^{\prime} \Delta_{\Omega}^{-it} .
\end{equation}
where $\Delta_{\phi}$ is the modular operator associated with the vector $\ket{\phi}$ and the algebra $\mathcal{A}$. If the vector $\ket{\phi} \in \mathcal{P}^{1/4}_{\Omega}(\mathcal{A})$, then
\begin{equation}
    J_{\phi, \Omega} = J_{\Omega} = J_{\phi} \ .
\end{equation}
In this work, to be more compact in the notation, we will denote $J_{\Omega} = J$ and $\Delta_{\Omega} = \Delta$. For a more detailed explanation of the properties of the modular operator, see \cite{bratteli1979operator,haag2012local}, and for the relative modular operator, including a definition when the vectors $\ket{\Omega}$ and $\ket{\phi}$ are not cyclic and separating, see \cite{PositiveConesLSpaces}, Appendix C (also see \cite{Faulkner:2020iou}).

\subsection{Algebraic work}
\label{ap:AlgebraicWork}

In this appendix, we  make a derivation of \eqref{eq:relHQ}.

Using that $P_{\alpha}^{\prime} = -i (J \Delta^{\frac{1}{2} - 2 \alpha} H \Delta^{\frac{1}{2}} J - \Delta^{2 \alpha - \frac{1}{2}} Q_{\alpha})$, we have that
\begin{equation}
\begin{split}
    P_{\alpha}^{\prime} \Delta^{2 \alpha - 1} &= -i \left( J \Delta^{\frac{1}{2} - 2 \alpha} H \Delta^{\frac{3}{2} - 2 \alpha} J - \Delta^{2 \alpha - 1} Q_{\alpha} \Delta^{2 \alpha - 1} \right) \ , \\
    \Delta^{2 \alpha - 1} P_{\alpha}^{\prime} &= -i \left( J \Delta^{\frac{3}{2} - 4 \alpha} H \Delta^{\frac{1}{2}} J - \Delta^{4 \alpha - 2} Q_{\alpha} \right) \ .
\end{split}
\end{equation}
Also, we have by \eqref{eq:derp2}
\begin{equation}
    P_{\alpha}^{\prime} \Delta^{2 \alpha - 1} + \Delta^{2 \alpha - 1} P_{\alpha}^{\prime}
    = -i J (\Delta^{\frac{1}{2} - 2 \alpha} H \Delta^{\frac{3}{2} - 2 \alpha} -  \Delta^{\frac{3}{2} - 2 \alpha} H \Delta^{\frac{1}{2} - 2 \alpha})  J  \ .
\end{equation}
Then, reordering the terms, we arrive at
\begin{equation}
    J \left( \Delta^{\frac{3}{2} - 4 \alpha} H \Delta^{\frac{1}{2}} + \Delta^{\frac{3}{2} - 2 \alpha} H \Delta^{\frac{1}{2} - 2 \alpha} \right) J
    =  \Delta^{2 \alpha - 1} Q_{\alpha} \Delta^{2 \alpha - 1} + \Delta^{4 \alpha - 2} Q_{\alpha} \ .
\end{equation}
Multiplying by $\Delta^{\frac{3}{2} - 3 \alpha}$ on the left and $\Delta^{\frac{1}{2} - \alpha}$ on the right of both expressions, we arrive at
\begin{equation}
    J \left( \Delta^{-\alpha} H \Delta^{\alpha} + \Delta^{\alpha} H \Delta^{-\alpha} \right) J
    =  \Delta^{\frac{1}{2} - \alpha} Q_{\alpha} \Delta^{-(\frac{1}{2} - \alpha)} + \Delta^{-(\frac{1}{2} - \alpha)} Q_{\alpha} \Delta^{\frac{1}{2} - \alpha} \ .
\end{equation}

\subsection{Modular flow at complex time}
\label{A:ModularFlowImaginary}

Suppose that we have a Hilbert space $\mathcal{H}$, a von Neumann algebra $\mathcal{A} \subset \mathcal{B}(\mathcal{H})$, and a vector $\ket{\Omega} \in \mathcal{H}$ which is cyclic and separating for $\mathcal{A}$. Then, we have the modular operator $\Delta$ associated with $\mathcal{A}$ and $\ket{\Omega}$, and therefore we have the modular automorphism $\Delta^{it} \mathcal{A} \Delta^{-it}$ of $\mathcal{A}$.

Suppose that we have a function $f(t)$ which has a Fourier transform $\hat{f}(\omega)$ such that
\begin{equation}
    f(t) = \frac{1}{\sqrt{2 \pi}} \int_{-\infty}^{\infty} d\omega \ \hat{f}(\omega) e^{i \omega t} \ .
\end{equation}
Let $A \in \mathcal{A}$. We can construct a smeared operator $A(f)$ as
\begin{equation}
    A(f) = \int_{-\infty}^{\infty} dt \ \Delta^{it} A \Delta^{-it} f(t) \ .
\end{equation}
Now, if we use the spectral decomposition $g(\Delta) = \int_{0}^{\infty} dE(\lambda) g(\lambda)$, where $dE(\lambda)$ is a projection-valued measure associated with the operator $\Delta$, we can rewrite $A(f)$ as
\begin{equation}
\begin{split}
    A(f) &=  \int_{0}^{\infty} \int_{0}^{\infty} \int_{-\infty}^{\infty} dt \ dE(\rho) A dE(\lambda) \left( \frac{\lambda}{\rho} \right)^{-it} f(t) \\
    &=  \sqrt{2 \pi} \int_{0}^{\infty} \int_{0}^{\infty} \ dE(\rho) A dE(\lambda) \ \hat{f}(\ln(\lambda / \rho)) \ .
\end{split}
\end{equation}
If we evolve $A(f)$ with a modular evolution with a complex parameter $\Delta^{z}$, where $z \in \mathbb{C}$, we will get
\begin{equation}
\begin{split}
    \Delta^{z} A(f) \Delta^{-z} 
    &= \int_{-\infty}^{\infty} dt \ \Delta^{it + z} A \Delta^{-it - z} f(t) \\
    &= \int_{0}^{\infty} \int_{0}^{\infty} \int_{-\infty}^{\infty} dt  \  dE(\rho) A dE(\lambda) \left( \frac{\lambda}{\rho} \right)^{-(it + z)} f(t) \\
    &= \sqrt{2 \pi} \int_{0}^{\infty} \int_{0}^{\infty} \  dE(\rho) A dE(\lambda) \ \left( \frac{\lambda}{\rho} \right)^{-z} \hat{f}(\ln(\lambda / \rho)) \ . 
\end{split}
\end{equation}
Therefore,
\begin{equation}
    \Delta^{z} A(f) \Delta^{-z} = A(f_{z}) \quad \text{where} \quad
    f_{z}(t) = \frac{1}{\sqrt{2 \pi}}\int_{-\infty}^{\infty} d\omega \ e^{-z \omega} \hat{f}(\omega) e^{i \omega t} = f(t + iz) \ .
\label{eq:Aevolvedimg}
\end{equation}
The function $e^{-z \omega} \hat{f}(\omega)$ has to be good enough to ensure the convergence of the integral in \eqref{eq:Aevolvedimg}. For example, if $f_{n}(t) = \sqrt{n / \pi } \ e^{-n t^{2}}$, then $\hat{f}_{n}(\omega) = 1/ \sqrt{2 \pi} \ e^{- \omega^{2} / (4 n)}$, and therefore $e^{-z \omega} \hat{f}(\omega)$ is a continuous $L^{1}$ function on $\omega$, and the integral converges.

\subsection{Inversion formula}
\label{A:InversionFormula}
Suppose that we have a Hilbert space $\mathcal{H}$, a von Neumann algebra $\mathcal{A} \subset \mathcal{B}(\mathcal{H})$, and a vector $\ket{\Omega} \in \mathcal{H}$ which is cyclic and separating for $\mathcal{A}$. Then, we have the modular operator $\Delta$ associated with $\mathcal{A}$ and $\ket{\Omega}$.

We start with the following equation for operators $Q, A \in \mathcal{A}$
\begin{equation}
    \Delta^{\alpha} Q \Delta^{-\alpha} + \Delta^{-\alpha} Q \Delta^{\alpha} = A \ .
\end{equation}
To simplify the notation, we will call
\begin{equation}
    D_{\alpha}(Q) = \Delta^{\alpha} Q \Delta^{-\alpha} + \Delta^{-\alpha} Q \Delta^{\alpha} \ .
\end{equation}
Then, we can rewrite the equation as $D_{\alpha}(Q) = A$. A problem that arises in this work is: given the operator $A$, which operator $Q$ fulfills this equation? The answer is the following inversion formula
\begin{equation}
    Q = I_{\alpha}(A) = \frac{1}{4 \alpha} \int_{-\infty}^{\infty} \frac{\Delta^{it} A \Delta^{-it}}{\cosh(\frac{\pi t}{2 \alpha})} \ .
\end{equation}
This inversion formula was proved in \cite{bratteli1979operator}, Theorem 2.5.14, for the case $\alpha = 1/2$. Here, we generalize it for $\alpha \geq 0$ with the same methods used in \cite{bratteli1979operator}. Following the notation in \ref{A:ModularFlowImaginary}, if we define
\begin{equation}
    f(t) = \frac{1}{4 \alpha} \frac{1}{\cosh(\frac{\pi t}{2 \alpha})}
    \Rightarrow
    \hat{f}(\omega) = \frac{1}{2 \sqrt{2 \pi}} \frac{1}{\cosh(\alpha \omega)} \ .
\end{equation}
If we perform $D_{\alpha}(I_{\alpha}(A))$ and use the result from \ref{A:ModularFlowImaginary}, Eq.~\eqref{eq:Aevolvedimg}, we arrive at
\begin{equation}
    D_{\alpha}(I_{\alpha}(A)) = \int_{-\infty}^{\infty} dt \ \Delta^{it} A \Delta^{-it} g(t), \quad
\end{equation}
where
\begin{equation}
\begin{split}
    \hat{g}(\omega) &= \frac{1}{2 \sqrt{2 \pi}} 
    \left( 
        \frac{e^{-\alpha \omega}}{\cosh(\alpha \omega)} 
        + \frac{e^{\alpha \omega}}{\cosh(\alpha \omega)} 
    \right)
    = \frac{1}{\sqrt{2 \pi}} 
    \Rightarrow g(t) = \delta(t). \\
\end{split}
\end{equation}
In conclusion, we have
\begin{equation}
    D_{\alpha}(I_{\alpha}(A)) = A \ .
\end{equation}

\section{Coherent representative}
\label{A:CoherentRepresentative}
In section \ref{S:Scalar}, we argued that, for the inner automorphism defined by the unitary $U(v) = e^{i v \phi(F)} \in \mathcal{A}$, the representative on the different cones is also a coherent operator. In this appendix, we present arguments supporting this. We know from section \ref{sS:RepInner} that, for inner automorphisms, the representative in the cone $\alpha$ can be constructed by performing the polar decomposition of
\begin{equation}
    J\Delta^{1/2 - 2 \alpha} U(v) \Delta^{1/2}J = W_{\alpha}(v) P_{\alpha}(v) \ .
\label{eq:polarcoherent}
\end{equation}
where $\Delta$ is the modular operator associated with the vacuum vector $\ket{\Omega}$ and the algebra $\mathcal{A}$ of the sphere of radius $R$ centered at the origin.
 
We showed that the unitary $U(v)$ belongs to the $l = 0$ theory because the smearing function $F$ in \eqref{eq:smearingfunction} is spherically symmetric. From \cite{Huerta:2022tpq}, we know that the modular Hamiltonian $K = -\log(\Delta)$ decomposes over the angular modes as $K = \sum_{l,m} K_{l,m}$. Therefore, in \eqref{eq:polarcoherent}, we only need to apply the modular operator corresponding to the $l = 0$ theory.

For the scalar theory, the modular Hamiltonian $K = - \log(\Delta)$ takes the form \cite{Casini:2009sr}
\begin{equation}
    K = \int_{\mathbb{B}} dx^{d-1} dy^{d-1} (\phi(x) M(x,y) \phi(y) + \pi(x) N(x,y) \pi(y))
    = \phi \cdot M \cdot \phi + \pi \cdot N \cdot \pi \ ,
\end{equation}
where we use the same notation as in \ref{sS:bestvar}. To inspect the form of the product $\Delta^{1/2 - 2 \alpha} U(v) \Delta^{1/2}$ in terms of the modular Hamiltonian $K$ and the fields $\phi(h)$ and $\pi(g)$, we have to use the Baker-Campbell-Hausdorff formula
\begin{equation}
    \exp(A) \exp(B) = \exp(A+B+ \frac{1}{2} [A,B] + \frac{1}{12} \left( [A,[A,B]] - [B,[A,B]] \right) + \cdots) \ .
\end{equation}
We note that
\begin{equation}
    \left[K, \phi(h) \right] = - 2 i h \cdot N \cdot \pi = -i \pi(\widetilde{h}), \quad
    \left[K, \pi(g) \right] = 2 i g \cdot M \cdot \phi = i \phi(\widetilde{g}) ,\\
\end{equation}
\begin{equation*}
    \left[\phi(h), \pi(g) \right] = i g \cdot h , \quad \left[K, K \right] = 0  \ .
\end{equation*}
where $g(x)$ and $h(x)$ are some functions. Therefore, after applying the Baker-Campbell-Hausdorff formula, we can only have terms that contain $K$, $\phi(g)$, $\pi(h)$, and complex constants. Thus, we can write
\begin{equation}
\begin{split}
    \Delta^{1/2 - 2 \alpha} U(v) \Delta^{1/2}
    &= e^{- (\frac{1}{2}  - 2 \alpha) K} e^{i v \phi(f)} e^{- \frac{1}{2} K} \\
    &= \exp \left[- (1 - 2 \alpha) K + i v ( a_{\alpha} \phi(g_{\alpha}) + b_{\alpha} \pi(h_{\alpha})) + c_{\alpha} v^{2} \right] \ ,
\end{split}
\end{equation}
where the terms linear in $\phi$ and $\pi$ have only one power of $v$ because only the commutators with $K$ give terms linear in the fields. The last term is quadratic in $v$ because it comes from commutators $\left[\phi, \pi \right]$ that arise from terms linear in $v$. The functions $g_{\alpha}(x)$ and $h_{\alpha}(x)$, and the constants $a_{\alpha}, b_{\alpha}, c_{\alpha} \in \mathbb{C}$, come from the re-summation of all the terms in the Baker-Campbell-Hausdorff formula and obviously depend on $\alpha$. This result is further explained and generalized in \cite{balian_nonunitary_1969}. If we calculate the positive part of the polar decomposition, we obtain that it must take the form
\begin{equation}
\begin{split}
    P_{\alpha}(v) &= J \sqrt{(\Delta^{1/2} U(v) \Delta^{1/2 - 2 \alpha})^{\dagger} \Delta^{1/2} U(v) \Delta^{1/2 - 2 \alpha}} J \\
    &= J \exp\left[ - (1 - 2 \alpha) K + i v ( \widetilde{a}_{\alpha} \phi(\widetilde{g}_{\alpha}) + \widetilde{b}_{\alpha} \pi(\widetilde{h}_{\alpha})) + \widetilde{c}_{\alpha} v^{2} \right] J \ ,
\end{split}
\end{equation}
where the tildes indicate that the constants and functions may differ. Then, the unitary $W_{\alpha}(v)$ takes the form
\begin{equation}
    W_{\alpha}(v) = J \Delta^{1/2} U(v) \Delta^{1/2 - 2 \alpha} J P_{\alpha}(v)^{-1}
    = \exp \left[i v ( d_{\alpha} \phi(H_{\alpha}) + e_{\alpha} \pi(G_{\alpha})) + l_{\alpha} v^{2}  \right] \ ,
\end{equation}
where again $d_{\alpha}, e_{\alpha}, l_{\alpha} \in \mathbb{C}$, and $H_{\alpha}(x), G_{\alpha}(x)$ are functions. The terms that contain the modular Hamiltonian disappear in the expression for $W_{\alpha}(v)$ because, in the inversion of $P_{\alpha}(v)$, the terms of $K$ acquire a minus sign, yielding a coherent operator. Moreover, the terms linear in the fields remain linear in the transformation parameter $v$. This fact is important because, in the discussion of the inner one-parameter group in Section \ref{sS:InnerOneparametergroup}, we found an explicit form in terms of modular tools for $\frac{d}{dv} W_{\alpha}(v) \big|_{v=0} = Q_{\alpha}$, and here it is directly $Q_{\alpha} = d_{\alpha} \phi(F_{\alpha}) + e_{\alpha} \pi(G_{\alpha})$.

The term quadratic in $v$ contributes only as a phase because $W_{\alpha}$ is a partial isometry. Therefore, $l_{\alpha}$ must be purely imaginary. Since it is only a phase and the operators applying the automorphism are defined up to a phase, we can set $l_{\alpha} = 0$.

\bibliographystyle{utphys}
\bibliography{EE}

\providecommand{\href}[2]{#2}\begingroup\raggedright\begin{thebibliography}{10}

\bibitem{SomePropertiesModularConjugation}
H.~Araki, ``Some properties of modular conjugation operator of von neumann
  algebras and a non-commutative radon-nikodym theorem with a chain rule,''
  {\em Pacific Journal of Mathematics} {\bfseries 50} (1974) 309--354.
  \url{https://api.semanticscholar.org/CorpusID:59489383}.

\bibitem{Uhlmann:1975kt}
A.~Uhlmann, ``{The Transition Probability in the State Space of a* Algebra},''
  {\em Annalen Phys.} {\bfseries 42} (1985) 524.

\bibitem{casini2021entropic}
H.~Casini, M.~Huerta, J.~M. Mag{\'a}n, and D.~Pontello, ``Entropic order
  parameters for the phases of qft,'' {\em Journal of High Energy Physics}
  {\bfseries 2021} no.~4, (2021) 1--98.

\bibitem{Pedro}
H.~Casini, J.~M. Magan, and P.~J. Martinez, ``{Entropic order parameters in
  weakly coupled gauge theories},''
  \href{http://dx.doi.org/10.1007/JHEP01(2022)079}{{\em JHEP} {\bfseries 01}
  (2022) 079}, \href{http://arxiv.org/abs/2110.02980}{{\ttfamily
  arXiv:2110.02980 [hep-th]}}.

\bibitem{Cardy:2007mb}
J.~L. Cardy, O.~A. Castro-Alvaredo, and B.~Doyon, ``{Form factors of
  branch-point twist fields in quantum integrable models and entanglement
  entropy},'' \href{http://dx.doi.org/10.1007/s10955-007-9422-x}{{\em J.
  Statist. Phys.} {\bfseries 130} (2008) 129--168},
  \href{http://arxiv.org/abs/0706.3384}{{\ttfamily arXiv:0706.3384 [hep-th]}}.

\bibitem{doplicher1982local}
S.~Doplicher, ``Local aspects of superselection rules,'' {\em Communications in
  Mathematical Physics} {\bfseries 85} no.~1, (1982) 73--86.

\bibitem{Doplicher:1983if}
S.~Doplicher and R.~Longo, ``{Local aspects of superselection rules. II},''
  \href{http://dx.doi.org/10.1007/BF01213216}{{\em Commun. Math. Phys.}
  {\bfseries 88} (1983) 399--409}.

\bibitem{StandarandSplit}
S.~Doplicher and R.~Longo, ``Standard and split inclusions of von neumann
  algebras,'' \href{http://dx.doi.org/https://doi.org/10.1007/BF01388641}{{\em
  Inventiones mathematicae} {\bfseries 75} no.~3, (1984) 493--536}.
  \url{https://link.springer.com/article/10.1007/BF01388641}.

\bibitem{OnNoetherTheorem}
D.~Buchholz, S.~Doplicher, and R.~Longo, ``On noether's theorem in quantum
  field theory,''
  \href{http://dx.doi.org/https://doi.org/10.1016/0003-4916(86)90086-2}{{\em
  Annals of Physics} {\bfseries 170} no.~1, (1986) 1--17}.
  \url{https://www.sciencedirect.com/science/article/pii/0003491686900862}.

\bibitem{dutta2021canonical}
S.~Dutta and T.~Faulkner, ``A canonical purification for the entanglement wedge
  cross-section,'' {\em Journal of High Energy Physics} {\bfseries 2021} no.~3,
  (2021) 1--49.

\bibitem{Terhal:2002riz}
B.~M. Terhal, M.~Horodecki, D.~W. Leung, and D.~P. DiVincenzo, ``{The
  entanglement of purification},''
  \href{http://dx.doi.org/10.1063/1.1498001}{{\em J. Math. Phys.} {\bfseries
  43} no.~9, (2002) 4286--4298},
  \href{http://arxiv.org/abs/quant-ph/0202044}{{\ttfamily
  arXiv:quant-ph/0202044}}.

\bibitem{Nguyen:2017yqw}
P.~Nguyen, T.~Devakul, M.~G. Halbasch, M.~P. Zaletel, and B.~Swingle,
  ``{Entanglement of purification: from spin chains to holography},''
  \href{http://dx.doi.org/10.1007/JHEP01(2018)098}{{\em JHEP} {\bfseries 01}
  (2018) 098}, \href{http://arxiv.org/abs/1709.07424}{{\ttfamily
  arXiv:1709.07424 [hep-th]}}.

\bibitem{umemoto2018entanglement}
K.~Umemoto and T.~Takayanagi, ``Entanglement of purification through
  holographic duality,'' {\em Nature Physics} {\bfseries 14} no.~6, (2018)
  573--577.

\bibitem{haag2012local}
R.~Haag, {\em Local quantum physics: Fields, particles, algebras}.
\newblock Springer Science \& Business Media, 2012.

\bibitem{bratteli1979operator}
O.~Bratteli and D.~W. Robinson,
  \href{http://dx.doi.org/10.1007/978-3-662-02520-8}{{\em Operator Algebras and
  Quantum Statistical Mechanics 1: $C^*$- and $W^*$-Algebras, Symmetry Groups,
  Decomposition of States}}.
\newblock Texts and Monographs in Physics. Springer-Verlag, Berlin, Heidelberg,
  2~ed., 1979.

\bibitem{PositiveConesAssociated}
H.~KOSAKI, ``Positive cones associated with a von neumann algebra,'' {\em
  Mathematica Scandinavica} {\bfseries 47} no.~2, (1980) 295--307.
  \url{http://www.jstor.org/stable/24491398}.

\bibitem{alberti_note_1983}
P.~M. Alberti, ``A note on the transition probability over {C}*-algebras,''
  \href{http://dx.doi.org/10.1007/BF00398708}{{\em Letters in Mathematical
  Physics} {\bfseries 7} no.~1, (Jan., 1983) 25--32}.
  \url{http://link.springer.com/10.1007/BF00398708}.

\bibitem{ArakiMasuda1982}
H.~Araki and T.~Masuda, ``{Positive Cones and $L_p$‑Spaces for von Neumann
  Algebras},'' \href{http://dx.doi.org/10.2977/prims/1195183577}{{\em
  Publications of the Research Institute for Mathematical Sciences, Kyoto
  University} {\bfseries 18} no.~2, (1982) 339--411}.

\bibitem{PositiveConesLSpaces}
H.~KOSAKI, ``Positive cones and l p -spaces associated with a von neumann
  algebra,'' {\em Journal of Operator Theory} {\bfseries 6} no.~1, (1981)
  13--23. \url{http://www.jstor.org/stable/24713811}.

\bibitem{Takesaki:1970aki}
M.~Takesaki, \href{http://dx.doi.org/10.1007/bfb0065832}{{\em {Tomita's Theory
  of Modular Hilbert Algebras and its Applications}}}.
\newblock Lecture Notes in Mathematics. Springer-Verlag, 1970.

\bibitem{buchholz1986noether}
D.~Buchholz, S.~Doplicher, and R.~Longo, ``On noether's theorem in quantum
  field theory,'' {\em Annals of Physics} {\bfseries 170} no.~1, (1986) 1--17.

\bibitem{Hollands:2020owv}
S.~Hollands, ``{Variational approach to relative entropies with an application
  to QFT},'' \href{http://dx.doi.org/10.1007/s11005-021-01474- 2}{{\em Lett.
  Math. Phys.} {\bfseries 111} no.~6, (2021) 136},
  \href{http://arxiv.org/abs/2009.05024}{{\ttfamily arXiv:2009.05024
  [quant-ph]}}.

\bibitem{Faulkner:2020iou}
T.~Faulkner, S.~Hollands, B.~Swingle, and Y.~Wang, ``{Approximate Recovery and
  Relative Entropy I: General von Neumann Subalgebras},''
  \href{http://dx.doi.org/10.1007/s00220-021-04143-6}{{\em Commun. Math. Phys.}
  {\bfseries 389} no.~1, (2022) 349--397},
  \href{http://arxiv.org/abs/2006.08002}{{\ttfamily arXiv:2006.08002
  [quant-ph]}}.

\bibitem{Kosaki1981}
H.~Kosaki, ``{Remarks on positive cones associated with a von Neumann
  algebra},'' \href{http://dx.doi.org/10.2748/tmj/1178229358}{{\em Tohoku
  Mathematical Journal, Second Series} {\bfseries 33} no.~4, (1981) 587--591}.

\bibitem{Watrous_2018}
J.~Watrous, {\em The Theory of Quantum Information}.
\newblock Cambridge University Press, 2018.

\bibitem{Witten:2018zxz}
E.~Witten, ``{APS Medal for Exceptional Achievement in Research: Invited
  article on entanglement properties of quantum field theory},''
  \href{http://dx.doi.org/10.1103/RevModPhys.90.045003}{{\em Rev. Mod. Phys.}
  {\bfseries 90} no.~4, (2018) 045003},
  \href{http://arxiv.org/abs/1803.04993}{{\ttfamily arXiv:1803.04993
  [hep-th]}}.

\bibitem{StratilaZsido1979}
Şerban Strătilă and L.~Zsid\'o, {\em Lectures on von Neumann algebras}.
\newblock Editura Academiei and Abacus Press, Bucharest, Romania; Tunbridge
  Wells, England, 1979.
\newblock Revised translation of the original 1975 Romanian edition.

\bibitem{kadison_ringrose_vol2}
R.~V. Kadison and J.~R. Ringrose, {\em Fundamentals of the Theory of Operator
  Algebras, Volume II: Advanced Theory}, vol.~100 of {\em Pure and Applied
  Mathematics}.
\newblock Academic Press, 1986.

\bibitem{Nielsen:2012yss}
M.~A. Nielsen and I.~L. Chuang,
  \href{http://dx.doi.org/10.1017/cbo9780511976667}{{\em {Quantum Computation
  and Quantum Information}}}.
\newblock Cambridge University Press, 6, 2012.

\bibitem{alberti2000bures}
P.~M. Alberti and A.~Uhlmann, ``On bures distance and*-algebraic transition
  probability between inner derived positive linear forms over w*-algebras,''
  {\em Acta Applicandae Mathematica} {\bfseries 60} (2000) 1--37.

\bibitem{araki1972bures}
H.~Araki, ``Bures distance function and a generalization of sakai’s
  non‐commutative radon–nikodym theorem,''
  \href{http://dx.doi.org/10.2977/PRIMS/1195193113}{{\em Publications of the
  Research Institute for Mathematical Sciences} {\bfseries 8} no.~2, (Aug,
  1972) 335--362}.

\bibitem{araki1964neumann}
H.~Araki, ``Von neumann algebras of local observables for free scalar field,''
  {\em Journal of Mathematical Physics} {\bfseries 5} no.~1, (1964) 1--13.

\bibitem{Araki:1963klf}
H.~Araki, ``{A Lattice of Von Neumann Algebras Associated with the Quantum
  Theory of a Free Bose Field},''
  \href{http://dx.doi.org/10.1063/1.1703912}{{\em J. Math. Phys.} {\bfseries 4}
  no.~11, (1963) 1343}.

\bibitem{Casini:2019qst}
H.~Casini, S.~Grillo, and D.~Pontello, ``{Relative entropy for coherent states
  from Araki formula},''
  \href{http://dx.doi.org/10.1103/PhysRevD.99.125020}{{\em Phys. Rev. D}
  {\bfseries 99} no.~12, (2019) 125020},
  \href{http://arxiv.org/abs/1903.00109}{{\ttfamily arXiv:1903.00109
  [hep-th]}}.

\bibitem{Huerta:2022tpq}
M.~Huerta and G.~van~der Velde, ``{Modular Hamiltonian of the scalar in the
  semi infinite line: dimensional reduction for spherically symmetric
  regions},'' \href{http://dx.doi.org/10.1007/JHEP06(2023)097}{{\em JHEP}
  {\bfseries 06} (2023) 097}, \href{http://arxiv.org/abs/2301.00294}{{\ttfamily
  arXiv:2301.00294 [hep-th]}}.

\bibitem{Srednicki:1993im}
M.~Srednicki, ``{Entropy and area},''
  \href{http://dx.doi.org/10.1103/PhysRevLett.71.666}{{\em Phys. Rev. Lett.}
  {\bfseries 71} (1993) 666--669},
\href{http://arxiv.org/abs/hep-th/9303048}{{\ttfamily arXiv:hep-th/9303048
  [hep-th]}}.

\bibitem{Hislop:1981uh}
P.~D. Hislop and R.~Longo, ``{Modular Structure of the Local Algebras
  Associated With the Free Massless Scalar Field Theory},''
\href{http://dx.doi.org/10.1007/BF01208372}{{\em Commun. Math. Phys.}
  {\bfseries 84} (1982) 71}.

\bibitem{Casini:2011kv}
H.~Casini, M.~Huerta, and R.~C. Myers, ``{Towards a derivation of holographic
  entanglement entropy},''
  \href{http://dx.doi.org/10.1007/JHEP05(2011)036}{{\em JHEP} {\bfseries 05}
  (2011) 036},
\href{http://arxiv.org/abs/1102.0440}{{\ttfamily arXiv:1102.0440 [hep-th]}}.

\bibitem{HandbookIntegralEquations}
A.~D. Polyanin and A.~V. Manzhirov, {\em Handbook of Integral Equations: Second
  Edition}.
\newblock Chapman and Hall/CRC, Boca Raton, FL, 2008.
\newblock \url{https://doi.org/10.1201/9781420010558}.

\bibitem{Akers:2023obn}
C.~Akers, T.~Faulkner, S.~Lin, and P.~Rath, ``{Entanglement of purification in
  random tensor networks},''
  \href{http://dx.doi.org/10.1103/PhysRevD.109.L101902}{{\em Phys. Rev. D}
  {\bfseries 109} no.~10, (2024) L101902},
  \href{http://arxiv.org/abs/2306.06163}{{\ttfamily arXiv:2306.06163
  [hep-th]}}.

\bibitem{Chen:2025cga}
L.~Chen, ``{R{\'e}nyi entanglement of purification and half R{\'e}nyi reflected
  entropy in free scalar theory},''
  \href{http://dx.doi.org/10.1007/JHEP06(2025)045}{{\em JHEP} {\bfseries 06}
  (2025) 045}, \href{http://arxiv.org/abs/2501.10944}{{\ttfamily
  arXiv:2501.10944 [hep-th]}}.

\bibitem{Dutta:2019gen}
S.~Dutta and T.~Faulkner, ``{A canonical purification for the entanglement
  wedge cross-section},'' \href{http://dx.doi.org/10.1007/JHEP03(2021)178}{{\em
  JHEP} {\bfseries 03} (2021) 178},
  \href{http://arxiv.org/abs/1905.00577}{{\ttfamily arXiv:1905.00577
  [hep-th]}}.

\bibitem{Caputa:2018xuf}
P.~Caputa, M.~Miyaji, T.~Takayanagi, and K.~Umemoto, ``{Holographic
  Entanglement of Purification from Conformal Field Theories},''
  \href{http://dx.doi.org/10.1103/PhysRevLett.122.111601}{{\em Phys. Rev.
  Lett.} {\bfseries 122} no.~11, (2019) 111601},
  \href{http://arxiv.org/abs/1812.05268}{{\ttfamily arXiv:1812.05268
  [hep-th]}}.

\bibitem{Hirai:2018jwy}
H.~Hirai, K.~Tamaoka, and T.~Yokoya, ``{Towards Entanglement of Purification
  for Conformal Field Theories},''
  \href{http://dx.doi.org/10.1093/ptep/pty063}{{\em PTEP} {\bfseries 2018}
  no.~6, (2018) 063B03}, \href{http://arxiv.org/abs/1803.10539}{{\ttfamily
  arXiv:1803.10539 [hep-th]}}.

\bibitem{PhysRevD.111.L021902}
X.~Jiang, P.~Wang, H.~Wu, and H.~Yang, ``Alternative to purification in
  conformal field theory,''
  \href{http://dx.doi.org/10.1103/PhysRevD.111.L021902}{{\em Phys. Rev. D}
  {\bfseries 111} (Jan, 2025) L021902}.
  \url{https://link.aps.org/doi/10.1103/PhysRevD.111.L021902}.

\bibitem{Jiang:2024xqz}
X.~Jiang, P.~Wang, H.~Wu, and H.~Yang, ``{Realization of ''ER=EPR''},''
  \href{http://arxiv.org/abs/2411.18485}{{\ttfamily arXiv:2411.18485
  [hep-th]}}.

\bibitem{Casini:2009sr}
H.~Casini and M.~Huerta, ``{Entanglement entropy in free quantum field
  theory},'' \href{http://dx.doi.org/10.1088/1751-8113/42/50/504007}{{\em J.
  Phys.} {\bfseries A42} (2009) 504007},
\href{http://arxiv.org/abs/0905.2562}{{\ttfamily arXiv:0905.2562 [hep-th]}}.

\bibitem{balian_nonunitary_1969}
R.~Balian and E.~Brezin, ``Nonunitary bogoliubov transformations and extension
  of {Wick}’s theorem,'' \href{http://dx.doi.org/10.1007/BF02710281}{{\em Il
  Nuovo Cimento B Series 10} {\bfseries 64} no.~1, (Nov., 1969) 37--55}.
  \url{http://link.springer.com/10.1007/BF02710281}.

\end{thebibliography}\endgroup

\end{document}